%% file: LinScaloHesselink_JFM_2016.tex
\documentclass{jfm}
\pdfoutput=1

\usepackage{hyperref}

\usepackage{cancel}
\usepackage{multirow}
\usepackage{amsmath}
\usepackage{amssymb}
\usepackage{natbib}
\usepackage{float}
\usepackage{color}
\usepackage{pbox}

\usepackage{subcaption}
\newcommand{\labelphantom}[1]{
	\parbox{0pt}{\phantomsubcaption\label{#1}}
}

\usepackage{cleveref}
\crefformat{figure}{figure~#2#1#3}
\Crefformat{figure}{Figure~#2#1#3}
\crefmultiformat{figure}{figures~#2#1#3}{ and~#2#1#3}{, #2#1#3}{ and~#2#1#3}
\Crefmultiformat{figure}{Figures~#2#1#3}{ and~#2#1#3}{, #2#1#3}{ and~#2#1#3}
\crefformat{equation}{equation~#2#1#3}
\Crefformat{equation}{Equation~#2#1#3}
\crefformat{section}{\S#2#1#3}
\Crefformat{section}{\S#2#1#3}
\crefmultiformat{section}{\S#2#1#3}{ and \S#2#1#3}{, \S#2#1#3}{, and \S#2#1#3}
\Crefmultiformat{section}{\S#2#1#3}{ and \S#2#1#3}{, \S#2#1#3}{, and \S#2#1#3}

\usepackage{subfloat}
\usepackage[normalem]{ulem}

\newcommand{\charlesx}{CharLES$^X$}
\newcommand{\deltaec}{\textsc{DeltaEC}}
\newcommand{\numofoscillators}{n_o}
\newcommand{\Ztarget}{Z_{exp}}
\newcommand{\Wthtarget}{\widehat{\widetilde{W}}_{\omega, exp}}
\newcommand{\defaultgrid}{grid-resolution/stack-type C/I}
\newcommand{\verticallines}{Vertical dashed lines indicate locations of abrupt area change (\cref{fig:computational_setup}).}
\newcommand{\shadedareaexplanation}{The shaded area highlights the frequency interval $f>450$ Hz of negative resistance $\Real\left\{Z_{exp}(\omega)\right\}<0$ for the experimentally determined impedance \eqref{eq:experimental_Z}, deemed unphysical.}
\newcommand*\conj[1]{#1^{*}}

\usepackage{tikz}
\usetikzlibrary{shapes}
\usetikzlibrary{plotmarks}

\definecolor{darkgray}{rgb}{0.5, 0.5, 0.5}
\definecolor{lightgray}{rgb}{0.8, 0.8, 0.8}

\newcommand{\legendline}{(\tikz[baseline=-0.75ex]{ \draw[very thick]  (0,0) -- (2ex,0);})}

\newcommand{\legenddots}{(\tikz[baseline=-0.3ex]{
		\draw(0.25ex,0.25ex) circle (0.5ex);
		})}
\newcommand{\legenddotsblack}{(\tikz[baseline=-0.3ex]{
		\draw[fill=black](0.25ex,0.25ex) circle (0.25ex);
		})}

\newcommand{\legenddotskandv}{(\tikz[baseline=-0.4ex]{
		\hspace{-0.00cm}
		\draw[fill=black](0.0ex,0.25ex) circle (0.5ex);
		\node[rotate=0] at (0.0ex,-0.25ex) {\pgfuseplotmark{triangle*}};
	}\hspace{-0.0cm})}

\newcommand{\legenddashed}{(\hspace{-0.02cm}\tikz[baseline=-0.75ex, color=black, dashed, very thick]{
		\draw  (0,0) -- (3.9ex,0);
		}\hspace{-0.03cm})}
\newcommand{\legenddashedgray}{(\hspace{-0.02cm}\tikz[baseline=-0.75ex, color=lightgray, dashed, very thick]{
		\draw  (0,0) -- (3.9ex,0);
	}\hspace{-0.03cm})}

\newcommand{\legenddashedLong}{( \tikz[baseline=-0.75ex, color=black, dashed, very thick]{ \draw[dash pattern=on 10pt off 10pt]   (0,0) -- (7ex,0); } )}
\newcommand{\legenddashedMedium}{( \tikz[baseline=-0.75ex, color=black, dashed, very thick]{ \draw[dash pattern=on 7.5pt off 5pt]  (0,0) -- (5ex,0); } )}
\newcommand{\legenddashedShort}{( \tikz[	baseline=-0.75ex, color=black, dashed, very thick]{ \draw[dash pattern=on 4pt off 2.5pt]  (0,0) -- (4.0ex,0); } )}
\newcommand{\legenddashedShortest}{( \tikz[	baseline=-0.75ex, color=black, dashed, very thick]{ \draw[dash pattern=on 2pt off 0.5pt]  (0,0) -- (4.0ex,0); } )}

\newcommand{\legendtriangles}{(\tikz[baseline=-0.4ex]{
		\hspace{-0.00cm}
		\node[rotate=180] at (0.0ex,0.25ex) {\pgfuseplotmark{triangle*}};
		}\hspace{-0.00cm})}

\newcommand{\legendonewhitesquare}{(\tikz[baseline=-0.5ex]{
		\node[rotate=180,fill=blue,color=white,fill=black,draw=white,stroke=yellow] at (0.0ex,0.25ex) {\pgfuseplotmark{square*}};
	}\hspace{0.cm})}

\title{High-fidelity simulation of a standing-wave thermoacoustic-piezoelectric engine}

\shorttitle{High-fidelity simulation of a standing-wave thermoacoustic-piezoelectric engine}
\shortauthor{J. Lin, C. Scalo and L. Hesselink}

\author{Jeffrey Lin\aff{1}
	\corresp{\email{linjef@stanford.edu}},
	Carlo Scalo\aff{2}
	\and Lambertus Hesselink\aff{1}}

\affiliation{\aff{1}Department of Electrical Engineering,
	Stanford University, Stanford, CA 94305, USA
	\aff{2}School of Mechanical and Aeronautical Engineering,
	Purdue University, West Lafayette, IN 47907, USA}

\begin{document}

	\clearpage
	\newpage

	\maketitle

	\begin{abstract}

We have carried out wall-resolved unstructured fully-compressible Navier--Stokes simulations of a complete standing-wave thermoacoustic piezoelectric (TAP) engine model inspired by the experimental work of \citet{SmokerNAB_2012}. The model is axisymmetric and comprises a 51 cm long resonator divided into two sections: a small diameter section
enclosing a thermoacoustic stack, and a larger diameter section 
capped by a piezoelectric diaphragm tuned to the thermoacoustically amplified mode (388 Hz). 
The diaphragm is modelled with multi-oscillator broadband time-domain impedance boundary conditions (TDIBCs), providing higher fidelity over single-oscillator approximations. Simulations are first carried out to the limit cycle without energy extraction.
The observed growth rates are shown to be grid-convergent and are verified against a numerical dynamical model based on Rott's theory. 
The latter is based on a staggered grid approach and allows jump conditions in the derivatives of pressure and velocity in sections of abrupt area change and the inclusion of linearized minor losses. 
The stack geometry maximizing the growth rate is also found. At the limit cycle, thermoacoustic heat leakage and frequency shifts are observed, consistent with experiments. Upon activation of the piezoelectric diaphragm, steady acoustic energy extraction and a reduced pressure amplitude limit cycle are obtained. A heuristic closure of the limit cycle acoustic energy budget is presented, supported by the linear dynamical model and the nonlinear simulations. 
The developed high-fidelity simulation framework provides accurate predictions of thermal-to-acoustic and acoustic-to-mechanical energy conversion (via TDIBCs), enabling a new paradigm for the design and optimization of advanced thermoacoustic engines.
\\

	\end{abstract}


\input{./Introduction.tex}

\input{./ProblemDescription.tex}

\input{./LinearModel.tex}

\input{./Results_StartUp.tex}

\input{./Results_PiezoTDIBCModel.tex}

\input{./Results_EnergyExtraction.tex}

\section{Conclusions}

We have presented compressible unstructured Navier--Stokes simulations of a complete standing-wave thermoacoustic piezoelectric (TAP) engine model inspired by the experimental work of \citet{SmokerNAB_2012}. Thermal and viscous boundary layers are resolved everywhere in the model and piezoelectric acoustic energy absorption is introduced and modelled with a multi-oscillator time-domain impedance boundary condition (TDIBC). The complete numerical model demonstrates the first known attempt, to the authors' knowledge, at modelling piezoelectric energy extraction in a high-fidelity Navier--Stokes simulation of a thermoacoustic engine. The goal is to advance computational tools for the simulation of realistic thermoacoustic engines, capturing with high-fidelity both acoustic energy production/dissipation mechanisms and direct power extraction. These two components are crucial for design and optimization of thermoacoustic engines.

The TAP engine model is analysed first in the start-up phase without acoustic energy absorption. Linear growth rate during the start-up phase compares favourably with Rott's linear theory. A new set of linearized equations for the axisymmetric thermoacoustic stack geometry was derived and demonstrated very good matching of frequencies and growth rates despite inherent limitations and assumptions, such as the neglect of edge effects due to complex geometrical features and nonlinear effects such as flow separation.

The linear stability model was used to explore the parameter space of annular stack geometries. Very strong dependence on the stack geometrical parameters for both operating frequency and growth rate was found in the linear stability analysis and congruently verified in high-fidelity Navier--Stokes calculations.  For constant stack layer density, the frequency of the thermoacoustically amplified mode decreases with increasing porosity; however, an optimal porosity maximizing transient growth rates exists between the limits of zero and 100\% porosity. The maximal growth rate also increases as stack layer density increases, implying high thermal contact is favourable for high-amplitude thermoacoustic engines. 

Analysis of the stack has shown that in thermoacoustic excitation, the first mode is dominant. Issues of growth rate sensitivity to the grid resolution are analysed, showing lower-than-expected order of convergence of the growth rates. This is attributed to grid stretching and quality and the inherent accumulation of error in measuring the growth rate. The growth rate extracted on the finest grid adopted, as used for the presented results, is within the error band of the Richardson extrapolation to zero-grid spacing.

Simulations were carried out with hard-wall boundary conditions until a limit cycle is obtained. Entrance effects, particularly into the stack, and thermoacoustic streaming are observed. These effects increase the operating frequency and reduce the effective temperature gradient in the engine, explaining experimental temperature observations. TDIBC-based acoustic energy extraction is then introduced, leading to a second, reduced limit cycle. 

The time-domain impedance formulation by \citet{FungJ_2004} has been used to derive an appropriate causal impedance for both a single- and multi-oscillator impedance model. This approach does not correspond to the time-domain impedance boundary condition implementation proposed by \citet{TamA_1996}, which neglects issues of causality. A single-oscillator impedance model was shown to be insufficient in capturing the experimental value of the impedance. A multi-oscillator model has shown higher degrees of fidelity. Constraints on the impedance were discussed, and the increasing fidelity of fitting with a greater number of basis frequencies was demonstrated.

The adopted numerical model allows for both the evaluation of the nonlinear effects of scaling and the effect of a fully electromechanically-coupled impedance boundary condition (IBC), representative of a piezoelectric element with variable resistance and time-variable power production. The construction of a simulation-ready IBC from experimental data was completed to the best of the authors' ability and its limitations and restrictions are reported. Because the experimentally-measured coefficients may not have been fully broadband and because of heat leakage, a shift in the engine operating frequency in the simulations was found. The shift in frequency is understood to be partly due to an attraction to a more numerically-compliant domain of the TDIBC. Linear scaling of power input and power output show congruency with experimentally-reported pressure and velocity profiles and power extraction results. While the numerical stack design was chosen to correspond to experimentally-reported porosity and hydraulic radius, geometrical differences in the experimental stack and the numerical stack lead to differing critical temperature ratios and thus different power output to temperature input ratios. The TDIBC, as constructed, results in acoustic energy output values which are consistent with experimentally-published results by \citet{SmokerNAB_2012}.

Solution functionals, including growth rate, limit cycle operation, and energy distribution and extraction, are otherwise consistent with experimental results and are self-consistent between Navier--Stokes simulations and linear theory predictions, demonstrating the presented models' predictive capabilities. Optimization of scaling and the impedance can be simultaneously applied; the Navier--Stokes numerical technique as demonstrated is suitable for studying high-frequency, reduced-footprint engines, a regime traditionally difficult to model with linear thermoacoustics. The present work improves upon \citet{ScaloJFM2013} by resolving heat transfer and drag in the thermoacoustic core, allowing for acoustic energy extraction, demonstrating consistency with experimental results, and extending the modelling framework to standing-wave engines. The presented standing-wave engine model demonstrates behaviour indicating hysteresis, which was not observed in travelling-wave engine models. Expected future work include the use of the model for analysing micro-TAEs and the high frequency measurement of piezoelectric diaphragm transmittance coefficients, as the reference electromechanical response is unphysical at high frequencies. 
\\

\section*{Acknowledgements}
Jeffrey Lin and Carlo Scalo acknowledge the support of the Inventec Stanford Graduate Fellowship and the Precourt Energy Efficiency Center Seed Grant at Stanford University. The authors wish to thank Prateek Gupta for rederiving the linear stability analysis equations and contributing to the write-up of Appendix A. The authors also acknowledge the generous computational allocation provided to Dr. Scalo on Purdue's latest supercomputing architecture, Rice, and the technical support of Purdue's Rosen Center for Advanced Computing (RCAC).

\appendix
\input{./Appendix.tex}

\bibliographystyle{jfm}
\bibliography{references_2,references_Carlo}

\end{document}

%% file: Introduction.tex
\section{Introduction}
\label{sec:introduction}

Thermoacoustic engines (TAEs) are devices capable of converting external heat sources into acoustic power, which in turn can be converted to mechanical or electrical power. TAEs do not require moving parts and are inherently thermoacoustically unstable if supplied with a critical heat input -- past which, an initial perturbation is sufficient to start generating acoustic power. The acoustic nature of the wave energy propagation in TAEs guarantees close-to-isentropic stages in the overall energy conversion process, promoting high efficiencies. One of the most advanced TAE reported in the available literature achieves a thermal-to-acoustic energy conversion efficiency of 32\%, corresponding to 49\% of Carnot's theoretical limit \citep{TijaniS_2011}. There are a variety of TAEs in use, with varying sizes, and heat sources and energy extraction strategies~\citep{Swift_1988}.

In any TAE, two key energy conversion processes are involved: thermal-to-acoustic and acoustic-to-electric.
Thermal-to-acoustic conversion mechanisms are fluid dynamic in nature and are well-understood and predictable at various levels of fidelity, from quasi one-dimensional linear acoustics~\citep{Rott_1980_AdvApplMech} to fully compressible three-dimensional Navier--Stokes models~\citep{ScaloLH_JFM_2015}.
High-fidelity modelling of acoustic energy extraction in the context of Navier--Stokes simulations has received limited attention.
In the following, we demonstrate a computational modelling strategy to simulate both processes concurrently with high fidelity.

\citet{Sondhauss_AnnPhys_1850} was the first to experimentally investigate the spontaneous generation of sound in the process of glassblowing. \citet{Rijke_1859} showed that sound is produced when heating a wire gauze within a vertically-oriented tube open at both ends. \citet{Rayleigh_Nature_1878} qualitatively reasoned the criterion for the thermoacoustic production of sound to explain both the Sondhauss tube and the Rijke tube. Building upon Rayleigh's seminal intuition, it can be stated that an appropriate phasing between fluctuations of velocity, pressure, and heat release is at the core of thermoacoustic instability (and hence energy conversion): velocity oscillations, in the presence of a background mean temperature gradient (typically sustained by an external heat source), create fluctuations in heat release that, if in phase with pressure oscillations, lead to thermoacoustic energy production via a work-producing thermodynamic cycle.

Sondhauss and Rijke's work inspired research efforts aimed at the technological application of thermoacoustic energy conversion. \citet{Hartley_1951} patented a thermoacoustic generator using a telephone receiver as an energy extractor.
In particular, the adoption of a piezoelectric element was suggested together with electric timing to maintain the desired thermoacoustic phasing.
\citet{Marrison_1958} developed a TAE
aimed at increasing the effectiveness of telephone repeaters. \citet{FeldmanJr_JournalSoundVibration_1968} was the first to introduce the thermoacoustic stack, noting that it simultaneously serves as a thermal regenerator, an insulator, and an acoustic impedance,
helping obtain the optimal phasing between pressure and velocity oscillations for thermoacoustic energy production.

Modern research has been focused on achieving conversion efficiencies comparable to theoretical expectations. 
\citet{Ceperley_1979} realized that the thermodynamic cycle induced by purely travelling waves is composed of clearly separated stages of compression, heating, expansion, and cooling, which are instead partly overlapped in standing waves,
leading in the latter case to a lower energy conversion efficiency. However, Ceperley was unsuccessful in developing a working travelling-wave TAE; the first practical realisation can be attributed to  \citet{YazakiIMT_PhysRevLett_1998}.
TAEs can therefore largely be classified into standing-wave and travelling-wave configurations, the latter being typically more efficient but more complicated to build. Hybrid configurations are also possible, with the two concepts combined in a cascaded system~\citep{GardnerS_JASA_2003}. 

A theoretical breakthrough was made possible by Rott and co-workers, who developed a comprehensive analytical predictive framework based on linear acoustics \citep{Rott_ZAMP_1969,Rott_ZAMP_1973,Rott_ZAMP_1974,Rott_ZAMP_1975,Rott_ZAMP_1976,Rott_NZZ_1976,ZouzoulasR_ZAMP_1976,Rott_1980_AdvApplMech,MullerR_ZAMP_1983,Rott_JFM_1984}, improving upon pre-existing theories by \citet{Kirchhoff_PoggAnn_1868} and \citet{Kramers_Physica_1949}. This theoretical framework, augmented with experimentally-derived heuristics, is at the core of engineering software tools such as \deltaec{} ~\citep{WardCS_2012} and Sage~\citep{Gedeon_2014}, which provide reliable predictions at the limit cycle for large, traditional TAEs operating in a low-acoustic-amplitude regime. Natural limitations of this approach include not accounting for effects of complex geometries, start-up transient behaviour, and hydrodynamic nonlinearities such as turbulence and unsteady boundary layer separation. High-fidelity simulations, while requiring a much greater computational cost, can very accurately model all of the aforementioned phenomena, allowing the computational study, design, and optimization of a new generation of TAEs---as well as informing more advanced, companion low-order models.

Previous high-fidelity efforts by \citet{ScaloLH_JFM_2015} demonstrated a full-scale three-dimensional simulation of a large TAE, revealing the presence of transitional turbulence and providing support for direct low-order modelling of acoustic nonlinearities such as Gedeon streaming. However, the complex porous geometry of the heat exchangers and regenerator was not resolved but instead modelled with highly parametrized source terms in the momentum and energy equations. Furthermore, no explicit energy extraction was considered. The present work improves upon both shortcomings while abandoning a three-dimensional configuration (i.e. not accounting for transitional turbulence). To aid the design of a realistic electricity-producing engine, accurate modelling of the electric power output within the context of high-fidelity prediction capabilities needs to be developed.

The conversion from acoustic power to electrical power is a severe efficiency bottleneck and a technological challenge. One option is that of linear alternators, which often have high impedances and suffer from seal losses in the gaps between the cylinder and the piston~\citep{YuJB_AppliedEnergy_2012}. Furthermore, linear alternators are by nature bulky and heavy.
An alternative strategy is to couple a piezoelectric diaphragm to a TAE, providing a hermetic seal and reducing losses. Piezoelectric energy extraction is particularly attractive for small-scale TAEs;
the maximal power output of a typical piezoelectric generator scales cubically with the
operating frequency, which is inversely proportional to the wavelength of the  thermoacoustically amplified mode. 
On the other hand, MEMS-constructed piezoelectric materials can be sensitive to high-frequency vibrations
\citep{AntonS_SmartMaterStruct_2007, Priya_JElectroceram_2007, ChenXYS_NanoLett_2010}.
Early suggestions of using piezoelectric energy extraction date back to \citet{Hartley_1951}'s proposed electric power source, with more recent theoretical and experimental investigations by \citet{MatveevWRS_2007} and \citet{SmokerNAB_2012}.

In this paper we present a high-fidelity fully compressible Navier--Stokes simulation of a thermoacoustic heat engine with a piezoelectric energy extraction device. Previous modelling efforts of piezoelectric energy extraction have been limited to linear acoustic solvers with impedance boundary conditions in the frequency domain. In the present work, the piezoelectric diaphragm is modelled with a multi-oscillator time-domain impedance boundary condition (TDIBC), building upon \citet{FungJ_2001,FungJ_2004} and following the implementation by \citet{ScaloBL_PoF_2015}. This approach guarantees physical admissibility and numerical stability of the solution by enforcing constraints such as causality and representation of the boundary as a passive element. The TAE model is inspired by the standing-wave thermoacoustic piezoelectric (TAP) engine experimentally investigated by \citet{SmokerNAB_2012}. This engine was chosen due to its simple design and the availability of experimentally measured electromechanical admittances of the piezoelectric diaphragm. This is a key stepping stone for the development of computational tools to better predict and optimize energy generation and extraction of high-performance, realistic TAEs.

In the following, the adopted theoretical TAP engine model is first introduced, together with the
governing equations and computational setup (\cref{sec:methodology}). A linear thermoacoustic model predicting the onset and growth of oscillations is presented, supporting and complementing the results from the  Navier--Stokes simulations (\cref{sec:lsamodel,sec:engine_transient_results}).
The effects of acoustic nonlinearities at the limit cycle are then analysed (\cref{sec:hw_limitcycle}). Finally, the modelling of the piezoelectric diaphragm with multi-oscillator TDIBCs is described (\cref{sec:modeling_a_physical_piezoelectric}), and results from the Navier--Stokes simulations with energy extraction are discussed (\cref{sec:energyextraction}).

%% file: ProblemDescription.tex
\section{Problem Description}
\label{sec:methodology}

\subsection{Engine Model Design}
\label{subsec:computationalsetup}

\begin{figure}
    \centering
       \includegraphics[width=\textwidth,keepaspectratio]{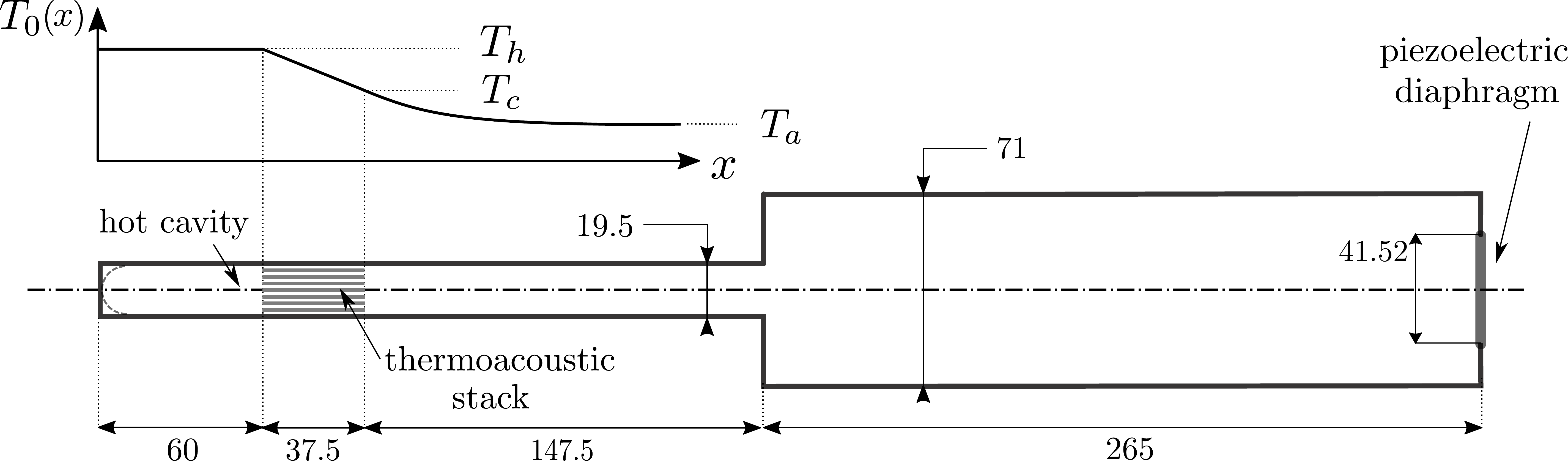}
    \caption{
    	Illustration of the axisymmetric TAP engine model (not to scale) inspired by the experimental work of ~\citet{SmokerNAB_2012}. All lengths are given in millimetres. The dashed lines in the hot cavity indicate the original experimental design. The axial distribution of mean temperature, $T_0(x)$, is qualitatively sketched. Different stack designs (table \ref{tab:stack_configurations}) and temperature settings (table \ref{tab:simulation_parameters}) have been considered in the simulations.
    }
    \label{fig:computational_setup}
\end{figure}

The TAP engine model (figure \ref{fig:computational_setup}) is 510 mm in length and is divided into two cylindrical, constant-area sections: one of 19.5 mm in diameter, enclosing an axisymmetric thermoacoustic stack (\cref{tab:stack_configurations}), and the other of 71 mm in diameter, capped by a piezoelectric diaphragm tuned to the thermoacoustically amplified mode ($388$~Hz) for maximization of acoustic energy extraction.
The TAP engine design is inspired by the design and experimental work of \citet{SmokerNAB_2012}.

An axisymmetric model cannot account for three-dimensional flow effects. 
The scope of this study is instead focused on the accurate modelling of thermoacoustic acoustic energy production (\cref{sec:linearmodeling} and \cref{sec:engine_transient_results}), nonlinear thermoacoustic transport (\cref{sec:hw_limitcycle}), and energy extraction (\cref{sec:energyextraction}), for which three-dimensional flow effects are secondary. Moreover, at the highest acoustic amplitude achieved in the present computations ($\simeq$ 6000 Pa), the Stokes Reynolds numbers based on the maximum centreline velocity amplitude in the device (at approximately $x$ = 245~mm) is $Re_{\delta_\nu}<$ 100, where $\delta_\nu$ is the Stokes boundary layer thickness \eqref{eq:viscous_thermal_bl}, falling well within the fully laminar regime of oscillatory boundary layers~\citep{Jensen1989JFM}. Even at significantly higher Reynolds numbers and acoustic amplitudes, such as the ones achieved in the three-dimensional calculations of a large travelling-wave engine by \citet{ScaloLH_JFM_2015}, hydrodynamic nonlinearities---such as Reynolds stresses associated with transition turbulence---were found to be negligible with respect to acoustic nonlinearities such as streaming and thermoacoustic transport.

In the experiments by \citet{SmokerNAB_2012}, a square-weave mesh-screen regenerator is used with porosity and hydraulic radius of $\phi=0.25$ and $r_h=0.34$ mm, respectively. The regenerator is heated on one side (in the hot cavity) by a resistive filament sustaining a hot temperature of $T_h = 790$ K, without a cold heat exchanger on the opposing side (Nouh, \emph{pers. comm.}). As a result, the mean axial temperature gradient weakens
throughout the course of the experiment due to conduction in the metal and thermoacoustic transport in the pore volume.

The thermoacoustic stack in our theoretical axisymmetric TAP engine model is composed of coaxial cylindrical annuli (table \ref{tab:stack_configurations}) with a linear axial wall-temperature profile (from $T_h$ to $T_c$) imposed via isothermal boundary conditions. This choice allows for the direct application of Rott's theory for verification of growth rates and frequencies observed during the start-up phase of the Navier--Stokes simulations, along with a clear definition of the geometrical parameter space for the exploration of the optimal stack design.

Three stack configurations have been investigated (\cref{tab:stack_configurations}) and were obtained by varying two parameters: 1) the number of coaxial solid annuli, $n_s$, surrounding a central solid rod of $h_s/2$ in radius; and 2) the ratio of the solid annulus thickness to the annular gap width,
$h_s/h_g$. Following the geometrical constraint
\begin{equation}
  R_{stk} = \left(n_s+1\right){h}_g + \left(\frac{1}{2}+n_s\right){h}_s
\end{equation}
where $R_{stk}=19.5$ mm (radius of the small-diameter section enclosing the thermoacoustic stack), unique values for $h_s$ and $h_g$ were determined for given values of $n_s$ and $h_s/h_g$. Volume porosity, $\phi$, and hydraulic radius, $r_h$, are calculated as
\begin{equation}
  \phi = \frac{A_g}{A_g + A_s},\qquad r_h = \frac{V_g}{S_{heat}}
  \label{eq:stack_design_constraints}
\end{equation}
where $V_g$ is the total gas-filled volume in the stack, $S_{heat}$ is the gas-solid contact surface through which wall-heat transfer occurs, $A_g$ is the cross-sectional area available to the gas, and $A_g+A_s=A_{stk}=\pi R_{stk}^2 $ where $A_s$ is the cross-sectional area occupied by the solid.

Stack I has been designed by selecting a combination of $n_s$ and ${h}_s/{h}_g$ resulting in a porosity and hydraulic radius close to the values of the mesh-wire regenerator of ~\citet{SmokerNAB_2012}. Stack II is characterized by a higher porosity with respect to Stack I, without significant differences in the hydraulic radius. Stack III has been designed by imposing ${h}_s = {h}_g$ and $n_s=3$, resulting in a more porous and regularly-spaced stack, and allowing for the formation of an inviscid acoustic core in the annular gap (missing in Stack I and II, see \cref{tab:stack_configurations}),
at the expense of thermal contact
(see discussion in \cref{sec:optimal_stack_design}).

\begin{table}
{

	\begin{minipage}{0.59\linewidth}
	\vspace*{-0.5cm}
		\includegraphics[width=\textwidth,keepaspectratio]{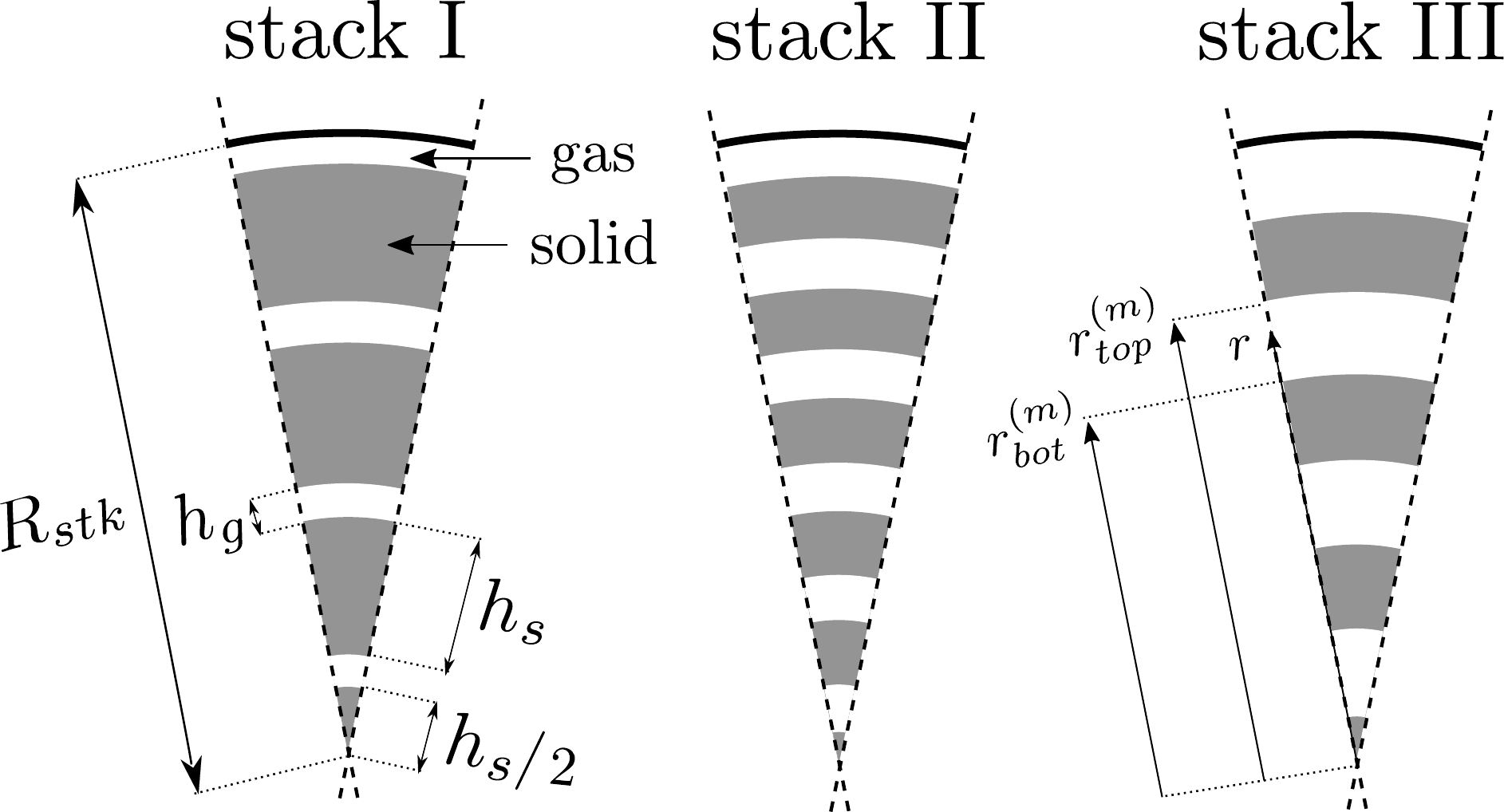}
	\end{minipage}
	\begin{minipage}{0.25\linewidth}
		\includegraphics[width=\textwidth,keepaspectratio]{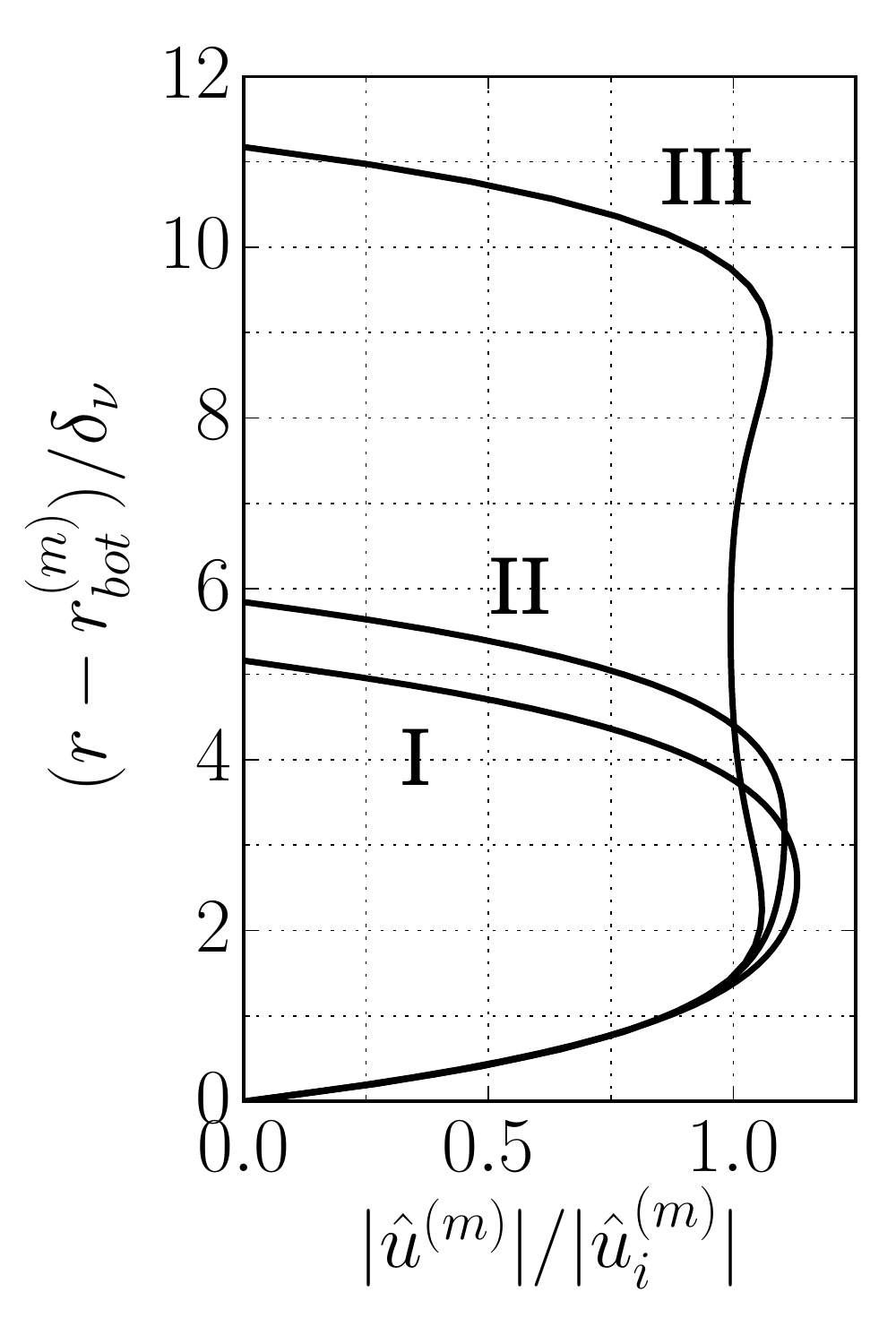}
	\end{minipage}
	\vspace*{-0.2cm}
}
	\centering
	\begin{tabular}{c|cc|cccc}
		\hline
		Stack Type & $\;\,n_s$  & ${h}_s/{h}_g$ & $\phi$ & $r_h$ (mm)  & ${h}_s$  (mm) & ${h}_g$   (mm) \\
		\hline
		I & 3 & 3.5 & 0.273  & 0.30 & 2.1 & 0.6 \\
		II & 5 & 1.5 & 0.443 & 0.342 & 1.03 & 0.68 \\
		III & 3 & 1 & 0.569  & 0.65 & 1.3 & 1.3  \\
	\end{tabular}
	  	\caption{
	  		Geometrical parameters for stack types I, II and III used in the Navier--Stokes simulations. Values of porosity, $\phi$, hydraulic radius, $r_h$, solid annuli thickness, ${h}_s$, and gap width, ${h}_g$, are calculated based on \cref{eq:stack_design_constraints} for a given ratio ${h}_s/{h}_g$ and a given number of solid annuli, $n_s$, surrounding a central rod of radius $h_s / 2$. Experimental values of porosity and hydraulic radius used by \citet{SmokerNAB_2012} are $\phi=0.25$, and $r_h=0.34$ mm. Profiles of axial-velocity magnitude as predicted by Rott's theory \eqref{eq:uhat_profile} normalized with their inviscid acoustic counterpart are plotted for reference values of density and viscosity.
	  	}
	\label{tab:stack_configurations}
\end{table}

Five different temperature settings have been considered in the Navier--Stokes simulations (table \ref{tab:simulation_parameters}), bracketing values observed in the experiments~\citep{SmokerNAB_2012,NouhAB_2014}, ranging from a close-to-critical (case 1) to a very strong thermoacoustic response (case 5), the latter corresponding to a temperature gradient that might be challenging to sustain experimentally. A linear temperature profile, ranging from hot, $T_h$, to cold, $T_c$, is imposed on the thermoacoustic stack walls; no-slip isothermal boundary conditions corresponding to ambient conditions $T_a=300$~K are imposed everywhere else, with the exception of the left and right end (including the piezoelectric diaphragm, if applicable), which are kept adiabatic.

\subsection{Governing Equations}

\label{subsec:governingequations}

The conservation equations for mass, momentum, and total energy, solved in the fully compressible Navier--Stokes simulations of the TAP engine model are, respectively, 
\begin{subequations}
	\label{eq:navierstokes}
	\begin{align}
		\frac{\partial}{\partial t} \left(\rho\right) &+ \frac{\partial}{\partial x_j} \left(\rho u_j \right)  = 0
		\label{subeq:ns1}
		\\
		\frac{\partial}{\partial t} \left(\rho u_i\right) &+ \frac{\partial}{\partial x_j} \left(\rho u_i u_j\right)  =  -\frac{\partial}{\partial x_i} p  +
		\frac{\partial}{\partial x_j} \tau_{ij}
		\label{subeq:ns2}
		\\
		\frac{\partial}{\partial t} \left(\rho \, E\right) &+ \frac{\partial}{\partial x_j} \left[ u_j \left(\rho \, E + p \right) \right] =
		\frac{\partial}{\partial x_j } \left(u_i \tau_{ij} - q_j\right)
		\label{subeq:ns3}
	\end{align}
\end{subequations}
where $x_1$, $x_2$, and $x_3$ (equivalently, $x$, $y$, and $z$) are axial and cross-sectional coordinates, $u_i$ are the velocity components in each of those directions, and $p$, $\rho$, and $E$ are respectively pressure, density, and total energy per unit mass. The gas is assumed to be ideal, with equation of state $p= \rho \,R_{gas}\, T$ and a constant ratio of specific heats, $\gamma$. The gas constant is fixed and calculated as $R_{gas} = p_{\textrm{ref}} \left(T_{\textrm{ref}}\,\rho_{\textrm{ref}}\right)^{-1}$, based on the reference thermodynamic density, pressure, and temperature, $\rho_{\textrm{ref}}$, $p_{\textrm{ref}}$, and $T_{\textrm{ref}}$, respectively.
The viscous and conductive heat fluxes are:
\begin{subequations}
	\label{eq:heatfluxes}
	\begin{eqnarray}
		\tau_{ij} &=& 2 \mu \left[S_{ij} - \frac{1}{3} \frac{\partial u_k}{\partial x_k} \delta_{ij} \right]\\
		\label{subeq:hf1}
		q_j &=& -\frac{\mu\,C_p}{\Pran} \frac{\partial}{\partial x_j} T
		\label{subeq:hf2}
	\end{eqnarray}
\end{subequations}
where $S_{ij}$ is the strain-rate tensor, given by $S_{ij}=(1/2) \left(\partial u_j/\partial x_i + \partial u_i /\partial x_j \right)$; $\Pran$ is the Prandtl number; and $\mu$ is the dynamic viscosity, given by $\mu = \mu_{\textrm{ref}}\left(T/T_\textrm{ref}\right)^n$, where $n$ is the viscosity power-law exponent and $\mu_{\textrm{ref}}$ is the reference viscosity. Simulations have been carried out with the following gas properties: $\gamma=1.4$, $\rho_{\textrm{ref}} = 1.2\,\textrm{kg m}^{-3}$, $p_{\textrm{ref}} = 101\,325\,\textrm{Pa}$, $T_{\textrm{ref}}=300\,\textrm{K}$, $\mu_{\textrm{ref}}=1.98\times10^{-5} \, \textrm{kg}\, \textrm{m}^{-1} \textrm{s}^{-1}$,  $\Pran=0.72$, and $n=0.76$, valid for air~\citep{DeYiB_InternationalJournalHeatMassTransfer_1990}.

No-slip and isothermal boundary conditions are used on all axial boundaries in the model. 
Direct acoustic energy extraction is only allowed from the piezoelectric diaphragm (figure~\ref{fig:computational_setup}),  modelled with impedance boundary conditions
\begin{equation}
	\hat{p}(\omega) = Z(\omega) \hat{u}(\omega)
	\label{eq:Z_definition}
\end{equation}
formulated in the time domain following the numerical implementation by \citet{ScaloBL_PoF_2015}, summarized in \cref{app:convolutionintegral}. The broadband (dimensional) impedance $Z(\omega)$ is derived by collapsing the experimentally-determined two-port electromechanical admittance matrix for the piezoelectric element and fitting the resulting impedance with a multi-oscillator approach \citep{FungJ_2001} as discussed in detail in \cref{sec:modeling_a_physical_piezoelectric}. In the present work, the characteristic specific acoustic impedance 
\begin{equation}\label{eq:base_impedance}
Z_0 = \rho_0\,a_0
\end{equation}
is absorbed within the value of the impedance in \eqref{eq:Z_definition}; hence, \eqref{eq:Z_definition} is treated as dimensional in implementing both single- and multi-oscillator impedance boundary conditions (\cref{sec:modeling_a_physical_piezoelectric}).
As described in appendix \ref{app:convolutionintegral}, the impedance boundary conditions \eqref{eq:Z_definition} are implemented via imposition of the complex wall softness coefficient, $\widehat{\widetilde{W}}_{\omega}$,
defined as
\begin{equation} \label{eqn:wth_z}
\widehat{\widetilde{W}}_{\omega} \left(\omega\right) \equiv \frac{2Z_0}{Z_0+Z\left(\omega\right)}
\end{equation}
which is related to the complex reflection coefficient, $\widehat{W}_{\omega}$, via
\begin{align} \label{eqn:wh_z}
\widehat{W}_{\omega} \left(\omega\right) \equiv \frac{Z_0-Z\left(\omega\right)}{Z_0+Z\left(\omega\right)} = \widehat{\widetilde{W}}_{\omega} \left(\omega\right) - 1.
\end{align}
Hard-wall (purely reflective) conditions correspond to the limit of infinite impedance magnitude $|Z|\rightarrow\infty$ and therefore
can be imposed by setting $\widehat{\widetilde{W}}_{\omega}=0$.
The subscript $\omega$ is introduced to avoid ambiguity when \cref{eqn:wth_z,eqn:wh_z} are extended to the Laplace space, via the transformation $s=i\, \omega$, yielding
\begin{align} \label{eqn:laplace}
\widehat{\widetilde{W}}_{\omega} \left(\omega\right) =\widehat{\widetilde{W}}_{\omega} \left(-i\,s\right) =\widehat{\widetilde{W}}_{s} \left(s\right) \, .
\end{align}
It is important to stress that $\widehat{\widetilde{W}}_{\omega} \left(\cdot\right)$ and $\widehat{\widetilde{W}}_{s} \left(\cdot\right)$ are different functional forms, and the latter is convenient for the implementation of the TDIBC, as in \cref{sec:modeling_a_physical_piezoelectric}. 

\subsection{Computational Setup}

\begin{figure}
	\centering
	\includegraphics[width=0.9\textwidth,keepaspectratio]{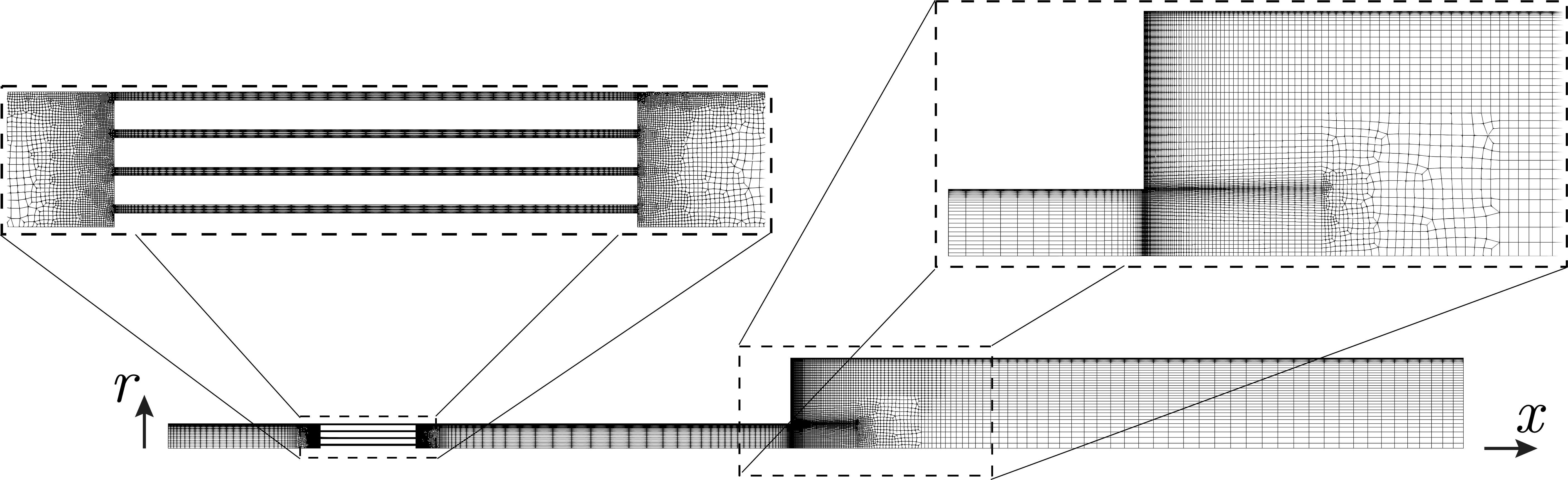}
	\caption{Computational grid for resolution/stack-type A/I (see tables \ref{tab:gridresolution} and \ref{tab:simulation_parameters}).
	}
	\label{fig:computational_grid}
\end{figure}

The three different stack types (I, II, and III) required different computational grids. For stack type I (figure \ref{fig:computational_grid}), three different levels of grid resolution were considered (A, B, and C, from coarse to fine). Stack types II and III were only meshed at the highest grid resolution level (table \ref{tab:gridresolution}).
Simulations with temperature settings 1 - 4 (table~\ref{tab:simulation_parameters}) have been performed on the finest grid resolution level C and only for stack type I. 
The viscous and thermal Stokes thicknesses at 300~K and 388~Hz (frequency of the thermoacoustically amplified mode) are $\delta_\nu \sim 0.25 $ mm and  $\delta_\kappa \sim 0.30 $ mm, respectively, and are resolved on all grids considered. The coarsest near-wall grid resolution considered is $\Delta r_w=0.06$ mm (table \ref{tab:gridresolution}). While the full three-dimensional Navier--Stokes equations are solved, azimuthal gradients are not captured on the adopted computational grid (figure \ref{fig:computational_grid}), which is extruded azimuthally with $1^{\circ}$ increments for a total of 5 cells, with rotational periodicity imposed on the lateral faces. The results from the numerical computations are, in practice, axisymmetric.

The governing equations are solved using \charlesx{}, a control-volume-based, finite-volume solver for the fully compressible Navier--Stokes equations on unstructured grids, developed as a joint-effort among researchers at Stanford University. \charlesx{} employs a three-stage, third-order Runge-Kutta time discretization and a grid-adaptive reconstruction strategy, blending a high-order polynomial interpolation with low-order upwind fluxes~\citep{HamMIM_2007_bookchpt}. The code is parallelized using the Message Passing Interface (MPI) protocol and highly scalable on a large number of processors \citep{BermejoBLB_IEEE_2014}. 

\begin{table}
	\centering
	\begin{tabular}{l|ccr}
		\hline
		\multirow{ 2}{*}{$N_{cv}$} & \multicolumn{3}{c}{Grid Resolution} \\
		& \multicolumn{1}{c}{A} & \multicolumn{1}{c}{B} & \multicolumn{1}{c}{C} \\
		\hline
		Stack I   & 16 000 	& 32 000 	& 66 000 \\
		Stack II  & $\cdot$ & $\cdot$ 	& 102 000 \\
		Stack III & $\cdot$ & $\cdot$ 	& 93 000 \\
		\hline
		$\Delta r_w $ (mm) & 0.06 & 0.04 & 0.02 \\
		\hline
	\end{tabular}
	\caption{
		Number of control volumes, $N_{cv}$, for available combinations of stack geometry types (I, II, and III) and grid resolution levels (A, B, and C). The wall-normal grid spacing at the wall, $\Delta r_w$, has been chosen independently from the stack type and is a function of the grid resolution level only.
	}
	\label{tab:gridresolution}
\end{table}

\begin{table}
	\centering
	\begin{tabular}{lcccrr}
		\hline
		Case &  $\Delta T$ (K) &  $T_c$  (K)  & $T_h$ (K) & Grid Resolution/Stack Type \\
		\hline
		1 & 340.0  & 450.0  & 790       & C/I \\
		2 & 377.5 & 412.5  & 790  & C/I \\
		3 & 415.0  & 375.0 & 790      & C/I \\
		4 & 452.5  & 337.5 & 790    & C/I \\
		5 & 490.0 & 300.0 & 790    & A/I, B/I, C/I, C/II, C/III \\
		\hline
	\end{tabular}
	\caption{
		Combinations of temperature settings (1, 2, 3, 4, and 5), stack geometry types (I, II, and III, illustrated in table \ref{tab:stack_configurations}), and grid resolution levels (A, B, and C) adopted in the Navier--Stokes simulations.
	}
	\label{tab:simulation_parameters}
\end{table}

%% file: LinearModel.tex
\section{System-Wide Linear Thermoacoustic Model}

\label{sec:lsamodel}
\label{sec:linearmodeling}

A system-wide linear dynamic model has been developed based on Rott's theory, to support the analysis of both the start-up phase (see \cref{sec:engine_transient_results}) and the low-acoustic-amplitude limit cycle (\cref{sec:energyextraction}). 
While the validity of Rott's theory is strictly limited to the former case, it is discussed later (\cref{sec:energybudgets}) how an extension to the limit cycle can inform the closure of acoustic energy budgets.

\begin{figure}
		\centering
		\includegraphics[width=0.95\linewidth]{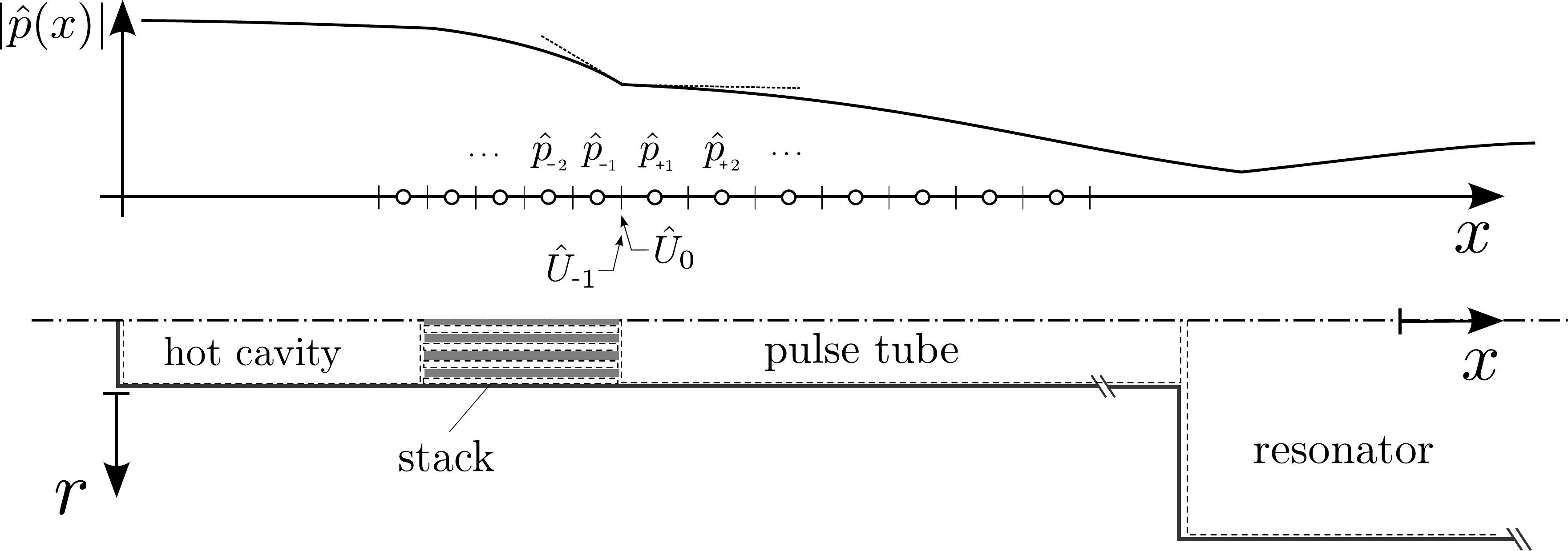}
		\caption{
			Partitioning of TAP engine model into constituent control volumes -- hot cavity, stack, pulse tube, and resonator -- for the formulation of the system-wide linear model (\cref{sec:lsamodel}). Illustration of staggered grid variable arrangement at the interface between adjacent control volumes where conditions \eqref{eqn:junction_conditions_fourier} are imposed.}
		\label{fig:LST_setup}
\end{figure}

The engine is divided into four Eulerian control volumes (\cref{fig:LST_setup}): the hot cavity, the gas-filled volume of the stack, and two constant-area sections. The governing equations have been linearized about the thermodynamic state $\{\rho_0,T_0,P_0\}$. The base pressure, $P_0$, is assumed to be uniform, and the mean density and temperature vary with the axial coordinate according to $P_0 = \rho_0(x)\,R_{gas}\,T_0(x)$. The base speed of sound is calculated as $a_0=\sqrt{\gamma R_{gas} T_0}$. All fluctuating quantities are assumed to be harmonic. The $e^{+i\sigma \, t}$ convention is adopted where $\sigma = -i\alpha + \omega$, with $\alpha$ and $\omega$ being the growth rate and angular frequency, respectively.

\subsection{Hot cavity, pulse tube, and resonator}

In the hot cavity, pulse tube, and resonator, a constant axial mean temperature distribution is assumed (figure~\ref{fig:computational_setup}), yielding the linearized equations
\begin{subequations}
	\begin{align}
	i \sigma \hat{p} &= -\frac{1}{1+\left(\gamma-1\right) f_\kappa} \frac{\rho_0 a_0^2}{A} \frac{d\hat{U}}{d x}  \label{eq:constant_T0_equations_masss_energy} \\
	i \sigma \hat{U} &= -\left(1-f_\nu\right) \frac{A}{\rho_0} \frac{d\hat{p}}{d x} \label{eq:constant_T0_equations_momentum}
	\end{align}
	\label{eq:constant_T0_equations}
\end{subequations}
enforcing the (combined) conservation of mass and energy \eqref{eq:constant_T0_equations_masss_energy} and momentum \eqref{eq:constant_T0_equations_momentum}, respectively. In these sections, the total cross-sectional area corresponds to the area available to the gas, $A=A_g$. The complex thermoviscous functions $f_\nu$ and $f_\kappa$ in \eqref{eq:constant_T0_equations} are
\begin{equation} \label{eq:complex_viscous_and_thermal_penetration_depth}
	f_\nu = \frac{2}{i\,\eta_w} \frac{J_1(i\eta_w)}{J_0(i\eta_w)},	\quad f_\kappa =  \frac{2}{i\,\eta_w\sqrt{\Pran}} \frac{J_1(i\eta_w\sqrt{\Pran})}{J_0(i\eta_w\sqrt{\Pran})}
\end{equation}
where $J_n(\cdot)$ are Bessel functions of the first kind and $\eta$ is the dimensionless complex radial coordinate
\begin{equation}
\eta \equiv \sqrt{\frac{i\omega}{\nu_0}}r =\sqrt{2\,i}\frac{r}{\delta_\nu}
\label{eq:dimensionlessradialcoordinate}
\end{equation}
where $\nu_0=\mu(T_0)/\rho_0$ is the kinematic viscosity based on mean values of density and temperature, and $\eta_w$ in \eqref{eq:complex_viscous_and_thermal_penetration_depth} is the dimensionless coordinate \eqref{eq:dimensionlessradialcoordinate} calculated at the radial location of the isothermal, no-slip wall.
The viscous, $\delta_\nu$, and thermal, $\delta_\kappa$, Stokes thicknesses are
\begin{equation} \label{eq:viscous_thermal_bl}
	\delta_\nu = \sqrt{\frac{2\,\nu_0}{\omega }}, \quad \delta_\kappa = \sqrt{\frac{2\,k}{\omega \rho_0 c_p}} \, 
\end{equation}
and are related via the Prandtl number, $\delta_\nu = \sqrt{\Pran} \; \delta_\kappa$. 
The effective laminar boundary layer thickness is approximately 3 times the Stokes thickness. 
For a Prandtl number below unity, $\Pran<1$, the thermal boundary layer is thicker than the viscous layer.

\subsection{Thermoacoustic Stack}

The analytical expression for the radial profile of the complex axial velocity amplitude within the $m$-th annular gap of the stack (\cref{tab:stack_configurations}) has been derived for a generic axial location $x$ by neglecting radial variations of pressure, i.e. $\hat{p}^{(m)}(x,r) = \hat{p}^{(m)}(x)$ (appendix~\ref{app:rotts_theory_derivation}), yielding
\begin{equation} \label{eq:uhat_profile}
\hat{u}^{(m)}(\eta) = \hat{u}^{(m)}_i\,\left[1- \left(\frac{J_0(i\eta)}{J_0(i\eta^{(m)}_\textrm{top})}+\frac{H_0^{(1)}(i\eta)}{H^{(1)}_0(i\eta^{(m)}_\textrm{bot})}\right) \right]
\end{equation}
where 
\begin{equation} \label{eq:uhat_profile_inviscid}
\hat{u}^{(m)}_i = \frac{i}{\omega\,\rho_0}\frac{d\hat{p}}{dx}
\end{equation}
is the inviscid acoustic velocity, which varies with the axial direction $x$; $H^{(1)}_n(\cdot)$ are Hankel functions of the first kind; and
\begin{equation}
\eta^{(m)}_{\textrm{top/bot}} = \sqrt{\frac{i\omega}{\nu_0}}\,r^{(m)}_{\textrm{top/bot}} = \sqrt{2\,i}\,\frac{r^{(m)}_{\textrm{top/bot}}}{\delta_\nu} \; .
\end{equation}

Rott's wave equations can be written for the $m$-th annular flow passage of cross-sectional area $A_g^{(m)}$ (where $m  \in \left\{ 1, \cdots, n_s+1\right\}$), in the diagonalized form:
\begin{subequations}
	\begin{align}
	i\sigma \hat{p}^{(m)} &=
		\left[ \frac{\rho_0 a_0^2}{A_g^{(m)}}\frac{1}{1+\left(\gamma-1\right) f_\kappa^{(m)}} \left( \frac{\left(f^{(m)}_\kappa - f^{(m)}_\nu\right)}{\left(1-f^{(m)}_\nu\right)\left(1-\Pran\right)}\frac{1}{T_0} \frac{dT_0}{dx} - \frac{d}{dx} \right)\right]
		\hat{U}^{(m)} 	\label{eq:LST_mass_and_energy} \\
	i\sigma \hat{U}^{(m)} &=
	  -  \left[ \frac{\left(1-f^{(m)}_\nu\right)A_g^{(m)}}{\rho_0} \frac{d}{dx} \right]
	  \hat{p}^{(m)} \label{eq:LST_momentum}
	\end{align}
\end{subequations}
where the complex thermoviscous functions, $f^{(m)}_\nu$ and $f^{(m)}_\kappa$, are, in this case (appendix~\ref{app:rotts_theory_derivation})
\begin{subequations}
\begin{equation}
\begin{split}
f^{(m)}_\nu = -\frac{\pi\,\delta_\nu^2}{A_g^{(m)}}
	\Big\{
	\frac{1}{J_0(i\,\eta^{(m)}_\textrm{top})}
		\left[ \eta^{(m)}_\textrm{top} J_1(i\eta^{(m)}_\textrm{top}) - \eta^{(m)}_\textrm{bot} J_1(i\eta^{(m)}_\textrm{bot}) \right] + \\
\frac{1}{H^{(1)}_0(i\,\eta^{(m)}_\textrm{bot})} \left[ \eta^{(m)}_\textrm{top} H^{(1)}_1(i\eta^{(m)}_\textrm{top}) - \eta^{(m)}_\textrm{bot} H^{(1)}_1(i\eta^{(m)}_\textrm{bot}) \right]
\Big\}
\end{split}
\end{equation}
\begin{equation}
\begin{split}
f^{(m)}_\kappa = -\frac{\pi\,\delta_\kappa^2\,\sqrt{\Pran}}{A_g^{(m)}}
	\Big\{
	\frac{1}{J_0(i\,\eta^{(m)}_\textrm{top}\sqrt{\Pran})}
		\left[ \eta^{(m)}_\textrm{top} J_1(i\eta^{(m)}_\textrm{top}\sqrt{\Pran}) - \eta^{(m)}_\textrm{bot} J_1(i\eta^{(m)}_\textrm{bot}\sqrt{\Pran}) \right] + \\
\frac{1}{H^{(1)}_0(i\,\eta^{(m)}_\textrm{bot}\sqrt{\Pran})} \left[ \eta^{(m)}_\textrm{top} H^{(1)}_1(i\eta^{(m)}_\textrm{top}\sqrt{\Pran}) - \eta^{(m)}_\textrm{bot} H^{(1)}_1(i\eta^{(m)}_\textrm{bot}\sqrt{\Pran}) \right]
\Big\} \; .
\end{split}
\end{equation}
\end{subequations}

Assuming that all annular flow passages share the same instantaneous pressure field, $\hat{p}^{(m)}(x)=\hat{p}(x)$, and mean density and temperature axial distribution, and considering that the thermoviscous functions $f^{(m)}_\nu$ and $f^{(m)}_\kappa$ differ at most by $2\%$ over all values of $m$, it is possible to collapse the $n_s+1$ equations in \eqref{eq:LST_mass_and_energy} via area-weighted averaging and to take the arithmetic sum of \eqref{eq:LST_momentum} over $m$, yielding a new set of approximate wave equations for the thermoacoustic stack,
\begin{subequations}
	\begin{align}
i \sigma \hat{p}	 &\simeq \sum_{m=1}^{n_s+1} \frac{A_g^{(m)}}{A_g}
		\left[ \frac{\rho_0 a_0^2}{A_g}\frac{1}{1+\left(\gamma-1\right) f_\kappa^{(m)}} \left( \frac{\left(f^{(m)}_\kappa - f^{(m)}_\nu\right)}{\left(1-f^{(m)}_\nu\right)\left(1-\Pran\right)}\frac{1}{T_0} \frac{dT_0}{dx} - \frac{d}{dx} \right)\right]
	\hat{U}
	\label{eq:LST_mass_and_energy_final} \\
	i\sigma \hat{U} &=
	  - \sum_{m=1}^{n_s+1}
	  \left[ \frac{\left(1-f^{(m)}_\nu\right)A_g^{(m)}}{\rho_0} \frac{d}{dx} \right]
	  \hat{p} 	\label{eq:LST_momentum_final}
	\end{align}
	\label{eq:LST_stack_final}
\end{subequations}
where the total cross-sectional area available to the gas, $A_g$, and flow rate, $\hat{U}$, are
\begin{equation}
		\quad A_g = \sum_{m=1}^{n_s+1} A_g^{(m)}, \quad A_g^{(m)} = \int_{r^{(m)}_\textrm{bot}}^{r^{(m)}_\textrm{top}} 2\pi\,r\,dr
\end{equation}
\begin{equation}
		\hat{U} = \sum_{m=1}^{n_s+1} \hat{U}^{(m)}, \quad \hat{U}^{(m)} = \int_{r^{(m)}_\textrm{bot}}^{r^{(m)}_\textrm{top}} 2\pi\,r\,\hat{u}(r)\,dr
	\label{eq:flow_rate_equality}
\end{equation}
and an area-weighted equipartitioning of the flow rates, $\hat{U}^{(m)} =A_g^{(m)}/A_g\; \hat{U}$, has been assumed in \eqref{eq:LST_mass_and_energy_final}.

\subsection{Discretization, Boundary and Inter-segment Conditions}
\label{sec:boundaryconditions}

Isolated-component eigenvalue problems for the cavity ($c$), thermoacoustic stack ($s$), pulse tube ($t$), and resonator ($r$) control volumes (figure~\ref{fig:LST_setup}) are first assembled in the form
\begin{align}
	\left(i\sigma \mathbf{I} -
	\begin{bmatrix}
	\mathbf{B_c} & 0 & 0 & 0\\
	0 & \mathbf{B_s} & 0 & 0\\
	0 & 0 & \mathbf{B_t} & 0\\
	0 & 0 & 0 & \mathbf{B_r}\\
	\end{bmatrix}
	\right) \mathbf{v} = 0
\label{eq:lineareigenvalueproblem}
\end{align}
with $\mathbf{v} = \left\{\mathbf{u}_c, \mathbf{u}_s,\mathbf{u}_t,\mathbf{u}_r \right\}$ where $\mathbf{u}_l=\left\{\hat{\mathbf{p}}_l, \hat{\mathbf{U}}_l\right\}$ is the collection of the discrete complex amplitudes of pressure and flow rate for the $l$-th segment where $l \in \left\{ c,s,t,r\right\}$, $\mathbf{I}$ is the identity matrix, and $\mathbf{B}$ is an operator discretizing the equations \eqref{eq:constant_T0_equations} for the hot cavity, pulse tube, and resonator, and \eqref{eq:LST_stack_final} for the thermoacoustic stack.

Equations for each segment are discretized on a staggered, uniform grid (\cref{fig:LST_setup}). The mass and energy equation is written at each pressure node while the one for momentum is written at each flow rate location, both with a second-order central discretization scheme. Inter-segment conditions are
\begin{subequations}
	\begin{align}
		-\hat{p}_{-2} + 3\hat{p}_{-1} &= 3 \hat{p}_{+1} - \hat{p}_{+2} + 2 \Delta \hat{p}_{ml} \label{eq:extrapolative_pressure_condition} \\
		\hat{U}_{-1} &= \hat{U}_{1}
		\label{eq:extrapolative_flow_condition}
	\end{align}
	\label{eqn:junction_conditions_fourier}
\end{subequations}
where subscripts $-2$ and $-1$ indicate the second-last and last point of a segment, subscripts $+1$ and $+2$ indicate the first and second point of the following one, and $\Delta \hat{p}_{ml}$ is the pressure drop due to minor losses \eqref{eq:lin_minorlosses}, if applied. 
The extrapolation \eqref{eq:extrapolative_pressure_condition} does not constrain the axial derivative of pressure to be continuous.  
The continuity and jumps in the acoustic power are correctly captured. 
Zero-flow-rate conditions (hard walls), $\hat{U} = 0$, are imposed on both ends of the device. The corresponding zero-Neumann condition for pressure $d\hat{p}/dx = 0$ does not need to be explicitly enforced numerically, as it is a natural outcome of the solution of the eigenvalue problem.
Inter-segment and boundary conditions are inserted into \eqref{eq:lineareigenvalueproblem}, yielding the complete eigenvalue problem. Several analytical results for variable-area duct acoustic systems have been reproduced to machine-precision accuracy \citep{DowlingW_1983} and excellent agreement with the Navier--Stokes calculations of the TAP engine model is found in both the linear and low-acoustic-amplitude nonlinear regime (as discussed later).

%% file: Results_StartUp.tex
\section{Transient Response}
\label{sec:engine_transient_results}

\begin{figure}
\centering
	\includegraphics[width=0.9\linewidth]{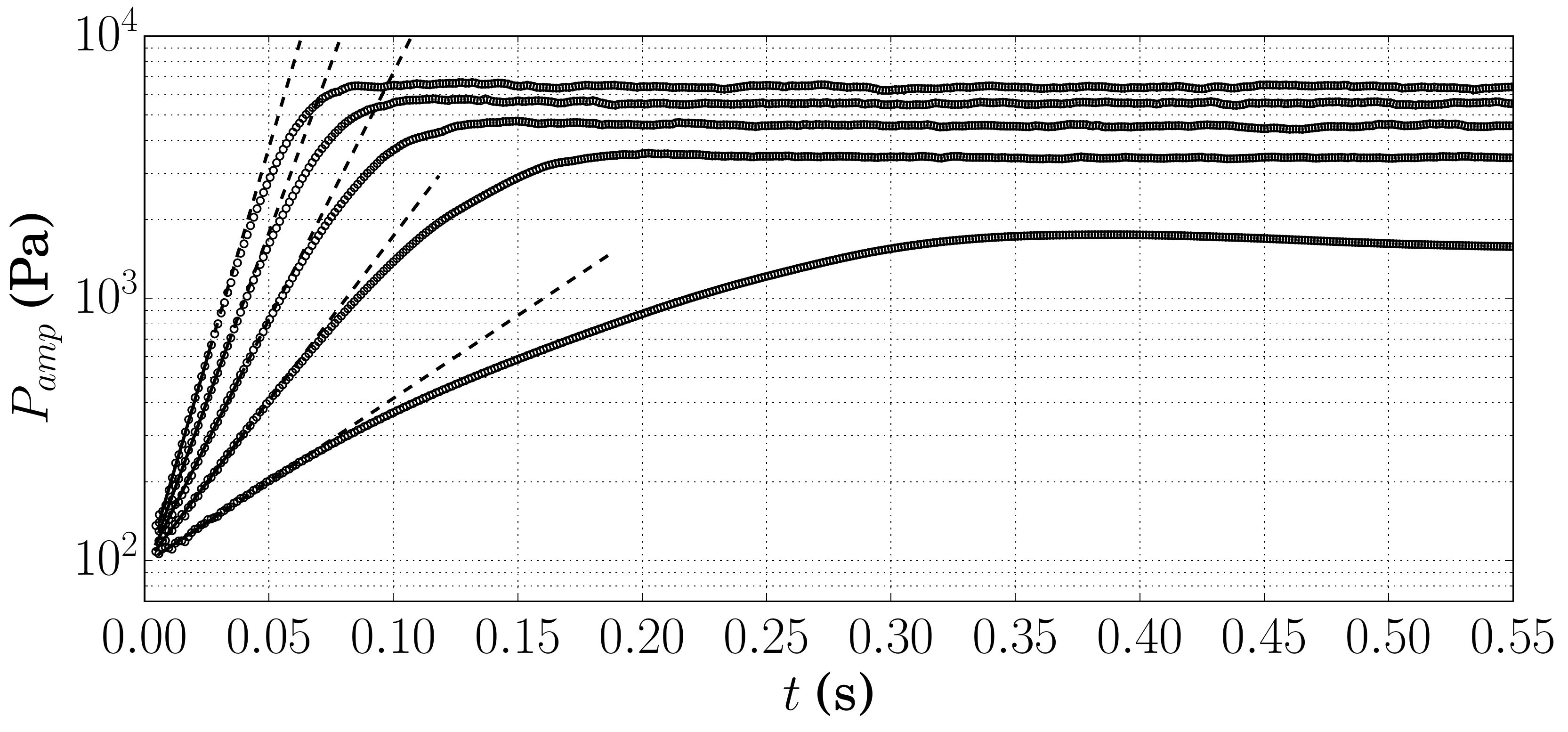}
	\caption{
		Time series of pressure amplitudes in the hot cavity for grid-resolution/stack-type C/I, for temperature settings 1 to 5 (\cref{tab:simulation_parameters}), corresponding to increasing growth rates and limit cycle pressure amplitudes.
	}
	\label{fig:transientprofiles}
\end{figure}

In this section, several aspects of the transient response of the TAP engine model are discussed. A comparison between the onset of instability as predicted by the linear thermoacoustic model derived in \cref{sec:linearmodeling} and the Navier--Stokes simulations is first carried out (\cref{sec:baseline_and_stack_effects}). 
A grid sensitivity study is then carried out, focusing on the effects of grid resolution on growth rates extracted from the Navier--Stokes simulations (\cref{sec:gridconvergence}). The performance of the three stack configurations in \cref{tab:stack_configurations} are compared and, with the aid of the linear thermoacoustic model, the criteria for the optimal stack design is inferred (\cref{sec:optimal_stack_design}). Finally, the natural (thermoacoustically unexcited) modes of the  TAP engine model are briefly discussed (\cref{subsec:damping_modes}) in the context of physical admissibility issues of time-domain impedance boundary conditions (TDIBCs) used to model piezoelectric energy absorption (discussed later in \cref{sec:energyextraction}).

\subsection{Engine Start-Up}
\label{sec:baseline_and_stack_effects}

Navier--Stokes simulations are carried out first without piezoelectric energy absorption (i.e. with hard-wall boundary conditions on the right end of the resonator) for all cases in table \ref{tab:simulation_parameters}. Initial conditions are that of zero velocity, ambient pressure, and temperature matching the expected mean axial distribution at equilibrium  (\cref{fig:computational_setup}). No initial velocity or pressure perturbations are prescribed. As also observed in \citet{ScaloLH_JFM_2015}, for a sufficiently large background temperature gradient, the simple activation of the heat source triggers a disturbance that is thermoacoustically amplified, initiating a transient exponential growth, followed by a saturation of the pressure amplitude (\cref{fig:transientprofiles}). During the late stages of energy growth the pressure amplitude overshoots its limit cycle value, especially for close-to-critical values of the temperature gradient. This behaviour was not observed in the travelling-wave engine investigated by \citet{ScaloLH_JFM_2015}.

\begin{figure}
	\centering
	\includegraphics[width=0.85\linewidth]{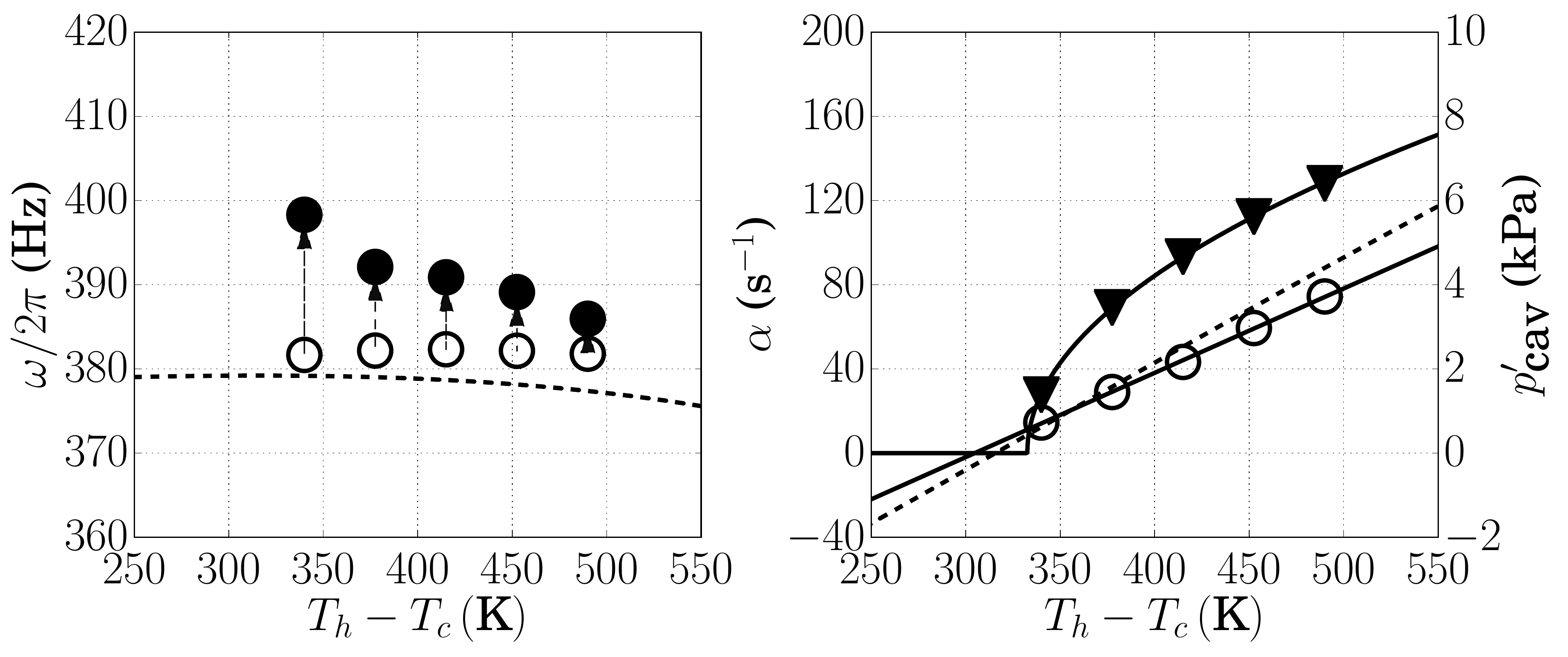}
	\put(-330,10){(a)}
	\put(-10,10){(b)}
	\caption[Frequency and growth rate of LSA versus numerical simulations.]{
		Frequency (a) and growth rate (b) versus temperature difference, $\Delta T=T_h - T_c$, for grid-resolution/stack-type C/I. Predictions from linear model \legenddashed{} and Navier--Stokes simulations at start-up \legenddots{}; limit cycle frequencies  \legenddotskandv{};
		square-root fit \eqref{eq:hopf} \legendline{} of limit cycle pressure amplitudes \legendtriangles{} in the hot cavity versus temperature difference; and linear fit of growth rates from Navier--Stokes calculations \legendline{}.
	}
	\label{fig:lsa_freqgrowth}

\end{figure}

Growth rates and frequencies predicted by the linear model developed in \cref{sec:lsamodel} are in good agreement with the nonlinear simulations (figure~\ref{fig:lsa_freqgrowth}). A linear fit of the growth rates extracted from the Navier--Stokes simulations against the temperature difference, $\Delta T$, yields a critical temperature difference of $\Delta T_{cr} = 305$ K. Linear theory predicts $\Delta T_{cr} = 315.7$ K, while fitting the limit cycle pressure amplitude, $p_{lc}$, with the equilibrium solution from a supercritical Hopf bifurcation model with dissipation term scaling as $p_{lc}^2$,
\begin{align}
	p_{lc} \propto  \sqrt{\frac{\Delta T-\Delta T_{cr}}{T_h}}
	\label{eq:hopf}
\end{align}
yields $\Delta T_{cr} = 332.7$ K. While small discrepancies between linear theory and Navier--Stokes simulations are expected, a difference of 30 K between the $\Delta T_{cr}$ calculated from the growth rates and the limit cycle pressures suggests that hysteresis effects associated with subcritical bifurcation may be present. This phenomenon was not observed in the numerical simulations of a travelling-wave engine by \citet{ScaloLH_JFM_2015}.

\begin{figure}
	\centering
	\includegraphics[width=0.85\linewidth]{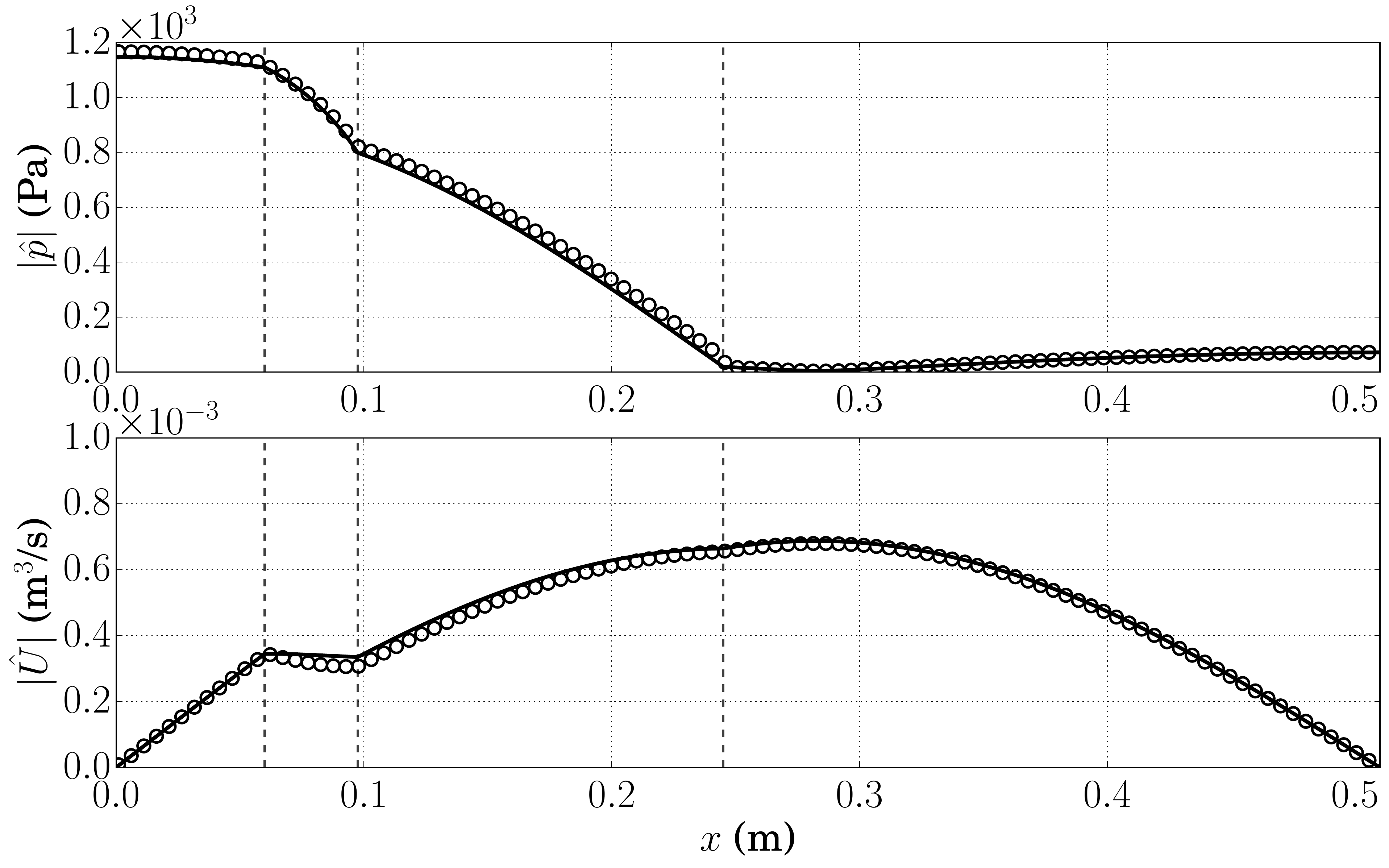}
		\put(-330,20){(b)}
		\put(-330,120){(a)}
	\caption[Eigenfunctions]{
		Axial distribution of pressure (a) and flow rate (b) amplitudes of the thermoacoustically unstable mode for temperature setting 5 and grid-resolution/stack-type C/I, as predicted by linear theory \legendline{}, rescaled to match amplitudes extracted from Navier--Stokes simulations \legenddots{} during the start-up phase. The frequency predicted by linear theory is $f=378.9\textrm{ Hz}$, while frequency extracted from the Navier--Stokes calculations yields $f=381.8\textrm{ Hz}$ (see figure~\ref{fig:lsa_freqgrowth}).
		\verticallines{}
	}
	\label{fig:lsa_eigenfunctions}

\end{figure}

Very good agreement is also found between pressure and flow rate eigenfunctions, and pressure and flow rate amplitudes extracted from the Navier--Stokes calculations via least squares fitting during the start-up phase (figure~\ref{fig:lsa_eigenfunctions}). Results were confirmed with peak-finding and windowed short-time Fourier transform (STFT). 
Minor discrepancies are present at locations of abrupt area change, where assumptions of quasi-one-dimensionality break down. Amplitude and phase detection were applied to time-series of cross-sectionally averaged pressure and cross-sectionally integrated axial flow velocity components. 

While good agreement is also retained in the nonlinear regime, and used to extract the axial distribution of acoustic power at the limit cycle with piezoelectric energy extraction (discussed later, figure \ref{fig:wdot_axial_distribution} in \cref{sec:energyextraction}), a frequency shift in the range $5 - 20$~Hz, from high to low temperature settings, is observed in the transient leading to the limit cycle (figures~\ref{fig:lsa_freqgrowth}, \ref{fig:freq_over_time}). This frequency change, as discussed in \cref{sec:energyextraction}, is enough to alter significantly the rate of energy extraction from a piezoelectric diaphragm -- indicating that, especially for complex geometries, its fine-tuning should be performed based on actual measurements or nonlinear calculations, and not by solely relying on linear theory. The observed frequency shift is also discussed in \cref{sec:frequencychange} in the context of acoustic streaming.

\subsection{Grid Sensitivity Study}
\label{sec:gridconvergence}

\begin{figure}
	\centering
	\includegraphics[width=0.85\linewidth]{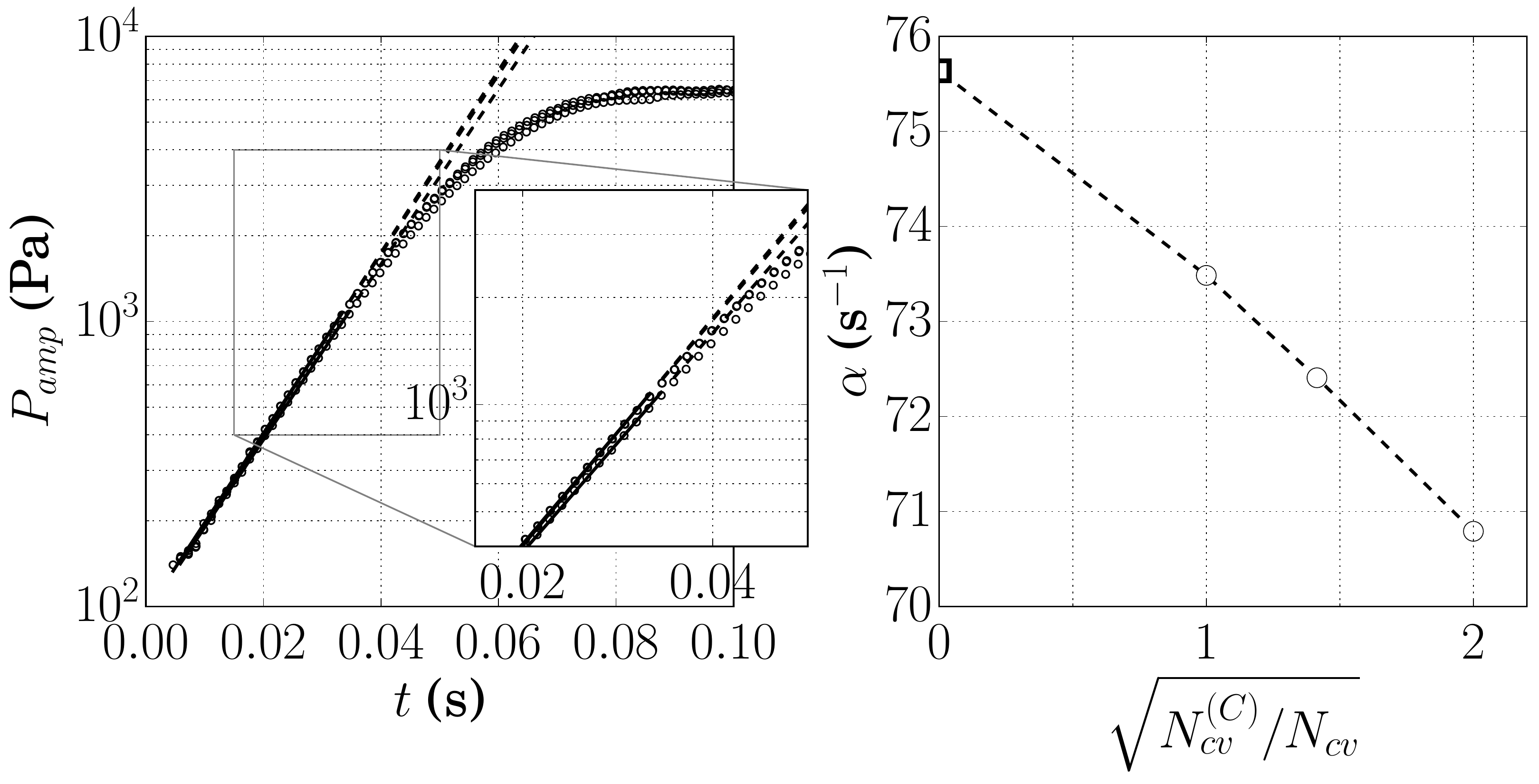}
	\put(-330,20){(a)}
	\put(5,20){(b)}
	\caption[Time series of pressure amplitudes within the hot cavity with semi-logarithmic linear fit]{
		Time series of pressure amplitudes within the hot cavity ($\circ{}$) with semi-logarithmic fit \legenddashed{} over initial start-up phase (a) and growth rates (b) for grid resolution levels A, B, and C, temperature setting 5 and stack type I (table \ref{tab:stack_configurations}, \ref{tab:gridresolution}, and \ref{tab:simulation_parameters}). An estimate of the growth rate \legendonewhitesquare{} at zero-grid spacing is derived via Richardson extrapolation.
	}
	\label{fig:transientprofiles_convergence}

\end{figure}

Nonlinear calculations for stack type I and temperature setting 5 have been carried out at all three available grid resolution levels -- A, B and C (table~\ref{tab:simulation_parameters}) -- with a successive linear grid refinement factor of approximately $r = \sqrt{N_{cv}^{(C)}/N_{cv}^{(B)}} \simeq \sqrt{N_{cv}^{(B)}/N_{cv}^{(A)}}\simeq \sqrt{2}$. The order of grid convergence estimated from the growth rates extracted from the Navier--Stokes calculations (figure \ref{fig:transientprofiles_convergence}b) is
	\begin{align}
	\label{eq:p_conv_factor}
	p &= \frac{\log\left[\frac{\alpha_A-\alpha_B}{\alpha_B-\alpha_C}\right]}{\log \left( r \right)}  \simeq 1.2
	\end{align}
where $\alpha_A$, $\alpha_B$, and $\alpha_C$ are the growth rates associated with the grid resolution levels $A$, $B$ and $C$, respectively (table~\ref{tab:simulation_parameters}).
Using Richardson extrapolation, the predicted growth rate in the limit of zero-grid spacing is $\alpha_{h=0} = 75.64\textrm{ s}^{-1}$ with an error band of $5.6\%$.

Grid-convergent values of the growth rate show that the amount of acoustic energy lost to the numerical scheme is related to its truncation error. The estimated order of grid convergence \eqref{eq:p_conv_factor} is, however, only slightly above first order, lower than the nominal order of spatial accuracy of the solver. This result demonstrates the inherent difficulties associated with the extraction of the growth rates from Navier--Stokes simulations of full-scale thermoacoustic devices on unstructured grids. Issues include: the arbitrary choice of the time window used for the semi-logarithmic fit of the pressure amplitude time-series (figure \ref{fig:transientprofiles_convergence}a); defining a systematic grid refinement criteria for complex unstructured grids (figure \ref{fig:computational_grid}) that compensates for changes in the effective order of the numerical discretization scheme due to regions of intense skewness and stretching; and the nature of the growth rate itself, which is an accumulation over several cycles of relatively small amounts of energy per cycle---modelling and/or numerical errors, which would otherwise be deemed negligible, accumulate in the same way. 

\begin{figure}
	\centering
		\labelphantom{fig:porosityeffects_a}
		\labelphantom{fig:porosityeffects_b}
	\includegraphics[width=0.9\linewidth]{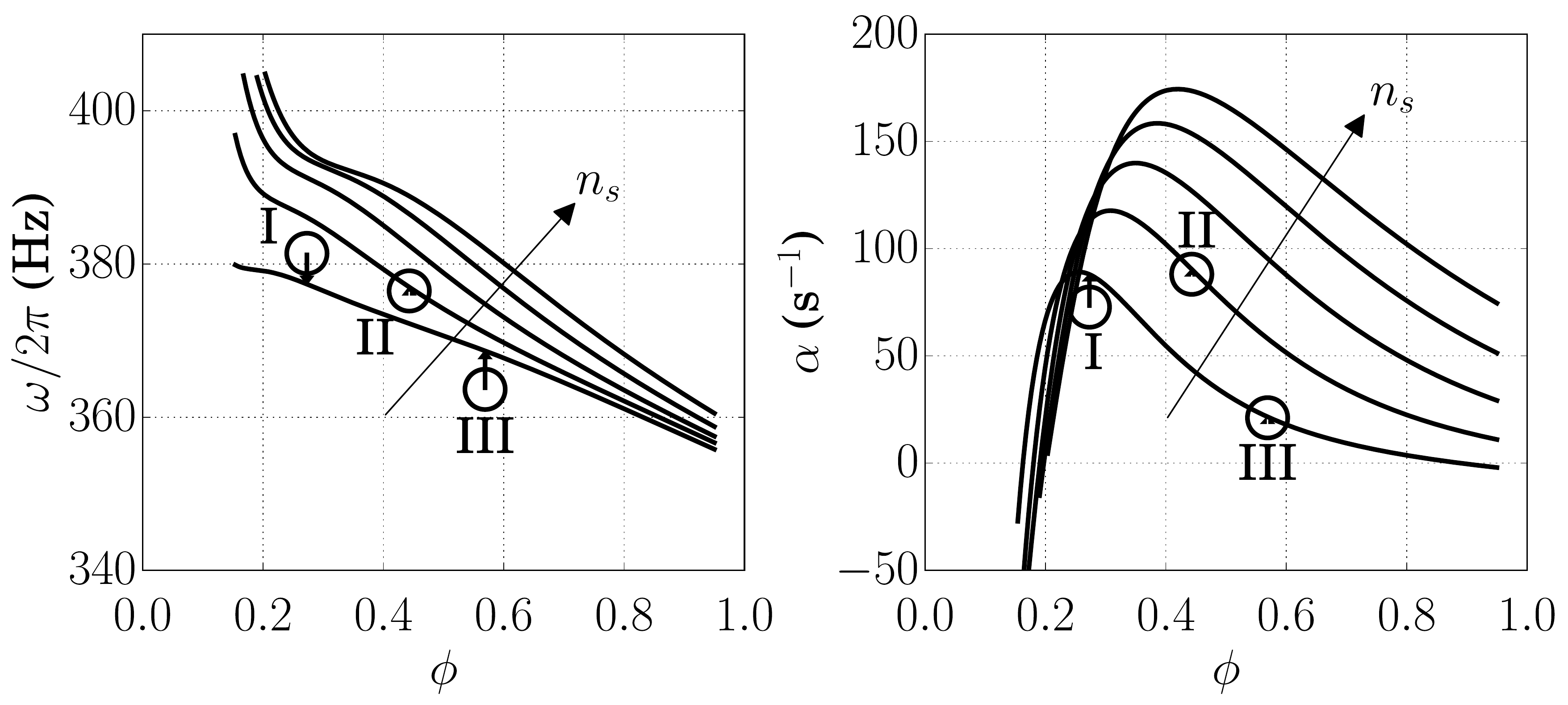}
	\put(-345,20){(a)}
	\put(5,20){(b)}
	\caption[frequency dependence on porosity]{
		Frequency (a) and growth rate (b) versus stack porosity during start-up phase. Results from the nonlinear Navier--Stokes simulations ($\circ{}$) for temperature setting 5 and grid-resolution/stack-type C/I, C/II and C/III; predictions from linear theory for $n_s = $ 3, 5, 7, 9, and 11 \legendline{}. Vertical arrows denote the difference between the linear theory prediction and Navier--Stokes simulations.
	}
	\label{fig:porosityeffects}
\end{figure}

\begin{figure}
	\centering
	\includegraphics[width=0.85\linewidth]{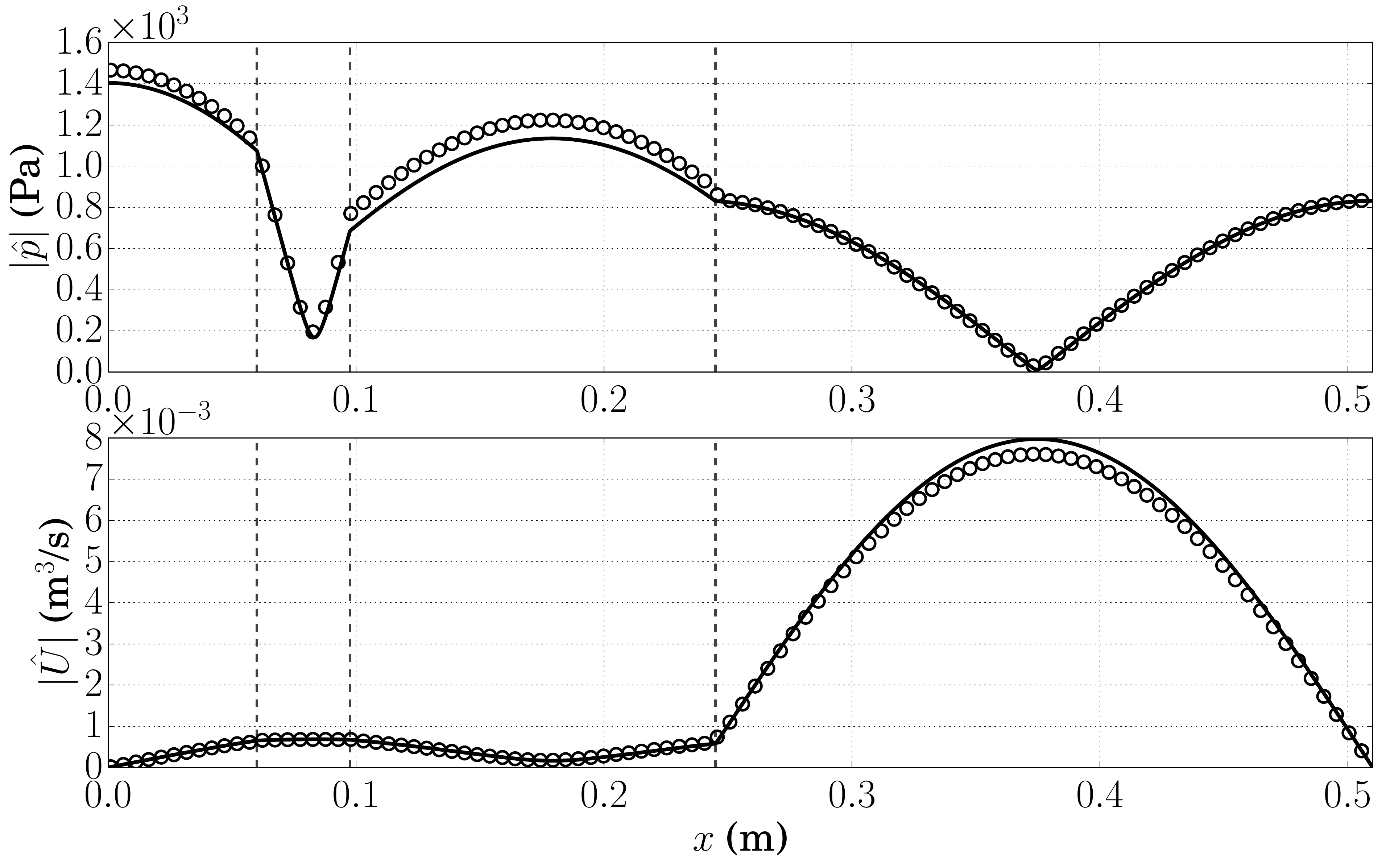}
		\put(-330,20){(b)}
		\put(-330,120){(a)}
	\caption[LSA eigenfunctions]{
		Axial distribution of pressure (a) and flow rate (b) amplitudes for the second (natural) resonant mode  predicted by linear theory \legendline{} rescaled to match pressure and flow rate amplitudes extracted from Navier--Stokes simulations in figure \ref{fig:dampinggraph} \legenddots{} at 30 cycles.
		\verticallines{}
	}
	\label{fig:lsa_eigendamp}

\end{figure}

Despite the aforementioned technical and conceptual issues related to the exact definition of the growth rate, results from Navier--Stokes simulations on the highest grid resolution level (grid C) are in very good agreement with linear theory (\cref{fig:lsa_freqgrowth,fig:lsa_eigenfunctions,fig:porosityeffects,fig:lsa_eigendamp}) and thus will be used for the remainder of the manuscript.

\subsection{Optimal Stack Design}
\label{sec:optimal_stack_design}

\begin{figure}
	\centering
	\includegraphics[width=0.75\linewidth]{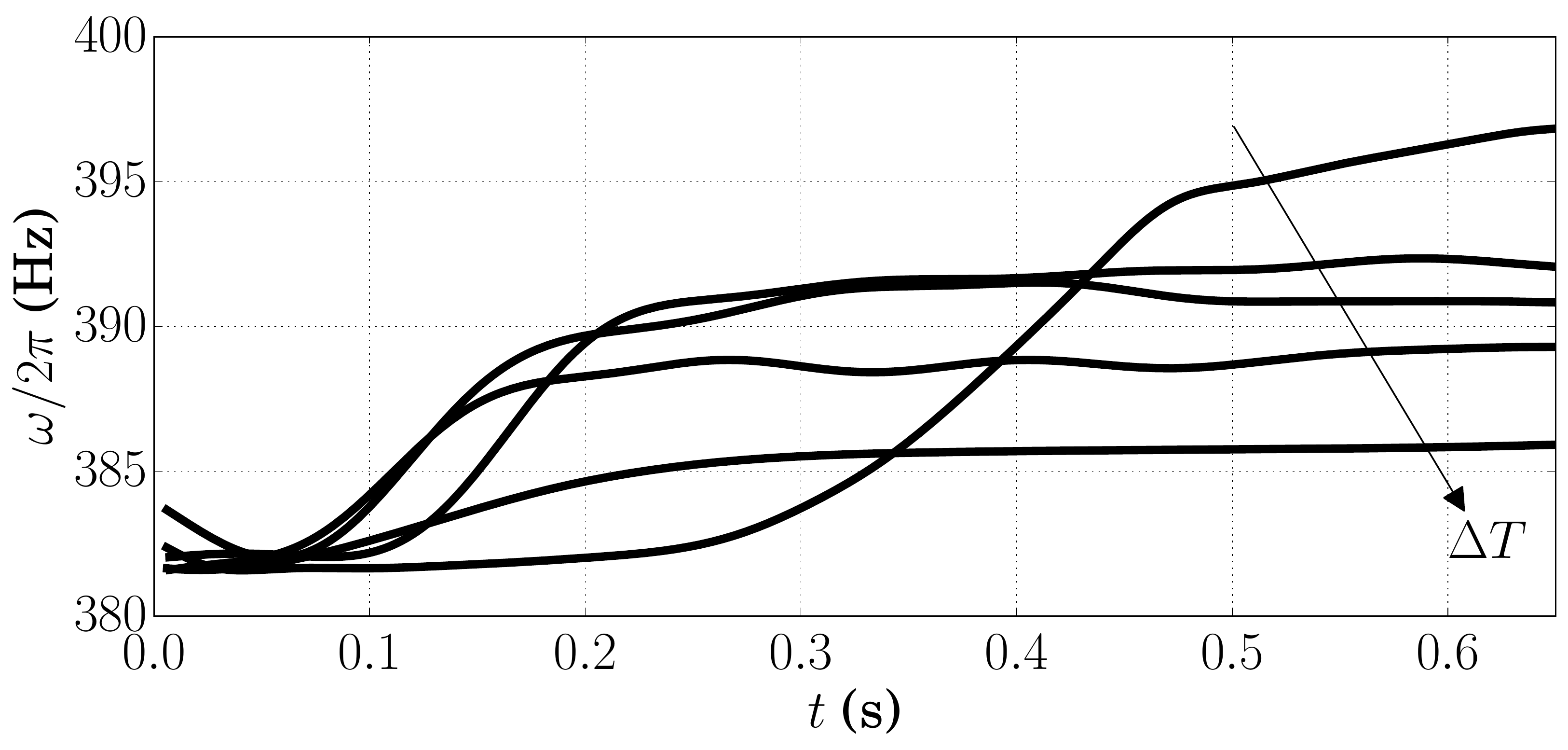}
	\caption{
		Temporal evolution of frequency of thermoacoustically amplified mode for temperature settings 1-5 and grid-resolution/stack-type C/I (table~\ref{tab:simulation_parameters}). Frequency is obtained via peak-finding and windowed over two acoustic periods. Higher temperature differences correspond to lower limit cycle frequencies, as shown by the arrow. 
	}
	\label{fig:freq_over_time}

\end{figure}

The frequency of the thermoacoustically amplified mode decreases with increasing porosity (\cref{fig:porosityeffects_a}), i.e. as a larger fraction of the cross-sectional area in the stack is made available to gas flow ($A_g=\phi\,A_{stk}$). For example, only by adopting stack I, which matches the porosity of the regenerator adopted in \citet{SmokerNAB_2012}'s experiments, and by carrying out the simulations to a limit cycle (figure \ref{fig:freq_over_time}), is it possible for the TAP engine model to operate close to the experimentally observed frequency of 388 Hz, at which the piezoelectric diaphragm was tuned.

The growth rate is dramatically affected by the stack porosity (\cref{fig:porosityeffects_b}). For any given number of solid annuli $n_s$, decreasing the porosity reduces the volume of gas available to thermoacoustic energy production ($V_g \to 0$), while increasing viscous blockage ($f_\nu \to 1$) leads to negative growth rates in the limit of $\phi \to 0$.
For example, for $n_s=11$, reducing the porosity from $\phi=0.42$ to $0.2$ reduces the growth rate from $\alpha_{max}=174.4\textrm{ s}^{-1}$ to zero.
On the other hand, increasing the porosity increases the cross-sectional area available to the gas ($A_g \to A$) at the expense of thermal contact, ultimately leading to an attenuation of the growth rate. For example, for $n_s=5$, a maximum growth rate of $\alpha_{max}=117.73\textrm{ s}^{-1}$ is obtained at $\phi=0.308$; this declines to $\alpha=90.24\textrm{ s}^{-1}$ for $\phi=0.443$ and to $\alpha=50\textrm{ s}^{-1}$ for $\phi=0.6$. 
Negative growth rates for $\phi>0.9$ are reached for $n_s=3$. 
The degree of thermal contact of the $m$-th annular flow passage is accounted for by the thermoacoustic gain term $(f^{(m)}_k-f^{(m)}_\nu)/(1-f^{(m)}_\nu)$ in \eqref{eq:LST_mass_and_energy_final} multiplying the mean temperature gradient, which is the driver of thermoacoustic instability. 
This term decays to a very small (but non-zero) value for $\phi \to 1$ \citep[p. 95]{Swift_2002}. The optimal value of porosity, $0<\phi_{opt}<1$, is therefore the result of a trade-off between thermal contact and available pore volume for themoacoustic energy production.

Increasing the growth rate for fixed values of porosity is possible by increasing $n_s$. This results in an increased solid-to-gas contact surface $S_{heat}$ and greater thermal contact without increasing flow obstruction. A higher stack density, however, requires a higher porosity to maintain the optimal growth rate, $\alpha_{max}$, to compensate for the increased viscous blockage. Moreover, the achievable $\alpha_{max}$ increases with $n_s$, demonstrating the importance of available surface area $S_{heat}$: for $n_s = 3$, the maximum growth rate achievable is $\alpha_{max}=89.1$ s$^{-1}$, while for $n_s = 11$ it increases to $\alpha_{max}=174.4$~s$^{-1}$.

Higher growth rates lead to higher limit cycle acoustic amplitudes (figure \ref{fig:transientprofiles}); for example, limit cycle pressure amplitude of approximately $6000\textrm{ Pa}$ are obtained for stack type I and temperature setting 5, while stack type II reaches $P_\textrm{amp}\simeq 11500\textrm{ Pa}$ (not shown) for the same imposed temperature gradient. This result reflects the increased thermal contact in stack II, which has almost twice the available solid-to-gas contact surface area of stack I. Stack III exhibits the lowest growth rate due to poor thermal contact, as suggested by the presence of an inviscid core (table~\ref{tab:stack_configurations}). Higher growth rates, however, may not straightforwardly be associated with higher thermal-to-acoustic efficiencies; in the case of increased $S_{heat}$, a higher external thermal energy input will be required.

\begin{table}
	\centering
	\begin{tabular}{l|cc|cc}
		\multirow{ 2}{*}{}  &  \multicolumn{2}{c}{Mode 1} &  \multicolumn{2}{c}{Mode 2} \\
		  & $\alpha$ & $f$ & $\alpha$ & $f$\\
		\hline
		$\Delta T= 0$ 	     & -103.1 s$^{-1}$& 335.4 Hz & -42.8 s$^{-1}$ & 633.6 Hz \\
		$\Delta T= 490$K  & $\;\;\,$88.0 s$^{-1}$ & 377.0 Hz & -23.6 s$^{-1}$ & 647.5 Hz \\
		\hline
	\end{tabular}
	\caption{Growth rates and frequencies predicted by linear theory for first and second modes at $\Delta T$ = 0 and $\Delta T$ = 490 K for stack type I.}
	\label{tab:growth_freq_of_decaying_modes}
\end{table}

\begin{figure}
	\centering
	\includegraphics[width=0.9\linewidth]{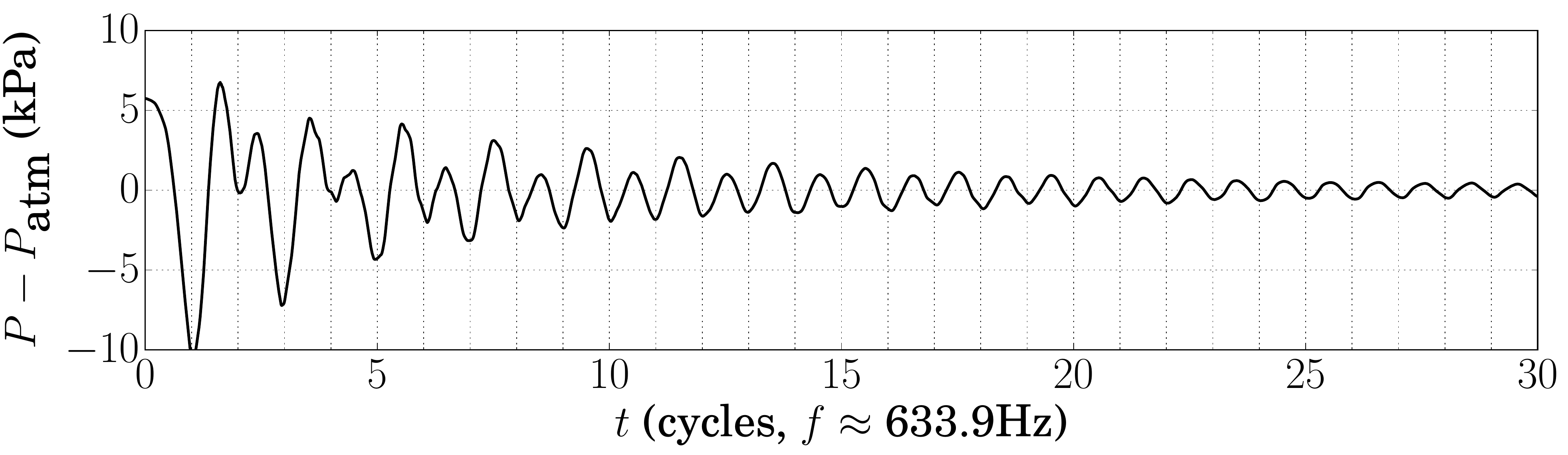}
	\caption{
		Time-series of pressure in the hot cavity for \defaultgrid{} at $\Delta T=0$ with initial quarter-wavelength pressure distribution of 6000 Pa in amplitude. Time is expressed in cycles of mode 2 with frequency $f=633.9\textrm{ Hz}$, in agreement with linear theory (table \ref{tab:growth_freq_of_decaying_modes}).
	}
	\label{fig:dampinggraph}

\end{figure}

\subsection{Unexcited Acoustic Modes}
\label{subsec:damping_modes}

Deactivating the temperature gradient in the stack allows for the analysis of the natural, unexcited acoustic modes of the TAP engine. Initiating the Navier--Stokes calculations with a large amplitude quarter-wavelength pressure distribution allows the observation of the simultaneous decay of the first two resonant modes (\cref{fig:dampinggraph}). The second mode (633.0 Hz) decays slower than the first mode (335.4 Hz), which is thermoacoustically amplified for $\Delta T>\Delta T_{cr}$. This is due to the structure of the second mode (figure \ref{fig:lsa_eigendamp}), which exhibits relatively low flow rate amplitudes in the stack, where the most intense viscous losses are concentrated.

The second mode is weakly thermoacoustically sustained by the temperature gradient, as shown by the increase in the growth rate with respect to the unexcited case (table \ref{tab:growth_freq_of_decaying_modes}). The persistence of a negative growth rate indicates that the associated thermoacoustic energy production (made inefficient by a pressure amplitude minima in the stack), is insufficient to overcome viscous dissipation.

In preliminary numerical trials at piezoelectric energy extraction, mode switching from the first mode to the second mode was mistakenly triggered. This was due to the erroneous application of a physically inadmissible impedance with negative resistance ($\Real(Z)<0$) at frequencies close to 633 Hz. While the second mode is not prone to being thermoacoustically amplified, the erroneously assigned impedance forced the device to operate at a frequency different from fundamental one, effectively controlling the thermoacoustic response. Admissibility issues arise, in particular, due to the fact that an impedance with negative resistance represents an active boundary element, i.e. it injects acoustic power into the system \citep{Rienstra_AIAA_2006}. 

\section{Thermoacoustic Transport and Streaming}
\label{sec:frequencychange}
\label{sec:hw_limitcycle}

During the transient evolution from the start-up phase to the limit cycle without acoustic energy absorption (\cref{fig:transientprofiles}), a gradual shift of the operating frequency of the engine is observed (\cref{fig:freq_over_time}). After an adjustment phase during the initial stages of acoustic energy growth, the frequency monotonically rises. In the case of temperature setting 5 and stack type I, the frequency approaches the experimentally reported value of 388 Hz, at which the piezoelectric diaphragm is tuned. In the case of (near-to-critical) temperature setting 1 and stack type I, a very long adjustment phase of the frequency is observed.

\begin{figure}
	\centering
	\includegraphics[width=0.9\linewidth]{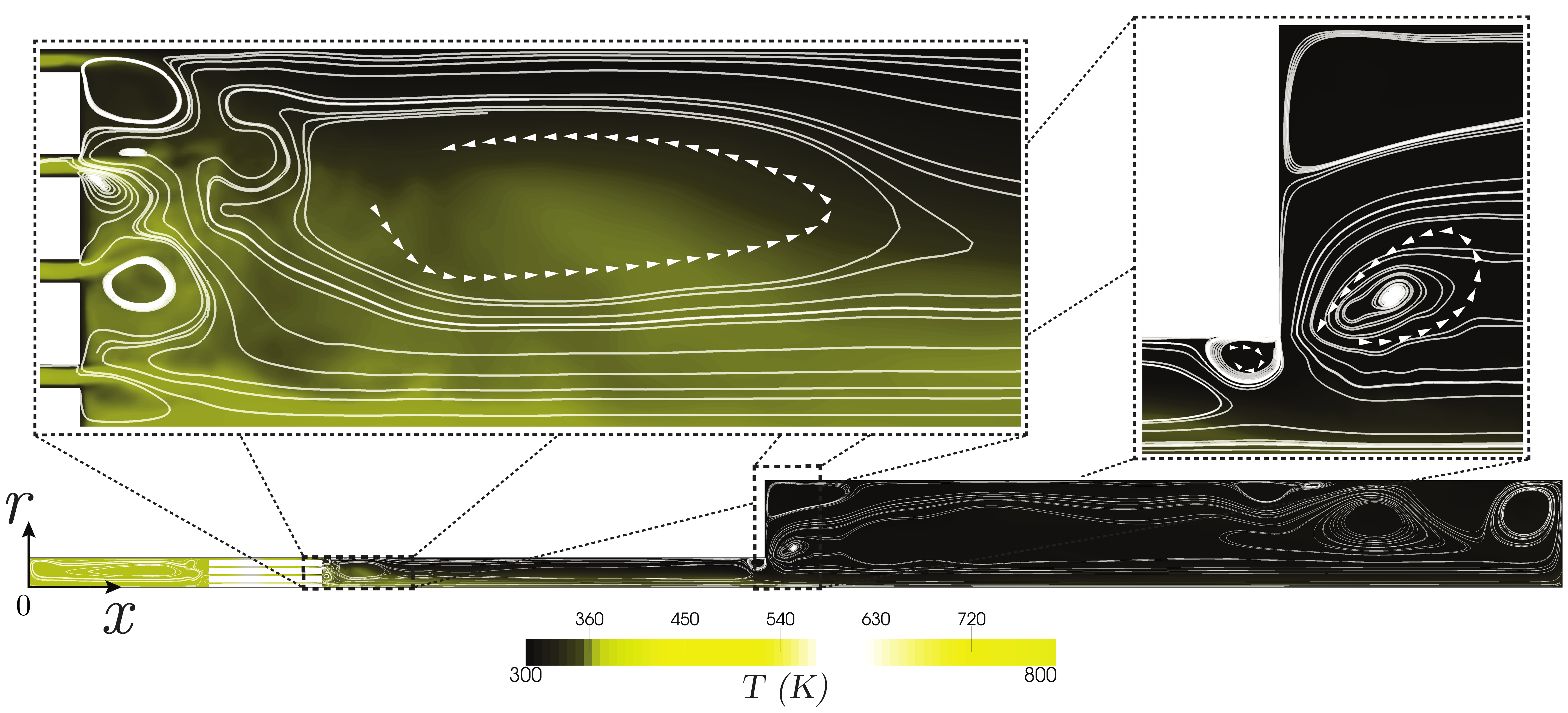}
	\caption{
		Velocity streamlines, with orientation of circulation (shown with white arrow heads), and temperature contours obtained by averaging over 2 acoustic cycles under limit cycle conditions for temperature setting 5, grid-resolution/stack-type C/I. Supplementary material available online show visualizations of instantaneous fluid temperature.
	}
	\label{fig:stackstreaming}
\end{figure}

As the pressure amplitude rises, acoustic nonlinearities become important, as shown by the cycle-averaged temperature and velocity fields in figure~\ref{fig:stackstreaming}. Periodic flow separation occurring at each location of abrupt area change creates wave-induced Reynolds stresses, as also analysed by \citet{ScaloLH_JFM_2015}, driving recirculations in the streaming velocity. At the edges of the thermoacoustic stack in particular, small scale flow separations (of the order of $h_s$, see table \ref{tab:stack_configurations}) associated with entrance effects alter the effective porosity and lead to \textit{vena contracta}, lowering the effective stack porosity at limit cycle and increasing the frequency, consistent with the results in \cref{fig:porosityeffects_a}.

In the present configuration, without an opposing ambient heat exchanger, thermoacoustic transport and streaming, typically a concern for travelling-wave engines, are expected to directly affect the thermal-to-acoustic efficiency. The streaming velocity near the centreline follows the direction of the acoustic power (discussed in \cref{sec:energyextraction}) along the positive axial direction, from the stack to the resonator, where it is collected and partly absorbed in the presence of piezoelectric energy extraction. This qualitatively explains temperature observations in the experiments, which show heat leakage downstream of the stack and a slow relaxation of the mean temperature gradient in the regenerator.

%% file: Results_PiezoTDIBCModel.tex
\section{Modelling of Piezoelectric Acoustic Energy Extraction via TDIBC}
\label{sec:modeling_a_physical_piezoelectric}

In this section, the general steps required for a causal multi-oscillator fit of a given impedance are outlined.
The specific goal of modelling a piezoelectric diaphragm as a purely acoustically absorbing element in time-domain Navier--Stokes calculations does not affect the generality of the procedure. 
A simple one-port model, derived by collapsing the experimentally measured two-port model for the piezoelectric diaphragm in the form of $\hat{p}\left(\omega \right) = Z_{exp}\left(\omega \right) \hat{u}\left(\omega \right)$ is first discussed (\cref{subsec:one-port_model}). 
Derivation of single-oscillator (\cref{subsec:singleoscillatorbroadbandimpedance}) and multi-oscillator (\cref{subsec:multioscillatorbroadbandimpedance}) approximations to $Z_{exp}(\omega)$ are then presented. Since values of $Z_{exp}$ above 450 Hz are deemed unphysical, an additional constraint to the multi-oscillator fitting strategy has been introduced and is discussed below.

\subsection{One-Port Electromechanical Impedance Model}
\label{subsec:one-port_model}

The experimental characterization of the electromechanical frequency response of a PZT-5A piezoelectric diaphragm has been carried out by \citet{SmokerNAB_2012}, resulting in the system of equations
\begin{equation} \label{eq:piezo_system_xqpV}
	\begin{Bmatrix}
		\hat{x}_\textrm{c} \left(i\omega\right) \\
		\hat{q}\left(i \omega \right)
	\end{Bmatrix}
	=
	\left[\begin{matrix}
	      T_{11} \left(i\omega\right) & T_{12} \left(i\omega \right) \\
		  T_{21} \left(i\omega\right) & T_{22} \left(i\omega \right)
		  \end{matrix}
	\right]
	\begin{Bmatrix}
		\hat{p} \left(i\omega\right) \\
		\hat{V}\left(i \omega \right)
	\end{Bmatrix}
\end{equation}
where $\hat{x}_\textrm{c}$ [m], $\hat{q}$ [C], $\hat{p}$ [Pa], and $\hat{V}$ [V]  are, respectively, the complex amplitudes of the fluctuating centreline displacement (positive along the $x$ direction), electric charge, pressure, and voltage. The electromechanical admittances, $T_{mn}$, in \eqref{eq:piezo_system_xqpV} have been measured for a broadband range of frequencies and fitted with the rational function
\begin{equation} \label{eq:fitted_Tmn_admittance}
	T_{mn} \left(i\omega\right) =
	\frac{a_{mn,10} \left(i\omega\right)^{10} +\cdots +  a_{mn,1} \left(i\omega \right) + a_{mn,0} }
	{b_{mn,10} \left(i\omega\right)^{10} +\cdots + b_{mn,1} \left(i\omega \right) + b_{mn,0} }
\end{equation}
where the fitting coefficients $a_{mn}$ and $b_{mn}$ are reported in \cref{appendix:transferfunctioncoefficients}.
Expressing the centreline displacement, $\hat{x}_\textrm{c}$, and charge, $\hat{q}$, in terms of velocity, $\hat{u}_c$, and current, $\hat{I}$, respectively, yields
\begin{equation} \label{eq:piezo_system_uIpV}
	\begin{Bmatrix}
		\hat{u}_c \left(i\omega\right) \\
		\hat{I}\left(i \omega \right)
	\end{Bmatrix}
	=
	\left[\begin{matrix}
	      i\omega\,T_{11} \left(i\omega\right) & i\omega\,T_{12} \left(i\omega \right) \\
		  i\omega\,T_{21} \left(i\omega\right) & i\omega\,T_{22} \left(i\omega \right) \end{matrix}
	\right]
	\begin{Bmatrix}
		\hat{p} \left(i\omega\right) \\
		\hat{V}\left(i \omega \right)
	\end{Bmatrix} \; .
\end{equation}
In the experiments, the piezoelectric diaphragm drives a load of resistance $R_L=3170\;\Omega$, relating voltage and current via $\hat{V} = R_L\, \hat{I}$. This allows \eqref{eq:piezo_system_uIpV} to be collapsed into a one-port model
\begin{equation}
		\hat{u}_c = \left[ i\omega \, T_{11}  + i\omega \,  T_{12}\, \frac{i\omega T_{21} }{1-i\omega \, T_{22}\, R_L } R_L \right] \hat{p}
\end{equation}
which corresponds to the (purely mechanical) impedance
\begin{equation}  \label{eq:experimental_Z}
	\Ztarget \left(\omega\right) = \left[ i\omega \, T_{11} + i\omega \,  T_{12}  \, \frac{i\omega T_{21} }{1-i\omega \, T_{22} \, R_L } R_L \right]^{-1}
	\end{equation}
and wall softness
\begin{equation}  \label{eq:experimental_Wth}
	\Wthtarget \left(\omega\right) =  \frac{2\,Z_0}{Z_0+\Ztarget\left(\omega\right)}
\end{equation}
where $Z_0=\rho_0 a_0$ is the characteristic specific acoustic impedance of the gas. While the impedance \eqref{eq:experimental_Z} is based on experimental measurements of the broadband frequency response of the diaphragm (measured at the centreline only), it is not necessarily computationally stable and/or physically admissible (as discussed below) and therefore cannot be applied directly in time-domain nonlinear Navier--Stokes simulations. 

\begin{figure}
    \centering
    \includegraphics[width=0.80\linewidth]{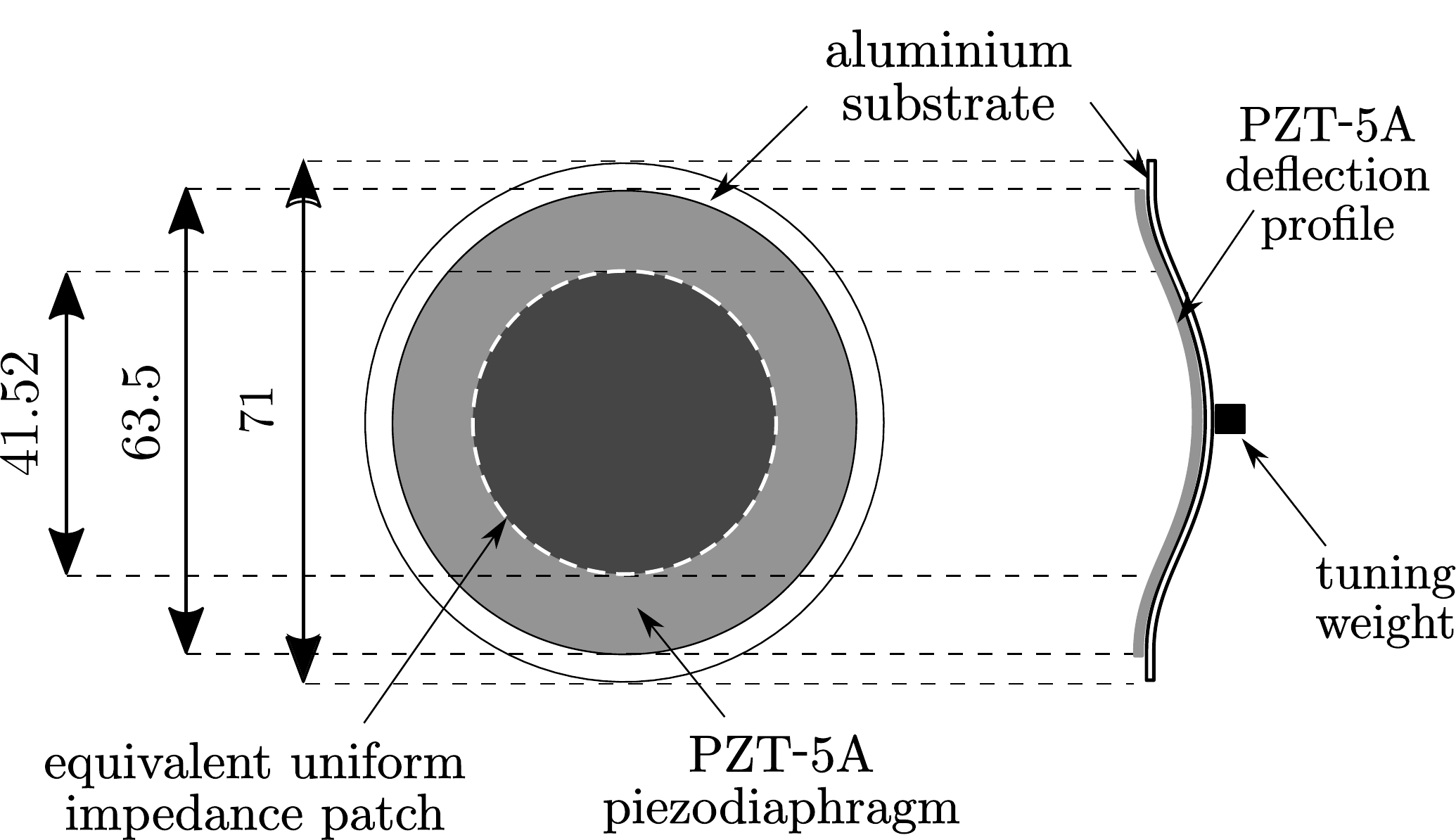}
	\caption[Conversion of experimentally-measured deflection profile $|\hat{x}(r)|$ to an equivalent top-hat profile preserving the volumetric flow rate amplitude. ]{Illustration of the PZT-5A piezoelectric diaphragm installation by {\protect \cite{SmokerNAB_2012}~} capping the resonator of the TAP engine (see figure \ref{fig:computational_setup}). All lengths are given in millimetres. An aluminium substrate backs the piezoelectric diaphragm with an added weight at the centreline used for tuning. Rational polynomial fit of impulse response measurements of the electromechanical admittances \eqref{eq:piezo_system_xqpV} has been performed solely based on centreline measurements. To model the piezoelectric diaphragm in the Navier--Stokes simulations, a patch of uniformly distributed impedance is used, with size scaled to match the acoustic power output of the actual PZT-5A diaphragm for the same pressure amplitude levels.
	}
    \label{fig:deflectionprofile}
\end{figure}

The Navier--Stokes calculations were carried out with a computationally and physically admissible impedance approximating \eqref{eq:experimental_Z} (approximations derived below), uniformly applied over a circular area scaled in size to preserve the surface-averaged displacement amplitude of the experimentally-measured deflection profile by \citet{SmokerNAB_2012}. This technique allows the matching of overall acoustic power output for the same pressure amplitude levels (\cref{fig:deflectionprofile}). Impedance boundary conditions impose a specific relationship between the Fourier transforms of velocity and pressure at a stationary boundary and should not be confused with the imposition of a moving boundary. The nature of the resulting power extraction is an acoustic-to-mechanical energy conversion, since it is associated with a mechanical deflection of a membrane driven by acoustic excitation. Mechanical-to-electric energy conversion is not directly accounted for.

\subsection{Single-Oscillator Approximation}
\label{subsec:singleoscillatorbroadbandimpedance}

A simple approach towards constructing a computationally admissible impedance approximating the experimental value \eqref{eq:experimental_Z} is to use a damped Helmholtz oscillator model~\citep{TamA_1996}, expressed as the three-parameter impedance
\begin{align} \label{eq:three_parameter_model}
	Z\left(\omega\right) &= Z_0 \left[ R + i \left(\omega X_{+1} - X_{-1} / \omega\right) \right]
\end{align}
where $R$, $X_{+1}$ and $X_{-1}$ are the resistance, acoustic mass and stiffness, respectively. Only one undamped resonant frequency,
\begin{equation} \label{eq:three_parameter_model_resonant_angular_frequency}
		\omega_0 = 2\pi f_0 =  \sqrt{\frac{X_{-1}}{X_{+1}}}
\end{equation}
is associated with \eqref{eq:three_parameter_model}, with corresponding wall softness, expressed in the Laplace domain, 
\begin{equation}
\widehat{\widetilde{W}}_s (s) = \frac{2\,s}{s^2 X_{+1} + s\left(1+R\right)+ X_{-1} } \, ,
\label{subeq:wth_z_s_three_parameter_model}
\end{equation}
where $s = i\omega$, with $\omega$ here being extended to the complex domain (via an abuse of notation). Computational admissibility requires the time-domain equivalent of \eqref{subeq:wth_z_s_three_parameter_model} to be causal, that is, the poles of the wall softness must lie in the left-half of the $s$-plane (negative real part) or, equivalently, in the upper-half of the complex $\omega$-plane (positive imaginary part). Poles of $\widehat{\widetilde{W}}_s (s)$ in the $s$-domain are in biunivocal correspondence with the poles of $\widehat{\widetilde{W}}_{\omega} (\omega)$ in the complex $\omega$-domain.

The wall softness of a generic oscillator with a single resonant frequency can be expressed via a decomposition in partial fractions in the Laplace domain, 
\begin{subequations}
	\begin{align}
		\widehat{\widetilde{W}}_s (s) &= \frac{\mu}{s-p}+\frac{\conj{\mu}}{s-\conj{p}} \label{subeq:wth_mulambda}\\
		&= \frac{2\left(as-C\right)}{s^2+\left(-2c\right)s+\left(c^2+d^2\right)} \, ,
		\label{subeq:wth_abcd}
	\end{align}
	\label{eqn:wth_polesresidues}
\end{subequations}
with one set of complex conjugate residues ($\mu,\mu^*$) and poles ($p,p^*$), where $\mu=a+i\,b$ and $p=c+i\,d$ with $a, b, c, d \in \mathbb{R}$.

In order for ($\mu,\mu^*$) and ($p,p^*$) to represent a single damped Helmholtz oscillator in the form of the three-parameter model \eqref{eq:three_parameter_model}, the following conditions, derived by comparing \eqref{subeq:wth_abcd} with \eqref{subeq:wth_z_s_three_parameter_model}, must be satisfied:
\begin{subequations} \label{eq:fitting_constraints_for_abcd}
	\begin{align}
		C &= 0 		\label{subeq:phaseparameter} \\
		1+R&= \frac{-2c}{a} 		\label{subeq:one_plus_R} \\
		X_{+1}&=\frac{1}{a} \label{subeq:Xplusone}\\
		X_{-1}&=\frac{c^2+d^2}{a}  \label{subeq:Xminusone}
	\end{align}
\end{subequations}
where $C=bd+ac$ is the phase parameter. Physical admissibility (boundary is a passive acoustic absorber) and causality require $R = -(1+2\,c/a)>0$  and $c=\Real \left(p\right) < 0 $, respectively. It is important to stress that a generic oscillator of the form \eqref{subeq:wth_abcd} cannot be equivalent to the single damped Helmholtz oscillator \eqref{subeq:wth_z_s_three_parameter_model} unless its phase parameter is zero \eqref{subeq:phaseparameter}. \citet{ScaloBL_PoF_2015} have demonstrated that it is possible to perform turbulent flow simulations with imposed wall-impedance of type \eqref{subeq:wth_z_s_three_parameter_model} without encountering numerical stability issues, confirming that \eqref{subeq:wth_z_s_three_parameter_model} is in fact physically and computationally admissible.

\citet{FungJ_2001} have suggested that it is not necessary for a single-oscillator model such as \eqref{eqn:wth_polesresidues} to have a zero phase parameter for its use in time-domain computations. However, in preliminary numerical trials, it was found that leaving the phase parameter unconstrained ($C\ne0$)
leads to unstable numerical simulations and causing, in our case, spurious mode switching and near-DC acoustic power extraction.

As seen from \eqref{subeq:wth_abcd}, the phase parameter is dominant in the low frequency limit ($s \to 0$), thus influencing the phase of $Z(\omega)$ over a broad range of near-DC frequencies. A non-zero phase parameter yields a purely real, non-zero, and finite $Z(\omega)$ at zero frequency. Because the experimentally-measured wall softness \eqref{eq:experimental_Wth} has a zero magnitude (infinite impedance magnitude) in the DC limit, a zero phase parameter $C=0$ is necessary in both the single- and multi-oscillator impedance approximations to \eqref{eq:experimental_Wth} (the latter discussed below) to retain physical admissibility.

\begin{figure}
	\centering
	\includegraphics[width=0.95\linewidth]{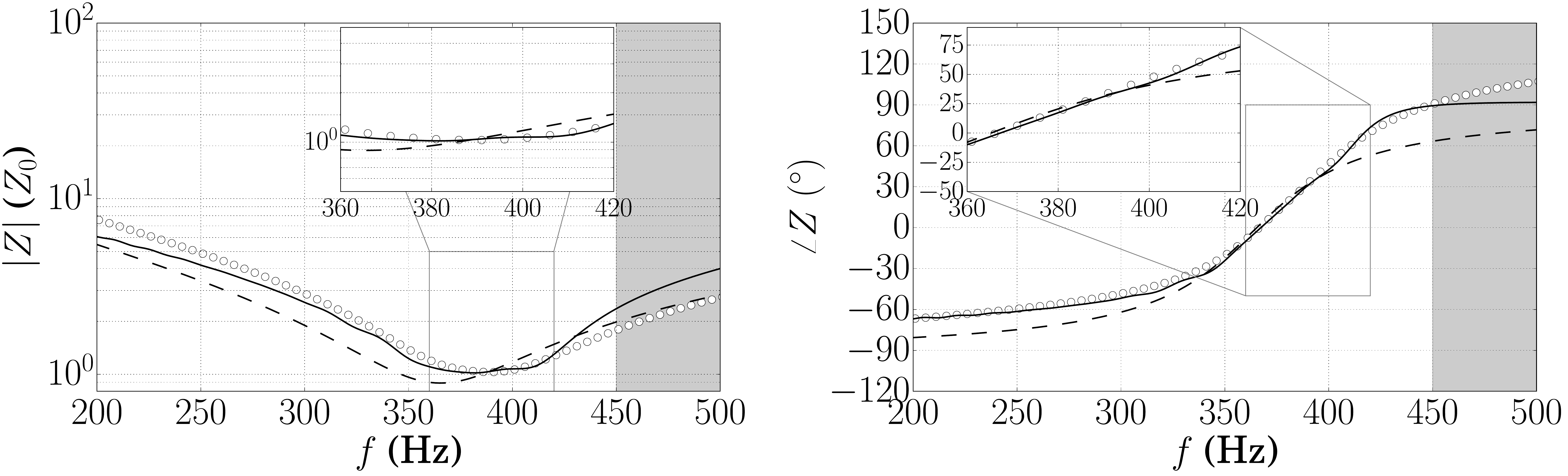}
	\put(-370,10){(a)}
	\put(-180,10){(b)}

	\includegraphics[width=0.95\linewidth]{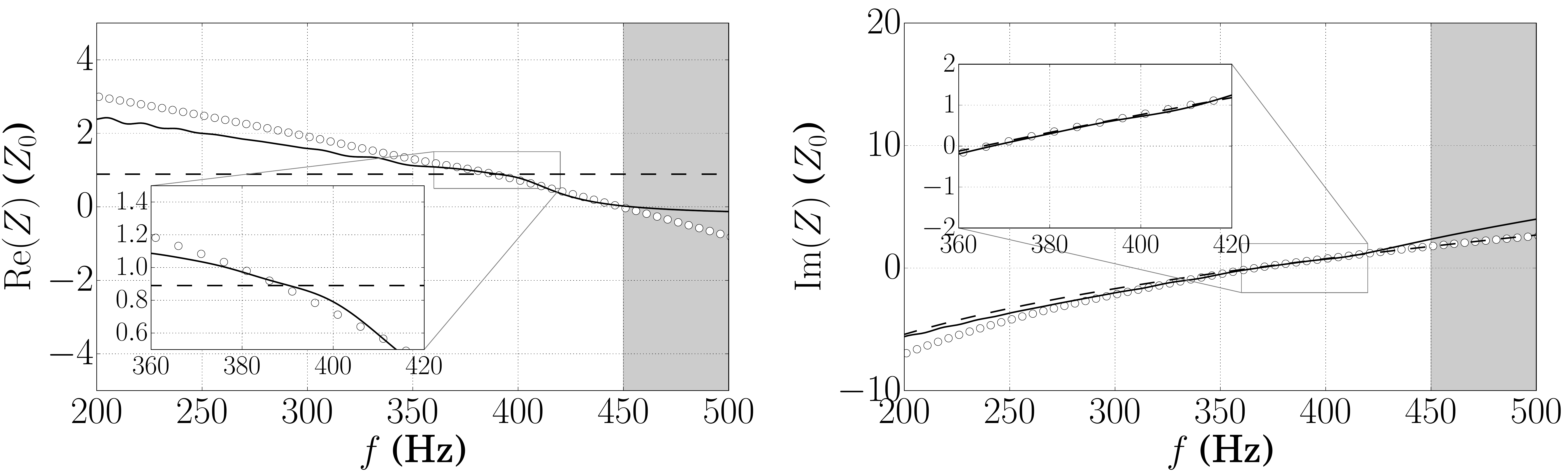}
	\put(-370,10){(c)}
	\put(-180,10){(d)}
		\caption[..]{Magnitude (a), phase (b), real part (c) and imaginary part (d) of the experimentally measured impedance \eqref{eq:experimental_Z} \legenddots{}, single-oscillator impedance model \eqref{eq:three_parameter_model},\eqref{subeq:wth_z_s_three_parameter_model} \legenddashed{} and multi-oscillator impedance fit \eqref{subeq:broadbandfungjuform} with $\bar{\alpha}=0.06$ and $n_o=18$ oscillators \legendline{}. The single-oscillator model is fitted in the range 388$\pm$10 Hz while the multi-oscillator model is fitted over the entire frequency range and with wall softness $\widehat{\widetilde{W}}$ constrained to have a positive real part, resulting in $\Real \left(Z\right) \geq -Z_0$ (see text). \shadedareaexplanation{}
		}
		\label{fig:fitting_broadband}

\end{figure}

Following the aforementioned considerations, the impedance \eqref{eq:experimental_Z} was first approximated by the three-parameter impedance model \eqref{eq:three_parameter_model} (guaranteeing a zero phase parameter) with $R$, $X_{+1}$, and $X_{-1}$ determined directly via least-squares fitting of $\Real (Z_{exp})$ and $\Imag (Z_{exp})$, where $Z_{exp}$ is the impedance corresponding to the collapsed two-port model \eqref{eq:experimental_Z}. The fitting window used is $f = 388\textrm{ Hz} \pm 10\textrm{ Hz}$
with resulting parameters reported in table \ref{tab:fitting_singleoscillator}. As expected, good agreement is found only for frequencies close to $f = 388\textrm{ Hz}$ (figure~\ref{fig:fitting_broadband}). The largest discrepancies are in the values of resistance $R$ (not constant in the experiments), which is responsible for differences in the location of the minima of $|Z|$. The latter is an attractor for the thermoacoustically unstable mode at the limit cycle. Negative values of resistance in the experimentally-measured impedance are observed for frequencies above 450 Hz, which is unphysical for a passive acoustic element and therefore are a challenge in the context of deriving a multi-oscillator impedance approximation.

\begin{table*}
	\small
	\footnotesize
	\centering

	\begin{tabular}{llll}
		\hline
		$R$   & $X_{+1}$ (rad$^{-1}$$\,$s)   & $X_{-1}$ (rad$\,$s$^{-1}$)  &   \\
		0.8909 &0.001842& 9703.2390 & \\
		\hline
		$a$  (rad$\,$s$^{-1}$) & $b$  (rad$\,$s$^{-1}$) & $c$  (rad$\,$s$^{-1}$) & $d$  (rad$\,$s$^{-1}$) \\
		542.9859 & 124.5988 & -513.3750 & 2237.2233\\
		\hline
		 $f_0$ (Hz)  & $\bar{\alpha}$  & $A_1$ (rad$\,$s$^{-1}$) & $B_1$ (rad$^{2}\,$s$^{-2}$) \\
		 365.3194 & 0.2237 & 1085.9718 & 0 \\
		\hline
	\end{tabular}
	\normalsize
	\caption{Parameters for equations \eqref{eq:three_parameter_model} (first row), \eqref{subeq:wth_abcd} (second row) and \eqref{subeq:broadbandfungjuform} (third row), all corresponding to the same single-oscillator impedance used to approximate the target measured impedance \eqref{eq:experimental_Z} -- both of which are plotted in figure \ref{fig:fitting_broadband}.}
		\label{tab:fitting_singleoscillator}
\end{table*}

\begin{figure}
	\centering
	\includegraphics[width=0.9\linewidth]{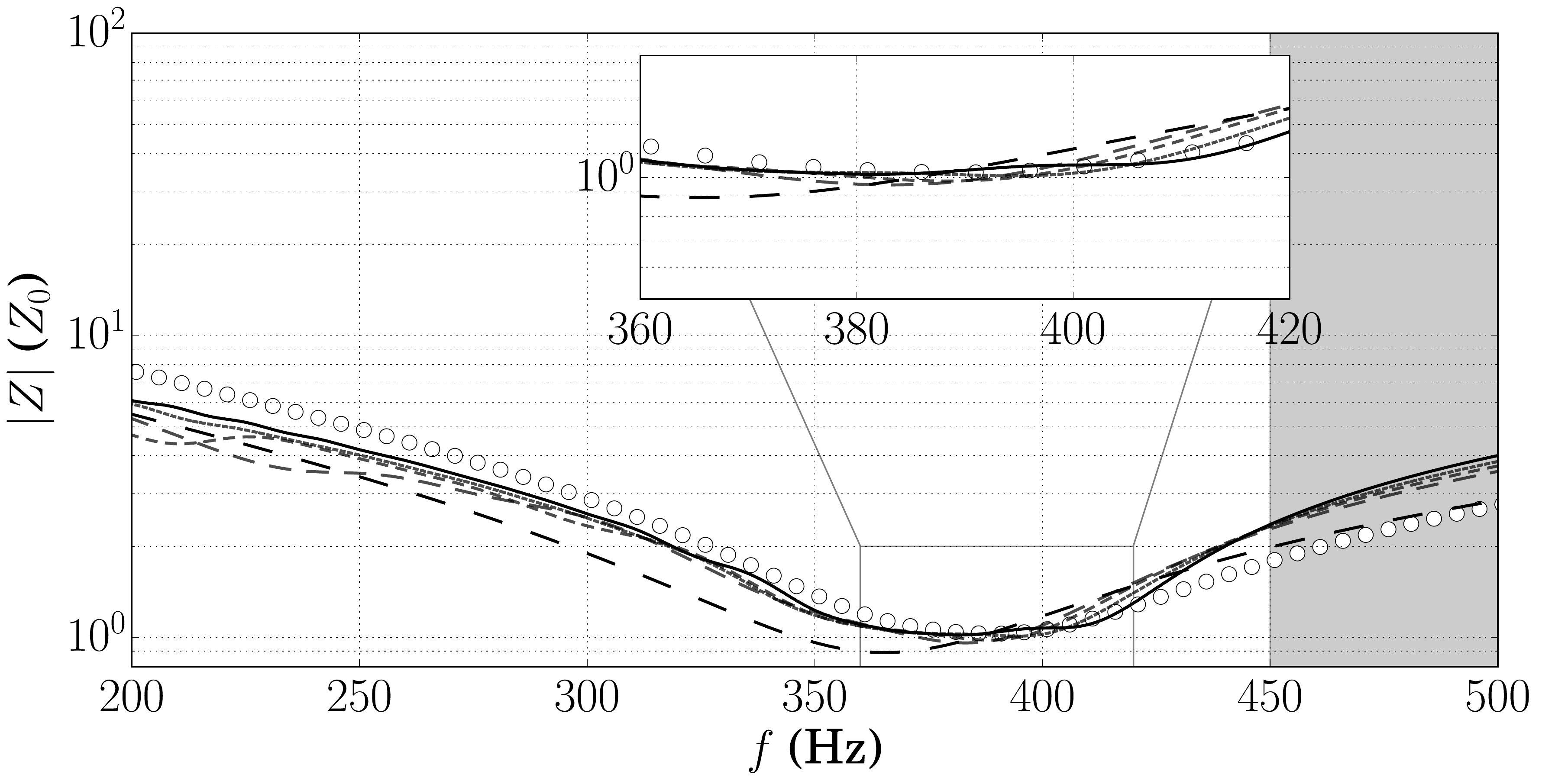}
	\put(-350,20){(a)}
	
	\includegraphics[width=0.9 \linewidth]{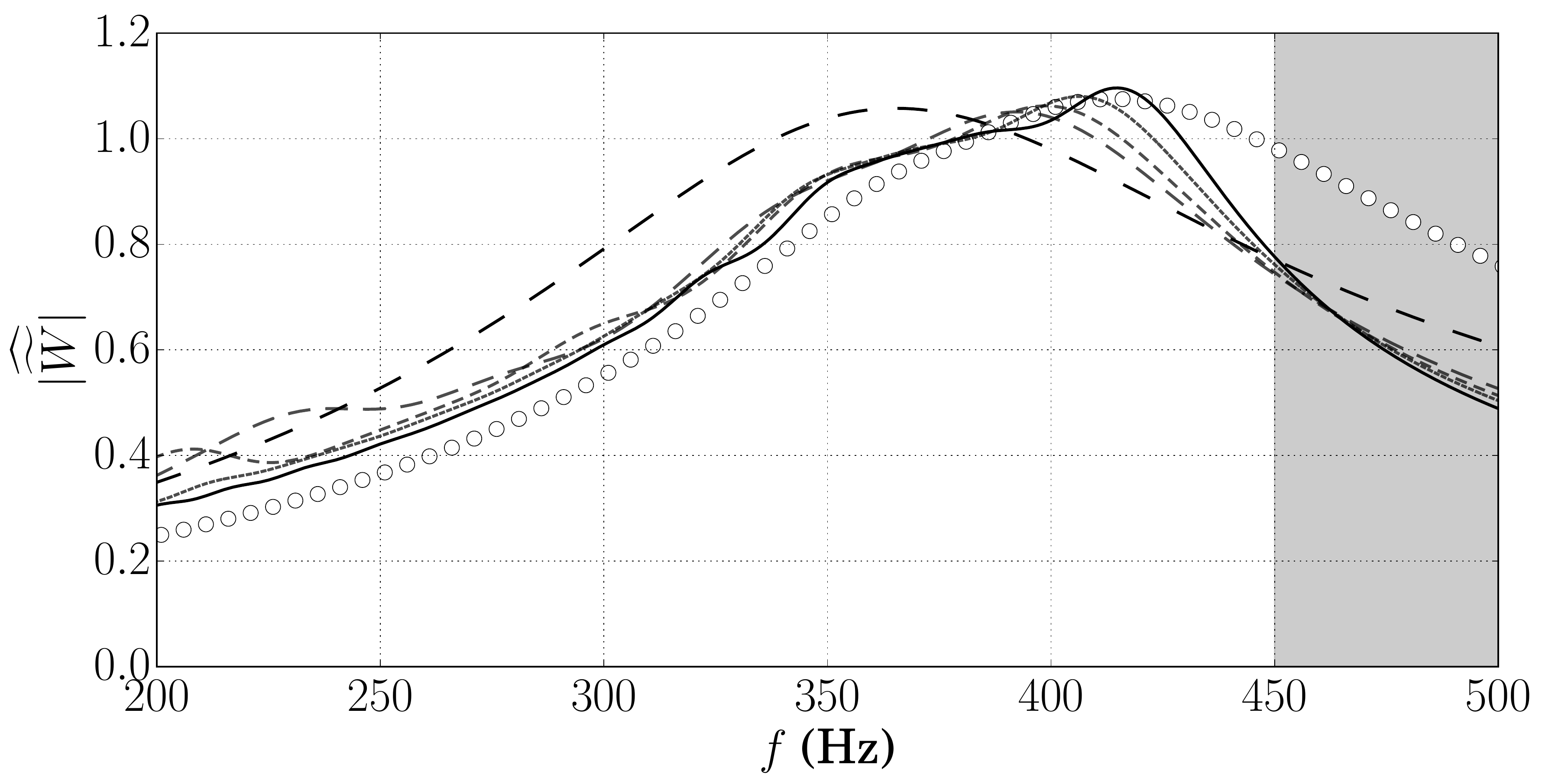}
	\put(-350,20){(b)}
	
	\caption[..]{Magnitude of experimentally measured impedance \eqref{eq:experimental_Z} (a) and (b) wall softness magnitude \legenddots{} compared with multi-oscillator model for $\bar{\alpha} = 0.2237$ ($\numofoscillators=1$) \legenddashedLong{}, $\bar{\alpha}=0.12$ ($\numofoscillators=4$)  \legenddashedMedium{}, $\bar{\alpha}=0.10$ ($\numofoscillators=6$) \legenddashedShort{}, $\bar{\alpha}=0.08$ ($\numofoscillators=13$) \legenddashedShortest{}, to $0.06$ ($\numofoscillators=18$) \legendline{}. \shadedareaexplanation
	}
	\label{fig:fitting_broadband_taylor}
\end{figure}

\subsection{Multi-Oscillator Approximation}
\label{subsec:multioscillatorbroadbandimpedance}

In order to fit \eqref{eq:experimental_Wth} over a broader frequency range, a linear superposition of the wall softness coefficients of $n_o$ oscillators, each decomposed in partial fractions with one conjugate pair of residues ($\mu_k,\mu_k^*$) and poles ($p_k,p_k^*$), is used, yielding
\begin{subequations}
	\begin{align}
		\widehat{\widetilde{W}}_{s, exp} (s) \simeq  \sum_{k=1}^{\numofoscillators} \widehat{\widetilde{W}}_{s, k} (s) &= \sum_{k=1}^{\numofoscillators} \left[ \frac{\mu_k}{s-p_k}+\frac{ {\mu_k^*}}{s-p_k^*} \right] \label{subeq:broadbandsummation} \\
		&= \sum_{k=1}^{\numofoscillators}  \frac{A_k \left(i\omega\right) + B_k}{\left(i\omega + \bar{\alpha} \omega_{0,k}\right)^2 + \omega_{0,k}^2 \left(1-\bar{\alpha}^2\right)}  \label{subeq:broadbandfungjuform}
	\end{align}
\end{subequations}
where \eqref{subeq:broadbandfungjuform} is an alternative form to \eqref{subeq:wth_abcd} adopted by \citet{FungJ_2004}, where $\omega_{0,k}$ is the resonant (or basis) frequency \eqref{eq:three_parameter_model_resonant_angular_frequency} of the $k$-th oscillator, $\bar{\alpha}$ is a damping parameter (common to all oscillators), and $A_k$ and $B_k$ are fitting coefficients corresponding to $2a$ and $-2\,C$ in the single-oscillator model in \eqref{subeq:wth_abcd}. 
The experimentally measured wall softness is expected to approach zero for $\omega \to \infty$ and $\omega \to 0$ (with implications on $B_k=0$, discussed below), making its functional form better suited for fitting than the impedance itself, which, in the case of a damped Helmholtz resonator, diverges for the same extremes. 
Moreover, fitting the wall softness as a linear superposition of oscillators is consistent with the numerical implementation in the time-domain, whereas linearly superimposing impedances is not.
Note that the linear superimposition of wall softnesses (as in \cref{subeq:broadbandsummation}), which is the approach used in the present work, is not equal to the wall softness resulting from the linear superposition of the corresponding single-oscillator impedances, that is
\begin{equation}
\sum_{k=1}^{\numofoscillators} \widehat{\widetilde{W}}_{\omega, k} (\omega)\ne \frac{2\,Z_0}{Z_0+\sum_{k=1}^{\numofoscillators}Z_k(\omega)}
\end{equation}
where
\begin{equation}
Z_k(\omega) = Z_0 \left( \frac{2}{\widehat{\widetilde{W}}_{\omega, k} (\omega)} - 1\right).
\end{equation}

The damping parameter $\bar{\alpha}$ in \eqref{subeq:broadbandfungjuform} -- common to all $n_o$ oscillators -- controls the bandwidth of the frequency response of each oscillator centred about its basis frequency; for low (high) values of $\bar{\alpha}$, each oscillator will exhibit a narrowband (broadband) response. Therefore, for a given fitting frequency window, a low (high) value of $\bar{\alpha}$ will require a larger (smaller) number of oscillators to approximate a given wall softness. A large number of narrowband oscillators results in a more accurate fit -- requiring, however, a closer spacing of basis frequencies.

The impedance $Z_{exp}$ has been fitted with the following numbers of oscillators: $n_o=1$, $n_o=4$, $n_o=6$, $n_o=13$, $n_o=18$ (see equation \eqref{subeq:broadbandfungjuform} and figure \ref{fig:fitting_broadband_taylor}). 
For the single-oscillator case, values of $f_{0,1} ( = \omega_{0,1}/2\pi)$, $\bar{\alpha}$, $A_1$, corresponding to the single-oscillator model in \eqref{subeq:wth_z_s_three_parameter_model}, are reported in the third row of \cref{tab:fitting_singleoscillator}. 
For the multi-oscillator case, $n_o>1$, values of $f_{0,k} ( = \omega_{0,k}/2\pi)$, $\bar{\alpha}$, $A_k$ are reported in \cref{tab:alphabar_selection}. For a given $\bar{\alpha}$, basis frequencies were selected through a gradient descent-based iterative method such that the approximate impedance $Z_\textrm{fit}$ is the least squares minimizer of $\log{\left|Z_{exp}\right|} -\log{\left|Z_\textrm{fit}\right|}$ and $\arg{Z_{exp}}-\arg{Z_\textrm{fit}}$. 

As $n_o$ increases, the basis frequencies are more closely spaced, corresponding to a decrease in $\bar{\alpha}$ and an increase in the accuracy of the fit in the frequency domain (\cref{fig:fitting_broadband_taylor}). For each $\bar{\alpha}$ and thus for a particular $n_o$, the impedance, as a function of frequency and basis frequencies, is fitted with least squares over frequencies $f\in[1, 440]\textrm{ Hz}$, with $5.5$-fold weighting on $f\in[360, 440]\textrm{ Hz}$ and $22$-fold weighting on $f\in[378, 398]\textrm{ Hz}$.

\begin{table*}
	\centering
	\begin{tabular}{r|cccc}
		\hline
		 & $\bar{\alpha}=0.12$   & $\bar{\alpha}=0.10$ &$\bar{\alpha}=0.08$ & $\bar{\alpha}=0.06$  \\
		&  ($\numofoscillators=4$)&($\numofoscillators=6$)& ($\numofoscillators=13$) &($\numofoscillators=18$)
		\\
		\hline
		\begin{tabular}[t]{@{}l@{}}
					$k\quad$  \\
			\hline
1 \\
2\\
3\\
4\\5\\6\\7\\8\\9\\10\\11\\12\\13\\14\\15\\16\\17\\18
		\end{tabular}
		 &
			\begin{tabular}[t]{@{}r@{\hspace{0.2cm}}r@{}}
				$f_{0,k}$ (Hz) & $A_k$ \\
					\hline
238.8950 & 62.3875 \\
283.3134 & 69.2957 \\
342.9779 & 230.9998 \\
391.7290 & 458.0321 \\
		\end{tabular}
		&
			\begin{tabular}[t]{@{}r@{\hspace{0.2cm}}r@{}}
				$f_{0,k}$ (Hz) & $A_k$ \\
								\hline
212.9031 & 42.1921 \\
246.4396 & 18.3780 \\
266.9725 & 28.7000 \\
302.2977 & 90.2265 \\
351.9786 & 238.5549 \\
397.4895 & 365.0711 \\
			\end{tabular}
		&
			\begin{tabular}[t]{@{}r@{\hspace{0.2cm}}r@{}}
				$f_{0,k}$ (Hz) & $A_k$ \\
				\hline
185.8443 & 24.7860\\
215.1331 & 13.2845 \\
220.0557 & 2.8828 \\
234.1201 & 11.1785 \\
245.2838 & 11.1954 \\
256.5088 & 9.2693 \\
268.4031 & 21.1408 \\
287.4474 & 28.9818 \\
301.5441 & 16.2862 \\
315.1012 & 52.7308 \\
345.6765 & 142.7597 \\
370.8757 & 144.4980 \\
404.7628 & 271.5407 \\
			\end{tabular}
		&
			\begin{tabular}[t]{@{}r@{\hspace{0.2cm}}r@{}}
				$f_{0,k}$ (Hz) & $A_k$ \\
				\hline
				153.2085 & 11.4054 \\
				171.8258 & 6.7512 \\
				187.2862 & 8.1940 \\
				203.0648 & 9.5276 \\
				219.2944 & 11.5028 \\
				235.5885 & 13.4362 \\
				250.6609 & 12.7558 \\
				260.5702 & 9.9235 \\
				271.8029 & 15.1965 \\
				278.7214 & 5.1049 \\
				287.4075 & 20.2923 \\
				299.8438 & 0.0241 \\
				302.9752 & 35.5632 \\
				324.9863 & 67.7324 \\
				349.9786 & 100.3191 \\
				367.5093 & 84.2461 \\
				385.6369 & 110.2340 \\
				412.6176 & 189.4710 \\
			\end{tabular}
		\\
		\hline
		\begin{tabular}[c]{@{}r@{}}
					fitting error \\
					(residual)
		\end{tabular}
		& 99.51
		& 84.96
		& 71.93
		& 57.33
		\\
		\hline
	\end{tabular}
	\caption{
		Collection of basis frequencies used in \cref{fig:fitting_broadband_taylor}, $f_{0,k} = \omega_{0,k}/2\pi$ for each number of oscillators $n_o$ and damping parameter $\bar{\alpha}$, and fitting coefficients $A_k$. In all cases, values of the phase parameter are set to zero, $B_k=0$.
	}
	\label{tab:fitting_broadband}
	\label{tab:alphabar_selection}
\end{table*}

As seen in \cref{fig:fitting_broadband}, at higher frequencies, the real component of the experimentally-measured impedance becomes negative, which is not consistent with a passive acoustic element, and may be the spurious result of the sampling rate used for the eigensystem realization algorithm as reported by \citet{SmokerNAB_2012} or simply the extrapolation of the rational polynomial fit beyond the tuned frequency of the piezoelectric diaphragm. To avoid unphysical values of the reconstructed impedance at high frequencies, $A_k$ in \eqref{subeq:broadbandfungjuform} is constrained to be positive, since negative $A_k$ can lead to unbounded negative resistance. By combining the constraints
\begin{subequations}
	\begin{align}
	A_k&\geq 0 \\
	B_k&=0 \, ,
	\end{align}
\end{subequations}
where $B_k$ is the phase parameter (see discussion in \cref{subsec:singleoscillatorbroadbandimpedance}), with equations~\ref{eq:experimental_Wth} and~\ref{subeq:broadbandfungjuform}, the real part of the resulting impedance $Z$ has a lower bound, i.e. $\Real \left(Z\right)>-Z_0$. Without such a constraint, a multi-oscillator impedance model could fit, with arbitrary accuracy, the given experimentally-measured impedance \eqref{eq:experimental_Z}, but would cause the impedance to inject energy into the system for $f>450$Hz, hence exciting its second mode (at $\simeq 640$ Hz, where $\Real \left(Z_{exp}\right)$ is large and negative). This causes unphysical mode switching (see \cref{subsec:damping_modes}), with the piezoelectric diaphragm no longer acting as a passive element but a (spurious) driver of oscillations.

In the following, results from Navier--Stokes calculations with piezoelectric energy absorption with the multi-oscillator model with $n_o=18$ and $\bar{\alpha} = 0.06$ are shown, since this model provides the highest level of fidelity over the frequency range of interest.

%% file: Results_EnergyExtraction.tex
\section{Acoustic Energy Extraction at Limit Cycle}
\label{sec:energyextraction}

\subsection{Thermal-to-Mechanical Efficiency}
\begin{figure*}
    \centering
    \includegraphics[width=0.95\linewidth]{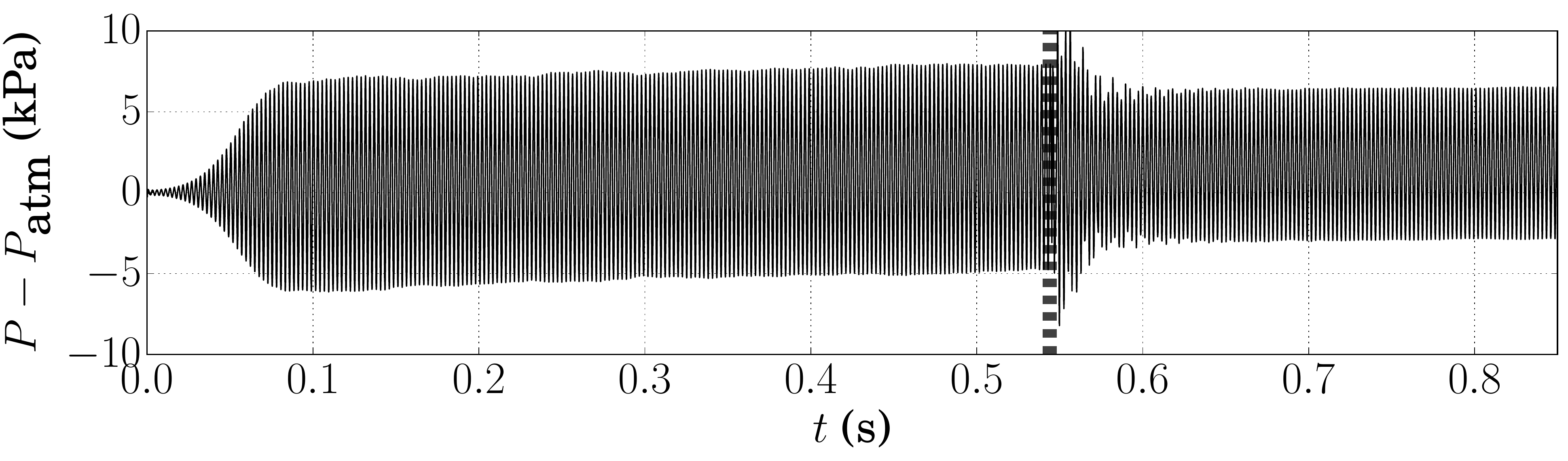}
    \caption{
    	Time series of pressure in the hot cavity for temperature setting 5 and grid-resolution/stack-type C/I from start-up to limit cycle, without ($t < 0.544\textrm{ s}$) and with ($t > 0.544\textrm{ s}$) piezoelectric energy absorption, as modelled by the multi-oscillator impedance model \eqref{subeq:broadbandfungjuform} with $n_o=18$ (table \ref{tab:fitting_broadband}).
   	}
    \label{fig:pressure-ibc}
\end{figure*}

Acoustic energy extraction, modelled via the TDIBCs designed in \cref{sec:modeling_a_physical_piezoelectric}, is applied once a limit cycle without energy absorption is achieved, and only for the TAP engine model with stack type I. The latter most closely matches the porosity and hydraulic radius of the regenerator used in the experiments (table \ref{tab:stack_configurations}) and, as a result, the operating frequency at which the piezoelectric diaphragm is tuned ($\sim$388 Hz).

\begin{figure}
		\centering
		\includegraphics[width=0.95\linewidth]{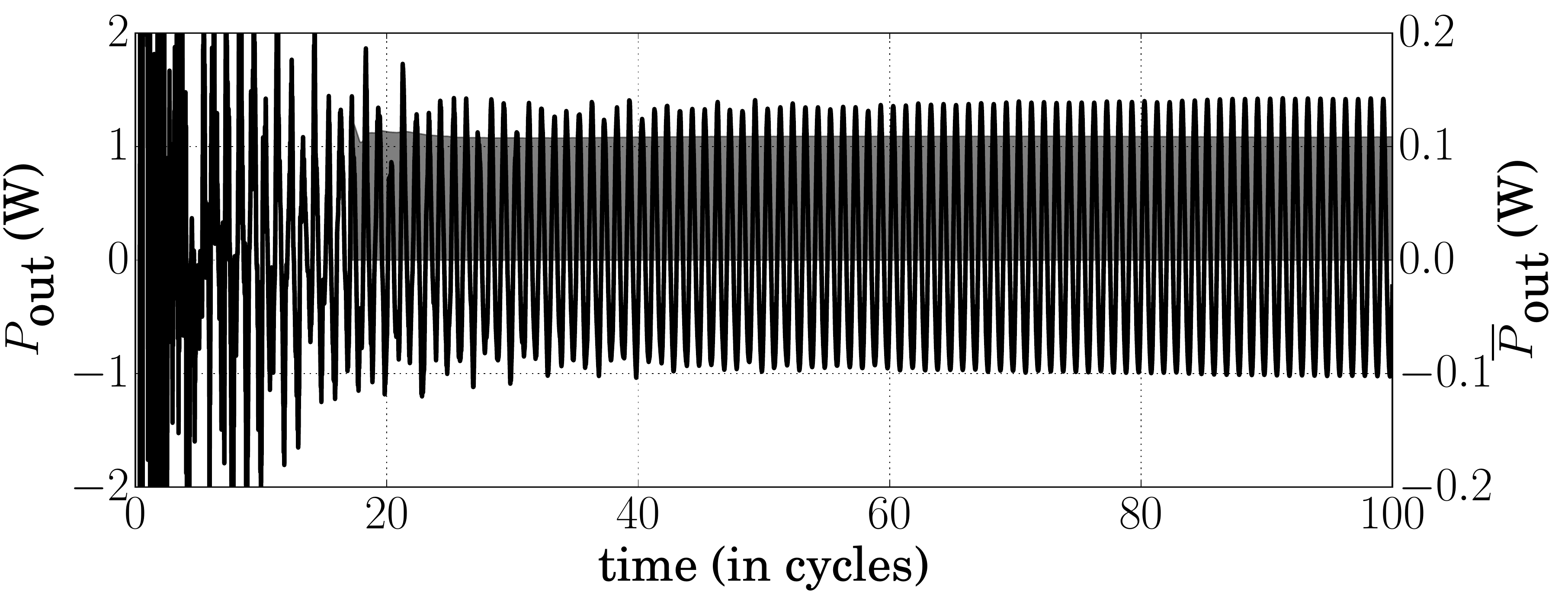}
		\caption[Time series of instantaneous acoustic power extracted by the piezoelectric diaphragm.]{
			Time series of instantaneous acoustic power \eqref{eq:instantaneous_acouPower} \legendline{} extracted at the limit cycle (left axis), cycle-averaged power \eqref{eq:cycle_average_acouPower} shown with the shaded area (right axis) for temperature setting 5, grid-resolution/stack-type C/I. 
			The beginning of the time series in this figure corresponds to the vertical dashed line in \cref{fig:pressure-ibc}.
		}
		\label{fig:poweroutput}
	\end{figure}

The imposition of the impedance boundary conditions designed in \cref{sec:modeling_a_physical_piezoelectric} results in a decrease in the pressure amplitude (\cref{fig:pressure-ibc}), corresponding to an extraction of acoustic energy (\cref{fig:poweroutput}), following an initial assessment phase with spurious high-frequency oscillations due to the abrupt initialization of the convolution integral \eqref{eq:convolutionintegral}. A new limit cycle is rapidly obtained with a slight frequency shift due to the resonance tuning of the piezoelectric diaphragm.

For temperature setting 5, acoustic energy absorption results in a pressure amplitude decrease of 10\%. The same acoustic energy absorption with temperature setting 1 (the close-to-critical temperature gradient) suppresses the thermoacoustic instability. The net power output per cycle
(\cref{fig:poweroutput}),
\begin{align} \label{eq:cycle_average_acouPower}
	\overline{P}_{\textrm{out}}\left(t\right) &= \int_{-\infty}^{+\infty} P_\textrm{out}\left(t+\tau\right) \frac{\sin\left(\pi f_c \tau \right)}{\pi \tau }d\tau \; ,
\end{align}
is extracted via sharp spectral filtering of the instantaneous acoustic power output,
\begin{equation} \label{eq:instantaneous_acouPower}
	P_\textrm{out}\left(t\right) = p'(t) U'(t) \; ,
\end{equation}
where $p'$ and $U'$ are the pressure and surface-averaged volumetric flow rate amplitudes at the diaphragm location. The convolution integral in \cref{eq:cycle_average_acouPower} is, in practice, limited to 
$\pm 17$ acoustic cycles with a cut-off frequency of 
$f_c=22.5\textrm{ Hz}$, 
lower than half that of the thermoacoustically amplified mode. 
The power extracted at the boundary is, at most, $111.25\textrm{ mW}$, corresponding to temperature setting 5. Thermal-to-mechanical efficiency $\eta$ is calculated for each case as the ratio of $\overline{P}_{\textrm{out}}$ and the cycle-averaged heat transfer rate through the stack walls, and is at most 1.3\% (\cref{tab:ibc_extraction}).

	\begin{figure}
		\centering
		\includegraphics[width=0.85\linewidth]{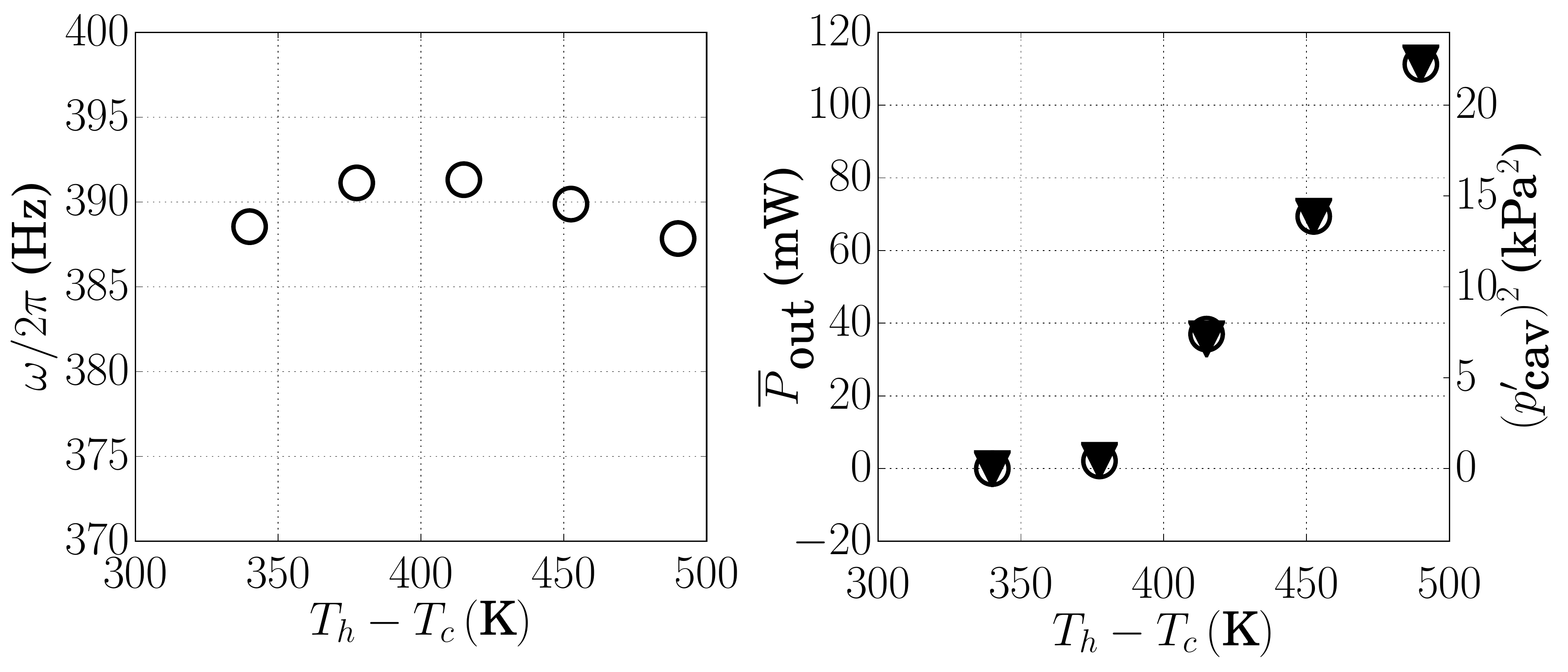}
			\put(-330,20){(a)}
			\put(5,20){(b)}
		\caption[Frequency and amplitude and power extraction numerical simulations.]{
			Limit cycle frequency, $\omega/2\pi$, versus temperature difference (a) with active piezoelectric energy extraction; cycle-averaged power output \eqref{eq:cycle_average_acouPower} \legenddots{} and the square of limit cycle pressure amplitudes in the hot cavity $p_{lc,cav}^2$ \legendtriangles{}  versus temperature difference (b). Results for grid-resolution/stack-type C/I. Note that limit cycle frequency values reported here differ from the ones in figure \ref{fig:lsa_freqgrowth}a, as the latter are obtained without imposition of piezoelectric energy extraction.
	}
		\label{fig:frequency_and_Pout_limitcycle}

	\end{figure}
\begin{table*}

	\centering
	\begin{tabular}{lrrrrr|rrr}
		\hline
		Case & $\Delta T$ (K)   &  $f_{lc}$ (Hz)   &  $p_{lc}$ (Pa) & $\overline{P}_\textrm{out}$ (mW) & $\quad$ $\eta$ (\%)   & $\overline{P}_{out,\textrm{ND}}$ & $\overline{P}_{out,\textrm{ND}}^{(\textrm{exp})}$ & err. in $\overline{P}_{\textrm{ND}}$ (\%) \\
		\hline
		1	& 340 & 388.55 & $--$ & 0 & $--$ & $--$ & $--$ & $--$\\
		2	& 377.5 & 391.13 & 672.76 & 2.09 & 0.1485 & 0.001407  & 0.001609 & 12.5\\

		3	& 415 & 391.32 & 2672.91 & 36.99  & 0.4289 & 0.001578 & 0.001609 & 1.9\\

		4	& 452.5 & 389.88 & 3726.58 & 69.38 & 0.7577 & 0.001523  & 0.001609 & 5.3 \\

		5     & 490 & 387.85 & 4724.45 & 111.25  & 1.3134 & 0.001519  & 0.001609  & 5.6 \\

		\hline
	\end{tabular}
	\normalsize
	\caption{
		Limit cycle frequency, $f_{lc}$, pressure amplitude, $p_{lc}$, acoustic energy extracted $\overline{P}_{out}$ and thermal-to-mechanical efficiency $\eta$ from Navier--Stokes calculations for grid-resolution/stack-type C/I, with piezoelectric energy extraction modelled by the multi-oscillator impedance model \eqref{subeq:broadbandfungjuform} with $n_o=18$ (table \ref{tab:fitting_broadband}).
	}
	\label{tab:ibc_extraction}
\end{table*}

In the experiments, an acoustic power output of $1.32\textrm{mW}$ is reported for conditions nominally meant to match the temperature setting 5 used in the present TAP engine model~\citep{SmokerNAB_2012}. However, results in \citet{NouhAB_2014} from the same engine show that thermoacoustic heat leakage and natural relaxation of the thermal gradient in the stack leads to unsteady temperature distributions in the stack, approaching temperature setting 1.
Due to differences in the regenerator/stack and uncertainties in the actual temperature gradient used in the experiments, numerical simulations with (steady) isothermal conditions cannot reproduce the experimentally observed limit cycle acoustic pressure amplitude. A normalized power output can be defined by compensating for the differences in pressure amplitude,
\begin{equation}\label{eq:normalized_power_out}
\overline{P}_{out,\textrm{ND}} = 	
\frac{Z_0  \, \overline{P}_\textrm{out}  
}{A_\textrm{piezo} p_{lc}^2}
,\quad\,
\overline{P}_{out,\textrm{ND}}^{(\textrm{exp})} = 
\frac{Z_0 \, \overline{P}_\textrm{out}^{(\textrm{exp})}
}{A_\textrm{piezo} \left(p_{lc}^{(\textrm{exp})}\right)^2} \; ,
\end{equation}
where $A_\textrm{piezo}$ is the area of the equivalent uniform impedance patch used in the present simulations 
and the superscript $(\textrm{exp})$ indicates experimental values. The good matching observed between the two non-dimensional powers (\cref{tab:ibc_extraction}) confirms that the impedance boundary conditions are imposing the correct phasing between pressure and velocity.

After the application of the TDIBC, the limit cycle operating frequency shifts (not shown) towards the frequency corresponding to the minimum impedance magnitude (maximum acoustic energy absorption). This is due to the increased compliance of the piezoelectric diaphragm at higher frequencies, corresponding to a reduction in the value of the resistance at higher frequencies (as seen in figure~\ref{fig:fitting_broadband} and discussed in \cref{subsec:multioscillatorbroadbandimpedance}). In the case of the single-oscillator impedance model \eqref{eq:three_parameter_model} with a constant value of resistance, the limit cycle frequency is controlled exclusively by the reactance. In all cases, an excessively large shift in frequency would disrupt the thermoacoustic phasing in the stack, leading to a suppression of the instability. 

\begin{subfigures}
	\begin{figure*}
		\centering
		\includegraphics[width=0.85\linewidth]{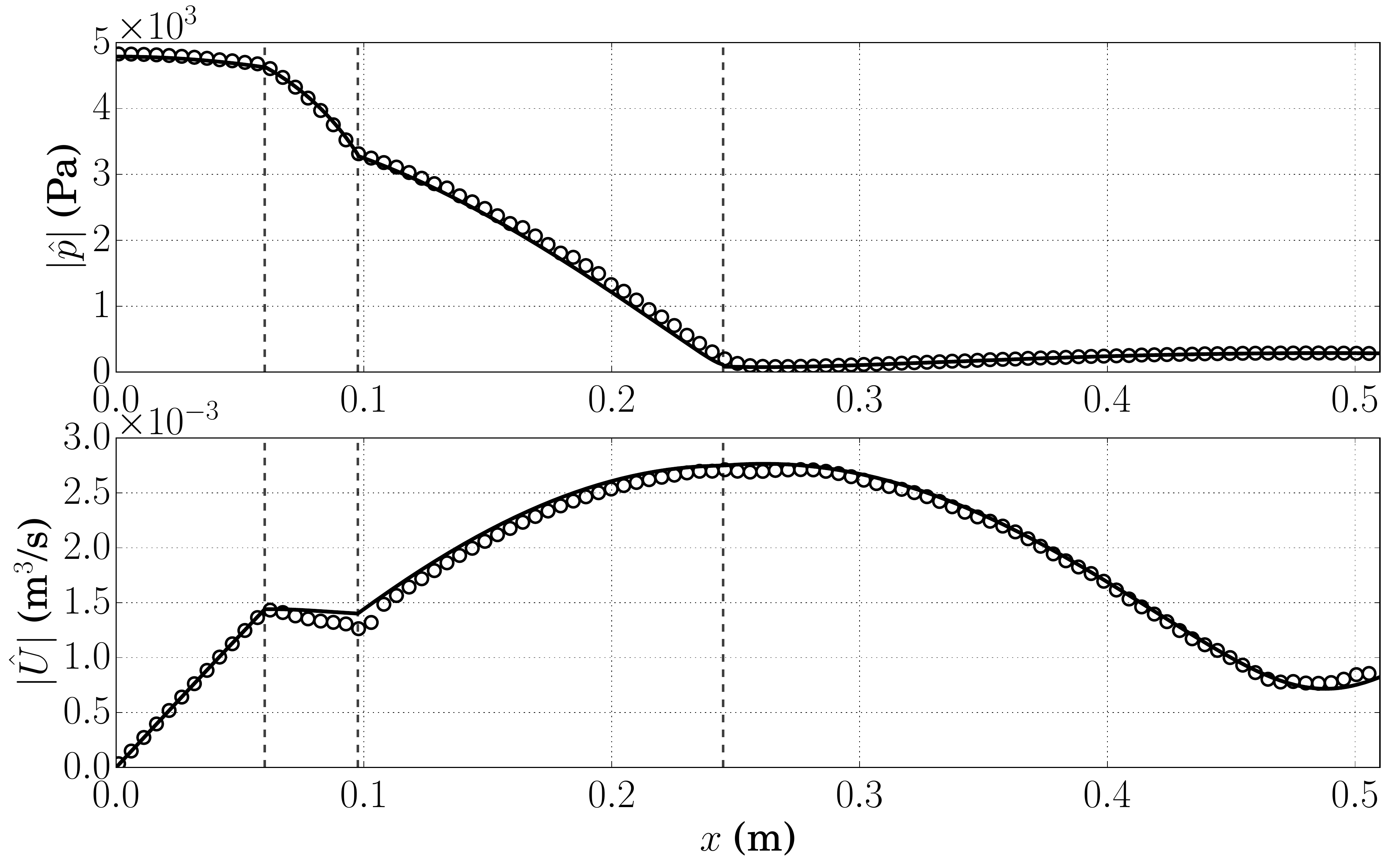}
				\put(-350,20){(b)}
				\put(-350,120){(a)}
		\caption[LSA IBC eigenfunctions]{
			Pressure (a) and flow rate (b) amplitudes of the thermoacoustically amplified mode predicted by linear theory \legendline{} rescaled to match amplitudes extracted from companion Navier--Stokes simulations \legenddots{} for temperature setting 5, grid-resolution/stack-type C/I, with active energy extraction at the limit cycle. Minor losses have been incorporated; however, the exclusion of minor losses (not shown) does not significantly alter the amplitudes predicted by linear theory.
			\verticallines{}
		}
		\label{fig:eigenfunctions_with_ibc}
		\includegraphics[width=0.85\linewidth]{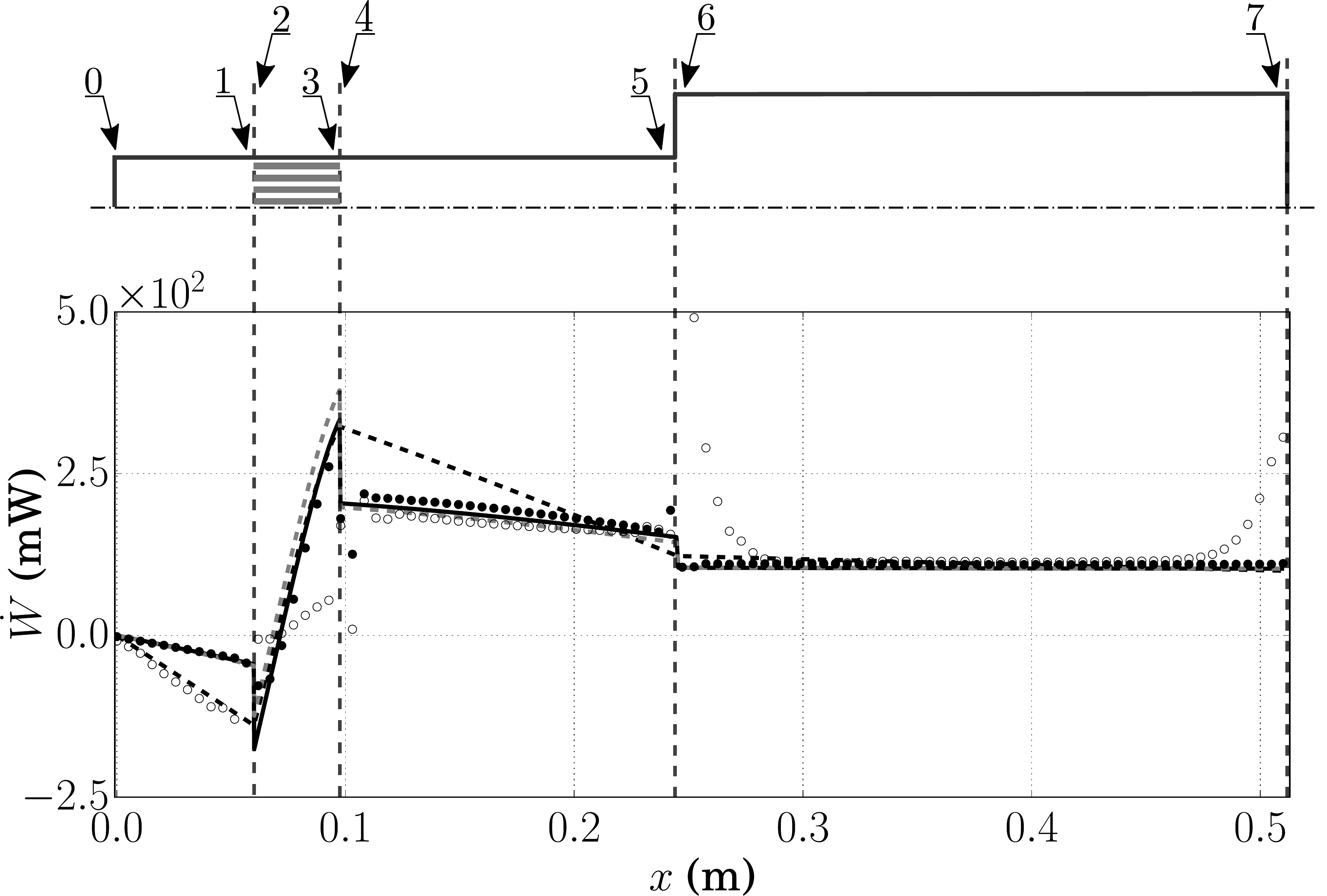}
		\caption[Energy balance diagram.]{
			Axial distribution of acoustic power \eqref{eq:wdot_from_uhatphat} from eigenfunctions predicted by linear theory (\cref{sec:lsamodel}), without minor losses \legenddashed{}, with linearized minor losses (\cref{eq:bordacarnot,eq:idelchik}) \legenddashedgray{}, and with minor losses calibrated from the Navier--Stokes simulations \legendline{}. Also shown are the acoustic power values extracted from the simulations, using only centreline \legenddots{} and using full cross-sectionally averaged \legenddotsblack{} values of axial velocity and pressure. Inter-segment locations are numbered above, referenced in \cref{tab:energybalancevalues}. Results correspond to temperature setting 5 and grid-resolution/stack-type C/I, with active energy extraction.
			\verticallines{}
		}
		\label{fig:wdot_axial_distribution}

	\end{figure*}
\end{subfigures}

The linear thermoacoustic model developed in \cref{sec:linearmodeling} has been augmented with minor losses (\cref{sec:energybudgets}) and is used here to reconstruct the axial distribution of acoustic power by applying a (constant) impedance, $Z\left(\omega_0\right)$ where $\omega_0 /2\pi = 388.0\textrm{ Hz}$ at $x=0.51\textrm{m}$. The axial distribution of acoustic power $\dot{W}$ can then be calculated as
\begin{align} \label{eq:wdot_from_uhatphat}
\dot{W}(x) & = \frac{1}{2} \Real \left\{ \hat{p} \left(x\right) \hat{U}^*\left(x\right) \right\}
\end{align}
where $\hat{p}$ and $\hat{U}$ are the eigenvectors predicted by the linear model and rescaled such that pressure and volume flow rate amplitudes match the Navier--Stokes calculations with TDIBCs, at limit cycle. The resulting eigenfunctions and axial power distribution are shown  in \cref{fig:eigenfunctions_with_ibc,fig:wdot_axial_distribution}. Results from the Navier--Stokes calculations were collapsed into axial time series of volume flow rate and pressure via surface integration, and the resulting quantities of $\hat{U}\left(x\right)$ and $\hat{p}\left(x\right)$ were fitted to complex phasors. Acoustic power along the axis is then calculated using \cref{eq:wdot_from_uhatphat}. The linear model predicts an acoustic power extraction of $111.09\textrm{ mW}$, in good agreement with the result of $111.25\textrm{ mW}$ from the Navier--Stokes calculations, as reported in table~\ref{tab:ibc_extraction}.

\subsection{Acoustic Energy Budgets}
\label{sec:energybudgets}

As expected, a positive slope in the acoustic power is present in the stack, while a negative slope in the pulse tube and the resonator volume indicates acoustic power dissipation due to viscous dissipation and thermal-relaxation. The acoustic power distribution is consistent with that predicted by engineering design software such as \deltaec{} in other standing-wave engines in literature~\citep{Swift_1992,WardCS_2012}.

The balance of acoustic energy at the limit cycle can be heuristically expressed as
\begin{align} \label{eq:energybudget}
\cancel{\frac{\partial E_a}{\partial t}}+\frac{d}{d x} \dot{W} &= S_{ta}-D_\mu-D_m \, ,
\end{align}
where the cycle-averaged acoustic energy per unit length, $E_a$, is assumed to be steady. The divergence of the acoustic energy flux is balanced by thermoacoustic source terms, $S_{ta}$, and viscous dissipation, $D_\mu$, which are accurately predicted by Rott's theory, as made evident by the predicted slope matching that of the data extracted from nonlinear calculations. Hydrodynamic minor losses due to abrupt area changes, $D_m$, manifest themselves as jumps in the value of $\dot{W}$ and will therefore be formally incorporated in the budgets above via Dirac functions (\cref{tab:energybalancevalues}).

\begin{table}
	\centering
\begin{tabular}{r|r|r}
	\begin{tabular}[x]{@{}c@{}} $ $ \\ $ $ \\  \end{tabular}
	& 
	\begin{tabular}[x]{@{}c@{}}acoustic power \\source/sink\\($\textrm{W/m}$)\end{tabular}
	& 
	\begin{tabular}[x]{@{}c@{}}acoustic power\\ contribution\\ ($\textrm{W}$) \end{tabular} \\
	
	segment  & 			\multicolumn{1}{|c|}{$S_{ta}-D_\mu$}   & 
	\multicolumn{1}{|c}{$\Delta \dot{W}$} \\ 

	\hline

	hot cavity    &  $-0.637$        &     $-0.0382$   \\
	
	stack      & $13.18$        &  $0.494$   \\
	
	tube    &  $-0.420$         &   $-0.0619$  \\
	
	resonator  &    $-0.00355$       &  $-0.00094$   \\
	
\end{tabular}
{
	\begin{minipage}{0.52\linewidth}
		\begin{tabular}{c|r|r}
			\begin{tabular}[x]{@{}c@{}}inter-\\segment\\ location \\$i$ \end{tabular}  &  \begin{tabular}[x]{@{}c@{}}acoustic power\\ change\\ ($\textrm{W}$) \\ 
				$\dot{W} \left(x_{i}\right)-\dot{W} \left(x_{i-1}\right)$
			\end{tabular}
			& 
			\begin{tabular}[x]{@{}c@{}}acoustic power,\\ cumulative\\
				($\textrm{W}$) \\
				$\dot{W} \left(x_{i}\right)$
			\end{tabular} \\

			\hline
			0     & 0     & 0 \\
			1     & $-0.0382$ & $-0.0382$ \\
			2     & $-0.126$ & $-0.164$ \\
			3     & $0.494$ & $0.330$ \\
			4     & $-0.111$ & $0.219$ \\
			5     & $-0.0619$ & $0.157$ \\
			6     & $-0.0454$ & $0.112$ \\
			7     & $-0.00094$ & $0.111$ \\
		\end{tabular}
	\end{minipage}
}
	\caption{Table of cycle-averaged acoustic power contributions by segment and thermoacoustic source and dissipation sink terms by segment. Table of changes in and cumulative acoustic power, for each inter-segment location in \cref{fig:wdot_axial_distribution}.  
	Acoustic power values are extracted from the simulations, using cross-sectionally averaged values of axial velocity and pressure. Results correspond to temperature setting 5 and grid-resolution/stack-type C/I, with active energy extraction. 
	}
	\label{tab:energybalancevalues}
\end{table}

Incorporating minor losses is necessary to reproduce first order features in the limit cycle acoustic power  distribution for the thermoacoustic piezoelectric engine. In the case of steady flow, minor losses due to abrupt area changes can be parameterized via the Borda-Carnot formula, 
\begin{align} \label{eq:bordacarnot}
\zeta_e &= \left(1-\frac{A_0}{A_1}\right)^2 \; ,
\end{align}
in the case of expansions or via the formula presented by \citet{Idelchik_2003},
\begin{align}  \label{eq:idelchik}
\zeta_c &= 0.5 \left(1-\frac{A_0}{A_1}\right)^{0.75}\, ,
\end{align}
in the case of sudden contractions, where $A_0$ and $A_1$ are the smaller and larger areas. In the present study, the two losses are combined, $\zeta = \zeta_e + \zeta_c$, and the pressure drop condition
\begin{align} \label{eq:lin_minorlosses}
\Delta \hat{p}_{ml} = -\frac{4}{3\pi} \rho \, \zeta \, u_{lc} \, \hat{u} \,,
\end{align}
linearized about the limit cycle axial velocity amplitude distribution $u_{lc}$, is incorporated in the inter-segment condition \eqref{eq:extrapolative_pressure_condition} in the linear thermoacoustic model in section \ref{sec:lsamodel}. While a similar approach to modelling minor losses in oscillatory flows has been adopted with great success by \citet{WardCS_2012}, a more accurate parameterization of $\zeta$ should be derived with ad hoc numerical or experimental investigations. Despite the strong assumptions made in deriving and introducing minor losses in the linear model, the agreement with the Navier--Stokes data is remarkable (figure \ref{fig:wdot_axial_distribution}). Axial amplitude profiles of pressure and flow rate are, however, not visibly altered by the addition of minor losses; the condition \eqref{eq:lin_minorlosses} primarily affects the pressure-velocity phasing in locations of sudden area change.

It is possible, however, to determine the value of $u_{lc}$ in \eqref{eq:lin_minorlosses} \emph{a priori} by imposing a zero-growth-rate condition in the eigenvalue solver in \cref{sec:linearmodeling}. This is indirectly done in modelling software such as \deltaec{}, where the limit cycle pressure and velocity amplitudes are found iteratively by assuming that acoustic energy budgets are balanced at steady-state conditions.

%% file: Appendix.tex
\section{Application of Rott's theory to axisymmetric thermoacoustic stack}
\label{app:rotts_theory_derivation}

Using the convention of \citet{Rott_ZAMP_1969}, the linearized equations of mass, momentum, and energy are
\begin{eqnarray}
	&&i\omega\hat{\rho} + \rho_0 \frac{\partial \hat{u}}{\partial x} + \hat{u} \frac{d\rho_0}{dx} + \rho_0 \frac{1}{r} \frac{\partial}{\partial r} \left(r\hat{v}\right)= 0 \\
	&&i\omega\,\hat{u} + \frac{1}{\rho_0}\frac{d\hat{p}}{dx} = \frac{\mu_0}{\rho_0} \frac{1}{r}\frac{\partial}{\partial r}\left(r\frac{\partial \hat{u}}{\partial r}\right) \\
	&&\rho_0\,C_p\left(i\omega\hat{T}+\hat{u}\frac{d\,T_0}{d\,x}\right) - i\omega\hat{p} = \frac{\mu_0\,C_p}{\Pran} \frac{1}{r}\frac{\partial}{\partial r} \left(r \frac{\partial}{\partial r} \hat{T}\right)
\label{eq: energy}
\end{eqnarray}
where the thermal conductivity is given by $k = \mu\,C_p/\Pran$ and $\Pran$ is the Prandtl number. Radial variations are neglected for pressure, $\hat{p}=\hat{p}(x)$; radial variations are retained for the axial and radial velocity components, $\hat{u}=\hat{u}(x,r)$ and $\hat{v}=\hat{v}(x,r)$, respectively, and temperature, $\hat{T}=\hat{T}(x,r)$. \\
The following constitutive equations are used:
\begin{eqnarray}
P_0 = \rho_0\,R_{gas}\,T_0 \\
\hat{T} = \hat{p}\frac{1}{\rho_0\,R_{gas}} - \hat{\rho}\frac{T_0}{\rho_0}
\label{app:eq:constitutive}
\end{eqnarray}
where $P_0$, $\rho_0$, $T_0$ correspond respectively to the base and constant pressure, density, and temperature.

In order to derive a local solution to the momentum equation, the application of the coordinate transformation
\begin{equation}
\xi = i\eta, \quad \eta = \sqrt{\frac{i\omega}{\nu}}r
\end{equation}
results in a momentum equation of
\begin{equation}
\xi^2\frac{\partial^2\hat{u}_*}{\partial \xi^2}+\xi\frac{\partial \hat{u}_*}{\partial \xi} + \xi^2\,\hat{u}_* = 0
\label{app:eq:transformed_momentum}
\end{equation}
where
\begin{equation} \label{app:eq:variable_transformation}
\widehat{u}_* = \frac{\hat{u}}{-\frac{1}{i\omega\rho_0}\frac{d\hat{p}}{dx}}-1 ,
\end{equation}
assuming that pressure does not vary radially.

Note that since $i\sqrt{2\,i} = i-1$, the dimensionless radial coordinate $\eta$ can also be written in the form
$\eta \equiv \sqrt{\frac{i\omega}{\nu}}r  =
\sqrt{2i} \sqrt{\frac{\omega}{2\nu}}\,r =
\frac{i-1}{i} \frac{r}{\delta_\nu}$
, which is useful in the following algebraic manipulations.

The general solution to \cref{app:eq:transformed_momentum} is
\begin{equation} \label{app:eq:general_solution}
	\widehat{u}_*(\xi) = a J_0(\xi) + b Y_0(\xi)
\end{equation}
where $a$ and $b$ are constants, and $J_0(\xi)$ and $Y_0(\xi)$ are Bessel functions of the first and second kind, respectively evaluating to purely real and imaginary values. 
Given the boundary conditions, using the Bessel function of the second kind results in a computationally singular solution.
Without loss of generality, \cref{app:eq:general_solution} can be re-written as
\begin{equation} \label{eq:general_solution2}
	\widehat{u}_*(\xi) = A J_0(\xi) + B H_0^{(1)}(\xi)
\end{equation}
where $H_0^{(1)}(\xi)=J_0(\xi)+i Y_0(\xi)$ is a Hankel function of the first kind, and $A$ and $B$ are constants. In an annular duct, for which no-slip and isothermal conditions at both upper and lower walls are imposed,
the conditions due to transformation \eqref{app:eq:variable_transformation} are $\widehat{u}_*(\xi_\textrm{top})=\widehat{u}_*(\xi_\textrm{bot})=-1$. 
Because $J_0\left(\xi\right)$ and $H^{(1)}_0(\xi)$ each diverge quickly for larger and smaller $\xi$, respectively, $H_0(\xi_\textrm{top})$ and $J_0(\xi_\textrm{bot})$ may be neglected in comparison with $H_0(\xi_\textrm{bot})$ and $J_0(\xi_\textrm{top})$, respectively. That is, the Bessel and Hankel functions diverge very rapidly for given $\xi$, such that 
\begin{subequations}\label{eq:approximations_H_J}
\begin{align}
H_0(\xi_\textrm{top}) & \ll H_0(\xi_\textrm{bot})\\
J_0(\xi_\textrm{bot}) & \ll J_0(\xi_\textrm{top})
\end{align}
\end{subequations}
The solution of \cref{eq:general_solution2}
is then
\begin{equation} \label{app:eq:general_solution3}
\widehat{u}_*(\xi) = - \frac{J_0(\xi)}{J_0(\xi_\textrm{top})}- \frac{H^{(1)}_0(\xi)}{H_0^{(1)}(\xi_\textrm{bot})}
\end{equation}
which yields
\begin{equation} \label{app:eq:uhat_of_xsi}
\hat{u}(\xi) = \frac{i}{\omega\,\rho_0} \frac{d\hat{p}}{dx}\left[1- \frac{J_0(\xi)}{J_0(\xi_\textrm{top})}- \frac{H^{(1)}_0(\xi)}{H_0^{(1)}(\xi_\textrm{bot})}\right]  \, \, .
\end{equation}
To verify that the expression above does indeed satisfy the boundary conditions, refer to the plotted velocity profiles in \cref{tab:stack_configurations},  which suggest that the approximations made in \eqref{eq:approximations_H_J} are satisfied.
The analytical integration in the annular cross-section, where $A_g$ is the annular area accessible to the gas, yields the relationship for the flow rate
\begin{equation} \label{app:eq:flow_rate_solution}
	i\omega\,\hat{U} = - \frac{A_g}{\rho_0} \frac{d\hat{p}}{dx}\left[ 1 - f_\nu \right]
\end{equation}
where
\begin{equation}
\begin{split}
	f_\nu = \frac{i\pi\delta_\nu^2}{A_g}\left\{ \frac{1}{J_0(\xi_\textrm{top})} \left( \xi_\textrm{top} J_1(\xi_\textrm{top})-\xi_\textrm{bot}\,J_1(\xi_\textrm{bot})\right)+ \right. \\ \left. \frac{1}{H^{(1)}_0(\xi_\textrm{bot})} \left( \xi_\textrm{top} H^{(1)}_1(\xi_\textrm{top})-\xi_\textrm{bot}\,H^{(1)}_1(\xi_\textrm{bot})\right)\right\}
	\, \, .
\end{split}
\end{equation}

Changing the acoustic variable $\hat{T}$ in (\ref{eq: energy}) to $\hat{p}$ and $\hat{\rho}$ using the constitutive equations~\ref{app:eq:constitutive}, the energy equation can be written in the following manner \citep{Rott_ZAMP_1969}:
\begin{equation}
i\omega\left[\left(\hat{\rho}-\rho_0\right)+\frac{\gamma-1}{a_0^2}\hat{p}\right]+\hat{u}\frac{d\rho_0}{dx}=\frac{\nu}{\Pran} \frac{1}{r}\frac{\partial}{\partial r} \left(r \frac{\partial}{\partial r} \left(\hat{\rho}-\rho_0\right)\right).
\label{eq: density}
\end{equation}
With the dimensionless variable $\xi$, the above equation can be recast as
\begin{equation}
\frac{\partial }{\partial \xi^2}\left(\hat{\rho}-\rho_0\right)+\frac{1}{\xi}\frac{\partial}{\partial\xi}\left(\hat{\rho}-\rho_0\right)+\Pran\left(\hat{\rho}-\rho_0\right)=-\frac{\Pran}{i\omega}\hat{u}\frac{d\rho_0}{dx}-\Pran\left(\frac{\gamma-1}{a_0^2}\right)\hat{p}
\, \, .
\end{equation}
Assuming a general solution of the form
\begin{equation}
\hat{\rho}-\rho_0=A J_0\left(\xi\sqrt{\Pran}\right)+B H^{(1)}_0\left(\xi\sqrt{\Pran}\right)+C \hat{u}\left(\xi\right)+D
\end{equation}
and utilizing the boundary conditions at $\xi_\textrm{bot}$ and $\xi_\textrm{top}$, the perturbation in density is given by a similar expression \citep{Rott_ZAMP_1969}:
\begin{equation}
\begin{split}
\hat{\rho}-\rho_0=\left(-\frac{\gamma-1}{a_0^2}\hat{p}+\frac{\theta}{\left(1-\Pran\right)\omega^2}\frac{d\hat{p}}{dx}\right)\left[1- \frac{J_0(\xi\sqrt{\Pran})}{J_0(\xi_\textrm{top}\sqrt{\Pran})}- \frac{H^{(1)}_0(\xi\sqrt{\Pran})}{H_0^{(1)}(\xi_\textrm{bot}\sqrt{\Pran})}\right]-\\
\frac{\Pran\,\theta}{\left(1-\Pran\right)\omega^2}\frac{d\hat{p}}{dx}\left[1- \frac{J_0(\xi)}{J_0(\xi_\textrm{top})}- \frac{H^{(1)}_0(\xi)}{H_0^{(1)}(\xi_\textrm{bot})}\right],
\end{split}
\label{app:eq:densityperturbation}
\end{equation}
where $\theta=(1/T_0)dT_0/dx$. Starting from \cref{eq: density}, substituting for the base state density gradient using the continuity equation and integrating over the annular cross-section results in
\begin{equation}
	i\omega\hat{p} + \frac{a_0^2\rho_0}{A_g}\frac{d\hat{U}}{dx} +
	\frac{2\pi\nu a_0^2}
	{\Pran A_g}
	\left[\xi_\textrm{top}\frac{\partial\hat{\rho}}{\partial\xi}\big|_{\xi_\textrm{top}} - \xi_\textrm{bot}\frac{\partial\hat{\rho}}{\partial\xi}\big|_{\xi_\textrm{bot}}\right]
	=0.
\end{equation}

Using the solution for density perturbation, \cref{app:eq:densityperturbation}, to link pressure and velocity disturbances, the radial gradient of density perturbations is then
\begin{equation}
\begin{split}
\frac{\partial\hat{\rho}}{\partial\xi} =
-\sqrt{\Pran}\frac{\gamma-1}{a_0^2}
	\left(\frac{J_1\left(\xi\sqrt{\Pran}\right)}{J_0\left(\xi_\textrm{top}\sqrt{\Pran}\right)}+\frac{H^{(1)}_1(\xi\sqrt{\Pran})}{H^{(1)}_1(\xi_\textrm{bot}\sqrt{\Pran})}\right)+\\
	\frac{\theta}{(1-\Pran)\omega^2}\frac{d\hat{p}}{dx}
		\left\{\left[\frac{\sqrt{\Pran}J_1\left(\xi\sqrt{\Pran}\right)}{J_0\left(\xi_\textrm{top}\sqrt{\Pran}\right)} + \frac{\sqrt{\Pran}H^{(1)}_1(\xi\sqrt{\Pran})}{H^{(1)}_1(\xi_\textrm{bot}\sqrt{\Pran})}\right] - \Pran\left[\frac{J_1\left(\xi\right)}{J_0\left(\xi_\textrm{top}\right)} + \frac{H^{(1)}_1(\xi)}{H^{(1)}_1(\xi_\textrm{bot})}\right]\right\}
\, \, .
\end{split}
\end{equation}
Evaluating the radial gradient of density perturbations at the radial boundaries and substituting in, the final linearized equation is
\begin{equation}
i\omega\hat{p}=\frac{1}{1+\left(\gamma-1\right)f_k}\left(\frac{\rho_0 a_0^2}{A_g}\right)\left[\frac{\theta\left(f_k-f_{\nu}\right)}{(1-f_{\nu})(1-\Pran)}-\frac{d}{dx}\right]\hat{U}
\end{equation}
where
\begin{equation}
\begin{split}
f_\kappa = \frac{i\pi\,\delta_\kappa^2\,\sqrt{\Pran}}{A_g}
\Big\{
\frac{1}{J_0(\xi_\textrm{top}\sqrt{\Pran})}
\left[ \xi_\textrm{top} J_1(\xi_\textrm{top}\sqrt{\Pran}) - \xi_\textrm{bot} J_1(\xi_\textrm{bot}\sqrt{\Pran}) \right] + \\
\frac{1}{H^{(1)}_0(\xi_\textrm{bot}\sqrt{\Pran})} \left[ \xi_\textrm{top} H^{(1)}_1(\xi_\textrm{top}\sqrt{\Pran}) - \xi_\textrm{bot} H^{(1)}_1(\xi_\textrm{bot}\sqrt{\Pran}) \right]
\Big\} \; .
\end{split}
\end{equation}

\section{Implementation of multi-oscillator TDIBCs}
\label{app:convolutionintegral}

For completeness, we continue discussion of the dimensional implementation of the time-domain impedance boundary condition, as was introduced in \cref{sec:modeling_a_physical_piezoelectric}.

A TDIBC, of the form proposed by \citet{FungJ_2004}, was coupled with the compressible flow solver \charlesx. The coupling strategy used here is proposed and described by \citet{ScaloBL_PoF_2015}, in which the implementation was demonstrated and validated using an impedance tube with a Helmholtz oscillator. The validation was performed using an incident broadband pulse; the numerical reflected wave was compared with the semi-analytical solution for a given impedance. Some concepts from \citet{ScaloBL_PoF_2015}, which used acoustics conventions for normalization with base density, speed of sound, and scaling parameters for channel flow normalization, are used here for illustration.
In this description, for clarity, we are instead reporting a dimensional derivation and implementation. 

A linear acoustic impedance boundary condition relates pressure and velocity at the boundary as:
\begin{align}
\label{eq:impedanceequation}
\hat{p}&=Z\left(\omega \right) \hat{u}
\end{align}
where $\hat{p}$ and $\hat{u}$ are complex pressure and velocity amplitudes, and $Z\left(\omega\right)$ is the dimensional/specific acoustic impedance, for which the characteristic specific acoustic impedance $\rho_0 a_0$ is a factor.

Relative to the boundary, incident ($+$) and reflected ($-$) travelling waves are:
\begin{subequations}
\begin{align}
\label{eq:travelingwaves}
u^{\pm} &= u' \pm \frac{p'}{\rho_0 a_0} \\
u'=\frac{u^+ + u^-}{2} \; , &\quad \; \frac{p'}{\rho_0 a_0}=  \frac{u^+ - u^-}{2}
\end{align}
\end{subequations}
where $u'$ and $p'$ are fluctuations in wall-normal velocity and pressure. Combining equations~\ref{eq:impedanceequation} and \ref{eq:travelingwaves} yields
\begin{subequations}
	\label{eq:reflectioncoefficient}
	\begin{align}
	\label{subeq:reflectedwave}
	\hat{u}^{-} \left(\omega\right) &=  \widehat{W}_{\omega} \left(\omega\right) \hat{u}^{+} \left(\omega\right)\\
	\label{subeq:reflectioncoefficient}
	\widehat{W}_{\omega} \left(\omega\right) &= \frac{\rho_0 a_0 -Z\left(\omega\right)}{\rho_0 a_0 +Z\left(\omega\right)}
	\end{align}
\end{subequations}
which correspond to the reflected wave $\hat{u}^{-} \left(\omega\right)$ and the reflection coefficient $\widehat{W}_{\omega} \left(\omega\right)$ in the frequency domain.

The direct term of a partial fraction expansion in the reflection coefficient can be removed by using the wall softness $\widehat{\widetilde{W}}_{\omega} \left(\omega\right)$ form to relate the incident wave and reflected wave:
\begin{subequations}
	\label{eq:softness}
	\begin{align}
	\label{subeq:reflectedsoftness}
	\hat{u}^{-} \left(\omega\right) &=  - \hat{u}^{+} \left(\omega\right) +  \widehat{\widetilde{W}}_{\omega} \left(\omega\right) \hat{u}^{+} \left(\omega\right)\\
\intertext{where}
	\label{subeq:softnesscoeff}
	\widehat{\widetilde{W}}_{\omega} \left(\omega\right) &= 	\widehat{W}_{\omega} \left(\omega\right) + 1 = \frac{2\rho_0 a_0 }{\rho_0 a_0 +Z\left(\omega\right)}
	\, \, .
	\end{align}
\end{subequations}

\Cref{subeq:reflectedsoftness} suggests that, provided the poles of $\widehat{\widetilde{W}}_{\omega} \left(\omega\right)$ are in the upper half of the complex $\omega$-plane, the reflected wave can be obtained from the causal convolution of the incident wave:
\begin{align}
\label{eq:convolutionintegral}
u^{-}\left(t\right) &= -u^{+} \left(t\right) + \int_{0}^{\infty} \widetilde{W} \left(\tau\right)  u^{+} \left(t-\tau\right) d\tau
\, \, .
\end{align}

Extending $\widehat{\widetilde{W}}_{\omega} \left(\omega\right)$ into the Laplace domain, based on the convention $\widehat{\widetilde{W}}_{\omega} \left(\omega\right) = \widehat{\widetilde{W}}_{\omega} \left(-i s\right) = \widehat{\widetilde{W}}_{s} \left(s\right)$, suggests that the softness function can be expanded with partial fractions and the linearity property of frequency-domain transforms can be used to obtain a solution for equation~\ref{eq:convolutionintegral}. 
Inverting the Laplace transform of 
\begin{align}
\widehat{\widetilde{W}}_{s} \left(s\right) &= \sum_{k=1}^{\numofoscillators} \left[\frac{\mu_k}{s-p_k}+\frac{\mu_k^*}{s-p_k^*}\right]
\end{align}
and discretizing and evaluating \eqref{eq:convolutionintegral} obtains
\begin{subequations}
	\label{eq:pferesult}
	\begin{eqnarray}
	\label{subeq:pfe_summation}
	u^{-}\left(t + \Delta t\right) &=&  -u^{+}\left(t + \Delta t\right) + \sum_{k=1}^{\numofoscillators} \left[ u_k^{-} \left(t+\Delta t\right) + 
	u_k^{-, *} \left(t+\Delta t\right)
	\right]\\
	\label{subeq:pfe_integral}
	u_k^{-} \left(t+\Delta t\right) &=& \int_0^\infty \mu_k e^{p_k \tau} u^{+} \left(t+\Delta t - \tau \right) d\tau \\
	\label{subeq:pfe_integral_conj}
	u_k^{-, *} \left(t+\Delta t\right) &=& \int_0^\infty \conj{\mu_k} e^{\conj{p_k} \tau} u^{+} \left(t+\Delta t - \tau \right) d\tau 
	\end{eqnarray}
\end{subequations}
where $u_k^{-} \left(t+\Delta t\right)$ and $u_k^{-, *} \left(t+\Delta t\right)$ are contributions to the convolution integral, $p_k$ and $\conj{p_k}$ are poles of $	\widehat{\widetilde{W}}_s \left(s\right)$, and $\mu_k = \textrm{Residue} \left[	\widehat{\widetilde{W}}_s \left(s\right), p_k \right]$ and similarly for $\conj{\mu_k}$.

The integral of equation~\ref{subeq:pfe_integral} can be recursively solved for a given $p_k$ and $\mu_k$:
\begin{align}
u_k^{-} \left(t\right) &= \int_0^{\infty} \mu_k e^{p_k \tau} u^{+} \left(t- \tau \right) d\tau \nonumber\\
&= e^{-p_k \Delta t} \int_{\Delta t}^{\infty} \mu_k e^{p_k \tau} u^{+} \left(t+\Delta t - \tau \right) d\tau \\
\therefore u_k^{-} \left(t+\Delta t\right) &= \int_0^{\Delta t} \mu_k e^{p_k \tau} u^{+} \left(t+\Delta t - \tau \right) d\tau + \int_{\Delta t}^{\infty} \mu_k e^{p_k \tau} u^{+} \left(t+\Delta t - \tau \right) d\tau \nonumber \\
&= z_k u_k^{-} \left(t\right) + \int_0^{\Delta t} \mu_k e^{p_k \tau} u^{+} \left(t+\Delta t - \tau \right) d\tau
\end{align}
where $z_k = e^{p_k \Delta t}$. The integral of equation~\ref{subeq:pfe_integral_conj} follows similarly. 

This integral can be evaluated with a trapezoid quadrature rule, resulting in:
\begin{align}
u_k^{-} \left(t+\Delta t\right) &= z_k u_k^{-} \left(t\right) + \mu_k \Delta t \left[w_{k0} u^{+} \left(t+\Delta t\right)+ w_{k1} u^{+} \left(t\right) \right]
\label{eq:integrandquadrature}
\end{align}
where
\begin{subequations}
	\begin{align}
	w_{k0} &= \frac{z_k -1}{p_k^2 \Delta t^2} - \frac{1}{p_k \Delta t} \\
	w_{k1} &=-\frac{z_k -1}{p_k^2 \Delta t^2} + \frac{z_k}{p_k \Delta t}
	\end{align}
\end{subequations}

In order to evaluate equations~\ref{subeq:pfe_summation} and \ref{eq:integrandquadrature}, $u^{+}\left(t+\Delta t\right)$ is required.
This is predicted at the boundary with a one-dimensional approximation, based on the spatial gradient of pressure and velocity at the boundary:
\begin{align}
	u^{+} \left(t+\Delta t\right) &\approx \left[\frac{1}{\rho_0 a_0} p'\left(x, t\right) + u'\left(x, t\right)\right] - a_0\Delta t \frac{\partial}{\partial x} \left[\frac{1}{\rho_0 a_0} p' \left(x, t\right) + u' \left(x, t\right)\right]
\, \, .
\end{align}

The fluctuation in pressure and wall-normal velocity at time step $t+\Delta t$ are then imposed as Dirichlet boundary conditions as
\begin{subequations}
	\begin{align}
		u' \left(t+\Delta t\right) &= \frac{1}{2} \left[u^{+} \left(t+\Delta t\right)+u^{-} \left(t+\Delta t\right)\right]\\
		p' \left(t+\Delta t\right) &= \frac{\rho_0 a_0 }{2} \left[u^{+} \left(t+\Delta t\right)-u^{-} \left(t+\Delta t\right)\right]
	\, \, .
	\end{align}
\end{subequations}

In the Navier--Stokes simulations, adiabatic conditions are imposed for boundary temperature.

\section{Impedance transfer function coefficients}
\label{appendix:transferfunctioncoefficients}

\begin{table*}
	\small
	\footnotesize
	\centering
	\scalebox{0.6}{
	\begin{tabular}{r|r|rrrrrrrrrrr}
		\hline
		\multicolumn{2}{c|}{$n$}   & 10    & 9     & 8     & 7     & 6     & 5     & 4     & 3     & 2     & 1     & 0 \\
		\hline
		\multirow{2}{*}{$T_{11}$}   & $a_n$   &   $1.5\times10^{-12}$    &   $-6.2\times10^{-7}$    &    $4.3\times10^{-2}$   &   $-3.006\times10^{3}$    &   $1.62\times10^{8}$    &    $-5.81\times10^{11}$   &   $-1.14\times10^{17}$    &    $6.3\times10^{20}$   &   $-6.1\times10^{25}$    &   $1.7\times10^{30}$    & $3.7\times10^{34}$  \\
		& $b_n$    &   $1\times10^{0}$    &   $8.3\times10^{4}$    &    $5.95\times10^{9}$   &   $2.16\times10^{14}$    &   $6.08\times10^{18}$    &   $1.05\times10^{23}$    &   $1.2\times10^{27}$    &   $9.21\times10^{30}$    &   $4.3\times10^{34}$    &   $5.9\times10^{37}$    &  $1.84\times10^{41}$ \\

		$T_{12}$   & $a_n$    &   $-2\times10^{-10}$    &     $4.6\times10^{-5}$  &    $-5.622\times10^{0}$   &    $4.76\times10^{5}$   &     $-2.82\times10^{10}$  &    $1.04\times10^{15}$   &   $-1.85\times10^{19}$    &     $-1.9\times10^{22}$  &    $4.4\times10^{27}$   &   $-1.6\times10^{31}$    &  $-2.48\times10^{35}$ \\
		& $b_n$    &   $1\times10^{0}$    &     $9.43\times10^{4}$  &    $5.44\times10^{9}$   &    $1.91\times10^{14}$   &     $4.61\times10^{18}$  &    $7.5\times10^{22}$   &   $7.9\times10^{26}$    &     $5.9\times10^{30}$  &    $2.5\times10^{34}$   &   $3.7\times10^{37}$    &   $1.06\times10^{41}$\\

		$T_{21}$   & $a_n$    &   $6.5\times10^{-13}$    &     $-2.5\times10^{-7}$  &    $2.357\times10^{-2}$   &    $-1.811\times10^{3}$   &     $1.11\times10^{8}$  &    $-3.54\times10^{12}$   &   $7.08\times10^{16}$    &     $-2.08\times10^{21}$  &    $3.7\times10^{25}$   &   $-1.3\times10^{28}$    &   $1.6\times10^{33}$\\
		& $b_n$    &   $1\times10^{0}$    &     $1.1\times10^{5}$  &    $8\times10^{9}$   &    $3\times10^{14}$   &     $8.7\times10^{18}$  &    $1.6\times10^{23}$   &   $1.9\times10^{27}$    &     $1.5\times10^{31}$  &    $8.5\times10^{34}$   &   $1.1\times10^{38}$    &   $3.8\times10^{41}$\\

		$T_{22}$   & $a_n$    &   $1.6\times10^{-7}$    &     $5.804\times10^{0}$  &    $-6.6\times10^{5}$   &    $5.2\times10^{10}$   &     $-3.5\times10^{15}$  &    $1.2\times10^{20}$   &   $-1.5\times10^{24}$    &     $7.2\times10^{27}$  &    $-2.4\times10^{32}$   &   $-5.2\times10^{34}$    &   $-1.4\times10^{39}$\\
		& $b_n$    &   $1\times10^{0}$    &     $3.7\times10^{7}$  &    $7.5\times10^{12}$   &    $6.32\times10^{17}$   &     $2.8\times10^{22}$  &    $6.8\times10^{26}$   &   $9.5\times10^{30}$    &     $8.5\times10^{34}$  &    $5.3\times10^{38}$   &   $6.15\times10^{41}$    &   $2.5\times10^{45}$\\
		\hline
	\end{tabular}
}
	\normalsize
	\caption{Transfer function coefficients used in this paper for the PZT-5A diaphragm.}
	\label{tab:ibc_coefficients}
\end{table*}

Transformed coefficients of the transfer functions as measured by \citet{SmokerNAB_2012} are reported in \cref{tab:ibc_coefficients}. To be consistent with the convention as used in \eqref{eq:piezo_system_xqpV}, numerator coefficients of $T_{11}$ and $T_{12}$ are negative values of those reported by \citet{SmokerNAB_2012}; the resulting transfer functions and impedance are consistent with an energy-extraction regime in the mode of interest.

%% file: LinScaloHesselink_JFM_2016.bbl
\begin{thebibliography}{47}
\expandafter\ifx\csname natexlab\endcsname\relax\def\natexlab#1{#1}\fi
\def\au#1{#1} \def\ed#1{#1} \def\yr#1{#1}\def\at#1{#1}\def\jt#1{\textit{#1}}
  \def\bt#1{#1}\def\bvol#1{\textbf{#1}} \def\vol#1{#1} \def\pg#1{#1}
  \def\publ#1{#1}\def\arxiv#1{#1}\def\org#1{#1}\def\st#1{\textit{#1}}

\bibitem[{Anton} \& {Sodano}(2007)]{AntonS_SmartMaterStruct_2007}
{\sc \au{{Anton}, S.~R.} \& \au{{Sodano}, H.~A.}} \yr{2007}  \at{A review of
  power harvesting using piezoelectric materials (2003{\textendash}2006)}.
  \jt{Smart Mater. Struct.}  \bvol{16}~(3),  \pg{R1}.

\bibitem[Bermejo-Moreno {\em et~al.\/}(2013)Bermejo-Moreno, Bodart, Larsson \&
  Barney]{BermejoBLB_IEEE_2014}
{\sc \au{Bermejo-Moreno, I.}, \au{Bodart, J.}, \au{Larsson, J.} \& \au{Barney,
  B.}} \yr{2013} {Solving the compressible Navier-Stokes equations on up to
  1.97 million cores and 4.1 trillion grid points}.  \bt{In {\em IEEE
  International Conference on High Performance Computing\/}}.

\bibitem[{Ceperley}(1979)]{Ceperley_1979}
{\sc \au{{Ceperley}, P.~H.}} \yr{1979}  \at{A pistonless {{Stirling}}
  engine{\textemdash}{{The}} traveling wave heat engine}.  \jt{J. Acoust. Soc.
  Am.}  \bvol{66}~(5),  \pg{1508--1513}.

\bibitem[{Chen} {\em et~al.\/}(2010){Chen}, {Xu}, {Yao} \&
  {Shi}]{ChenXYS_NanoLett_2010}
{\sc \au{{Chen}, X.}, \au{{Xu}, S.}, \au{{Yao}, N.} \& \au{{Shi}, Y.}}
  \yr{2010}  \at{1.6 {{V Nanogenerator}} for {{Mechanical Energy Harvesting
  Using PZT Nanofibers}}}.  \jt{Nano Lett.}  \bvol{10}~(6),  \pg{2133--2137}.

\bibitem[{De-Yi} \&
  {Bu-Xuan}(1990)]{DeYiB_InternationalJournalHeatMassTransfer_1990}
{\sc \au{{De-Yi}, S.} \& \au{{Bu-Xuan}, W.}} \yr{1990}  \at{Effect of variable
  thermophysical properties on laminar free convection of gas}.
  \jt{International Journal of Heat and Mass Transfer}  \bvol{33}~(7),
  \pg{1387--1395}.

\bibitem[{Dowling} \& {Williams}(1983)]{DowlingW_1983}
{\sc \au{{Dowling}, A.~P.} \& \au{{Williams}, J. E.~F.}} \yr{1983} {\em Sound
  and {{Sources}} of {{Sound}}\/}.  \publ{{Ellis Horwood Limited}}.

\bibitem[{Feldman Jr.}(1968)]{FeldmanJr_JournalSoundVibration_1968}
{\sc \au{{Feldman Jr.}, K.~T.}} \yr{1968}  \at{Review of the literature on
  {{Sondhauss}} thermoacoustic phenomena}.  \jt{Journal of Sound and Vibration}
   \bvol{7}~(1),  \pg{71--82}.

\bibitem[{Fung} \& {Ju}(2001)]{FungJ_2001}
{\sc \au{{Fung}, K.-Y.} \& \au{{Ju}, H.}} \yr{2001}  \at{Broadband time-domain
  impedance models}.  \jt{AIAA J.}  \bvol{39}~(8),  \pg{1449--1454}.

\bibitem[{Fung} \& {Ju}(2004)]{FungJ_2004}
{\sc \au{{Fung}, K.-Y.} \& \au{{Ju}, H.}} \yr{2004}  \at{Time-domain
  {{Impedance Boundary Conditions}} for {{Computational Acoustics}} and
  {{Aeroacoustics}}}.  \jt{Int. J. Comput. Fluid Dyn.}  \bvol{18}~(6),
  \pg{503--511}.

\bibitem[Gardner \& Swift(2003)]{GardnerS_JASA_2003}
{\sc \au{Gardner, D.} \& \au{Swift, G.~W.}} \yr{2003}  \at{{A cascade
  thermoacoustic engine}}.  \jt{J. Acoust. Soc. Am.}  \bvol{114}~(4),  \pg{1905
  -- 1919}.

\bibitem[{Gedeon}(2014)]{Gedeon_2014}
{\sc \au{{Gedeon}, D.}} \yr{2014}  \at{Stirling, {{Pulse-Tube}} and {{Low-T
  Cooler Model Classes}}} .

\bibitem[Ham {\em et~al.\/}(2007)Ham, Mattsson, Iaccarino \&
  Moin]{HamMIM_2007_bookchpt}
{\sc \au{Ham, F.}, \au{Mattsson, K.}, \au{Iaccarino, G.} \& \au{Moin, P.}}
  \yr{2007} {\em {Towards Time-Stable and Accurate LES on Unstructured
  Grids}\/},  \st{{Lecture Notes in Computational Science and Engineering}},
  \vol{vol.~56},  \pg{pp. 235 -- 249}.  \publ{Springer Berlin Heidelberg}.

\bibitem[{Hartley}(1951)]{Hartley_1951}
{\sc \au{{Hartley}, R. V.~L.}} \yr{1951} Electric power source. U.S.
  Classification 290/1.00R, 333/141, 60/39.77, 116/137.00A, 322/3, 116/DIG.220;
  International Classification F03G7/00, H02N11/00; Cooperative Classification
  F03G7/002, H02N11/002, Y10S116/22; European Classification H02N11/00B,
  F03G7/00B.

\bibitem[{Idelchik}(2003)]{Idelchik_2003}
{\sc \au{{Idelchik}, I.~E.}} \yr{2003} {\em Handbook of hydraulic
  resistance\/}, 3rd edn.  \publ{Boca Raton, FL: {CRC Press}}.

\bibitem[Jensen {\em et~al.\/}(1989)Jensen, Sumer \& Freds{\o}e]{Jensen1989JFM}
{\sc \au{Jensen, B.~L.}, \au{Sumer, B.~M.} \& \au{Freds{\o}e, J.}} \yr{1989}
  \at{Turbulent oscillatory boundary layers at high reynolds numbers}.  \jt{J.
  Fluid Mech.}  \bvol{206},  \pg{265--297}.

\bibitem[Kirchhoff(1868)]{Kirchhoff_PoggAnn_1868}
{\sc \au{Kirchhoff, G.}} \yr{1868}  \at{{{\"U}ber den Einfluss der
  W{\"a}rmeleitung in einem Gase auf die Schallbewegung}}.  \jt{Pogg. Ann.}
  \bvol{134},  \pg{177 -- 193}.

\bibitem[Kramers(1949)]{Kramers_Physica_1949}
{\sc \au{Kramers, H.~A.}} \yr{1949}  \at{{ Vibrations of a Gas Column}}.
  \jt{Physica}  \bvol{15}~(971),  \pg{971 -- 984}.

\bibitem[{Marrison}(1958)]{Marrison_1958}
{\sc \au{{Marrison}, W.~A.}} \yr{1958} Heat-controlled acoustic wave system.
  U.S. Classification 60/516, 116/137.00R, 340/384.7, 116/DIG.220, 310/27,
  290/1.00R, 116/137.00A, 60/531, 310/306; International Classification
  G08B17/04, F25B9/14, F03G7/00, H02N11/00; Cooperative Classification
  F03G7/002, F25B2309/1407, F02G2243/52, H02N11/002, G08B17/04, F25B2309/1403,
  F25B9/145, Y10S116/22; European Classification F25B9/14B, G08B17/04,
  H02N11/00B, F03G7/00B.

\bibitem[{Matveev} {\em et~al.\/}(2007){Matveev}, {Wekin}, {Richards} \&
  {Shafrei-Tehrany}]{MatveevWRS_2007}
{\sc \au{{Matveev}, K.~I.}, \au{{Wekin}, A.}, \au{{Richards}, C.~D.} \&
  \au{{Shafrei-Tehrany}, N.}} \yr{2007}  \at{On the {{Coupling Between
  Standing-Wave Thermoacoustic Engine}} and {{Piezoelectric Transducer}}}
  \pg{pp. 765--769}.

\bibitem[M{\"u}ller \& Rott(1983)]{MullerR_ZAMP_1983}
{\sc \au{M{\"u}ller, U.~A.} \& \au{Rott, N.}} \yr{1983}  \at{{Thermally driven
  acoustic oscillations, part VI: Excitation and power}}.  \jt{Z. Angew. Math.
  Phys.}  \bvol{34},  \pg{609 -- 626}.

\bibitem[{Nouh} {\em et~al.\/}(2014){Nouh}, {Aldraihem} \& {Baz}]{NouhAB_2014}
{\sc \au{{Nouh}, M.}, \au{{Aldraihem}, O.} \& \au{{Baz}, A.}} \yr{2014}
  \at{Transient characteristics and stability analysis of standing wave
  thermoacoustic-piezoelectric harvesters}.  \jt{J. Acoust. Soc. Am.}
  \bvol{135}~(2),  \pg{669--678}.

\bibitem[{Priya}(2007)]{Priya_JElectroceram_2007}
{\sc \au{{Priya}, S.}} \yr{2007}  \at{Advances in energy harvesting using low
  profile piezoelectric transducers}.  \jt{J Electroceram}  \bvol{19}~(1),
  \pg{167--184}.

\bibitem[{Rayleigh}(1878)]{Rayleigh_Nature_1878}
{\sc \au{{Rayleigh}}} \yr{1878}  \at{The {{Explanation}} of {{Certain
  Acoustical Phenomena}}}.  \jt{Nature}  \bvol{18},  \pg{319--321}.

\bibitem[Rienstra(2006)]{Rienstra_AIAA_2006}
{\sc \au{Rienstra, S.~W.}} \yr{2006} {Impedance Models in Time Domain,
  including the Extended Helmholtz Resonator Model}.  \bt{In {\em 12th
  AIAA/CEAS Aeroacoustics Conference\/}}.

\bibitem[{Rijke}(1859)]{Rijke_1859}
{\sc \au{{Rijke}, P.~L.}} \yr{1859}  \at{{{LXXI}}. {{Notice}} of a new method
  of causing a vibration of the air contained in a tube open at both ends}.
  \jt{Philos. Mag. Ser. 4}  \bvol{17}~(116),  \pg{419--422}.

\bibitem[Rott(1969)]{Rott_ZAMP_1969}
{\sc \au{Rott, N.}} \yr{1969}  \at{{Damped and Thermally Driven Acoustic
  Oscillations in Wide and Narrow Tubes}}.  \jt{Z. Angew. Math. Phys.}
  \bvol{20},  \pg{230 -- 243}.

\bibitem[Rott(1973)]{Rott_ZAMP_1973}
{\sc \au{Rott, N.}} \yr{1973}  \at{{Thermally driven acoustic oscillations,
  part II: Stability limit for helium}}.  \jt{Z. Angew. Math. Phys.}
  \bvol{24},  \pg{54 -- 72}.

\bibitem[Rott(1974)]{Rott_ZAMP_1974}
{\sc \au{Rott, N.}} \yr{1974}  \at{{The influence of heat conduction on
  acoustic streaming}}.  \jt{Z. Angew. Math. Phys.}  \bvol{25},  \pg{417 --
  421}.

\bibitem[Rott(1975)]{Rott_ZAMP_1975}
{\sc \au{Rott, N.}} \yr{1975}  \at{{Thermally driven acoustic oscillations,
  part III: Second-order heat flux}}.  \jt{Z. Angew. Math. Phys.}  \bvol{26},
  \pg{43 -- 49}.

\bibitem[Rott(1976)]{Rott_NZZ_1976}
{\sc \au{Rott, N.}} \yr{1976}  \at{{Ein 'Rudimentarer' Stirlingmotor}}.
  \jt{Neue Zurecher Ztg.}  \bvol{197}~(210).

\bibitem[Rott(1980)]{Rott_1980_AdvApplMech}
{\sc \au{Rott, N.}} \yr{1980}  \at{{Thermoacoustics}}.  \jt{Adv. Appl. Mech.}
  \bvol{20},  \pg{135--175}.

\bibitem[Rott(1984)]{Rott_JFM_1984}
{\sc \au{Rott, N.}} \yr{1984}  \at{{Thermoacoustic heating at the closed end of
  an oscillating gas column}}.  \jt{J. Fluid Mech.}  \bvol{145},  \pg{1 -- 9}.

\bibitem[Rott \& Zouzoulas(1976)]{Rott_ZAMP_1976}
{\sc \au{Rott, N.} \& \au{Zouzoulas, G.}} \yr{1976}  \at{{Thermally driven
  acoustic oscillations, part IV: Tubes with variable cross-section}}.  \jt{Z.
  Angew. Math. Phys.}  \bvol{27},  \pg{197 -- 224}.

\bibitem[Scalo {\em et~al.\/}(2015{\natexlab{{\em a\/}}})Scalo, Bodart \&
  Lele]{ScaloBL_PoF_2015}
{\sc \au{Scalo, C.}, \au{Bodart, J.} \& \au{Lele, S.~K.}}
  \yr{2015{\natexlab{{\em a\/}}}}  \at{Compressible turbulent channel flow with
  impedance boundary conditions}.  \jt{Phys. Fluids}  \bvol{27}~(035107).

\bibitem[Scalo {\em et~al.\/}(2015{\natexlab{{\em b\/}}})Scalo, Lele \&
  Hesselink]{ScaloLH_JFM_2015}
{\sc \au{Scalo, C.}, \au{Lele, S.~K.} \& \au{Hesselink, L.}}
  \yr{2015{\natexlab{{\em b\/}}}}  \at{{Linear and Nonlinear Modeling of a
  Theoretical Traveling-Wave Thermoacoustic Heat Engine}}.  \jt{J. Fuid Mech.}
  \bvol{766},  \pg{368 -- 404}.

\bibitem[Scalo {\em et~al.\/}(2013)Scalo, Piomelli \& Boegman]{ScaloJFM2013}
{\sc \au{Scalo, C.}, \au{Piomelli, U.} \& \au{Boegman, L.}} \yr{2013}
  \at{Self-similar decay of dissolved oxygen concentration in an oscillating
  boundary layer in the intermittently turbulent regime}.  \jt{J. Fluid Mech.}
  \bvol{726},  \pg{338--370}.

\bibitem[{Smoker} {\em et~al.\/}(2012){Smoker}, {Nouh}, {Aldraihem} \&
  {Baz}]{SmokerNAB_2012}
{\sc \au{{Smoker}, J.}, \au{{Nouh}, M.}, \au{{Aldraihem}, O.} \& \au{{Baz},
  A.}} \yr{2012}  \at{Energy harvesting from a standing wave
  thermoacoustic-piezoelectric resonator}.  \jt{J. Appl. Phys.}
  \bvol{111}~(10),  \pg{104901}.

\bibitem[{Sondhauss}(1850)]{Sondhauss_AnnPhys_1850}
{\sc \au{{Sondhauss}, C.}} \yr{1850}  \at{Ueber die {{Schallschwingungen}} der
  {{Luft}} in erhitzten {{Glasr{\"o}hren}} und in gedeckten {{Pfeifen}} von
  ungleicher {{Weite}}}.  \jt{Ann. Phys.}  \bvol{155}~(1),  \pg{1--34}.

\bibitem[{Swift}(1988)]{Swift_1988}
{\sc \au{{Swift}, G.~W.}} \yr{1988}  \at{Thermoacoustic engines}.  \jt{J.
  Acoust. Soc. Am.}  \bvol{84}~(4),  \pg{1145--1180}.

\bibitem[{Swift}(1992)]{Swift_1992}
{\sc \au{{Swift}, G.~W.}} \yr{1992}  \at{Analysis and performance of a large
  thermoacoustic engine}.  \jt{J. Acoust. Soc. Am.}  \bvol{92}~(3),
  \pg{1551--1563}.

\bibitem[{Swift}(2002)]{Swift_2002}
{\sc \au{{Swift}, G.~W.}} \yr{2002} {\em Thermoacoustics: a unifying
  perspective for some engines and refrigerators\/}.  \publ{Melville, NY:
  {Acoustical Society of America through the American Institute of Physics}}.

\bibitem[{Tam} \& {Auriault}(1996)]{TamA_1996}
{\sc \au{{Tam}, C. K.~W.} \& \au{{Auriault}, L.}} \yr{1996}  \at{Time-domain
  impedance boundary conditions for computational aeroacoustics}.  \jt{AIAA J.}
   \bvol{34}~(5),  \pg{917--923}.

\bibitem[{Tijani} \& {Spoelstra}(2011)]{TijaniS_2011}
{\sc \au{{Tijani}, M. E.~H.} \& \au{{Spoelstra}, S.}} \yr{2011}  \at{A high
  performance thermoacoustic engine}.  \jt{J. Appl. Phys.}  \bvol{110},
  \pg{093519}, 9.

\bibitem[{Ward} {\em et~al.\/}(2012){Ward}, {Clark} \& {Swift}]{WardCS_2012}
{\sc \au{{Ward}, B.}, \au{{Clark}, J.} \& \au{{Swift}, G.}} \yr{2012} {\em
  Design {{Environment}} for {{Low}}-amplitude {{Thermoacoustic Energy
  Conversion}}: {{Users Guide}}\/}.

\bibitem[{Yazaki} {\em et~al.\/}(1998){Yazaki}, {Iwata}, {Maekawa} \&
  {Tominaga}]{YazakiIMT_PhysRevLett_1998}
{\sc \au{{Yazaki}, T.}, \au{{Iwata}, A.}, \au{{Maekawa}, T.} \& \au{{Tominaga},
  A.}} \yr{1998}  \at{Traveling {{Wave Thermoacoustic Engine}} in a {{Looped
  Tube}}}.  \jt{Phys. Rev. Lett.}  \bvol{81}~(15),  \pg{3128--3131}.

\bibitem[{Yu} {\em et~al.\/}(2012){Yu}, {Jaworski} \&
  {Backhaus}]{YuJB_AppliedEnergy_2012}
{\sc \au{{Yu}, Z.}, \au{{Jaworski}, A.~J.} \& \au{{Backhaus}, S.}} \yr{2012}
  \at{Travelling-wave thermoacoustic electricity generator using an
  ultra-compliant alternator for utilization of low-grade thermal energy}.
  \jt{Applied Energy}  \bvol{99},  \pg{135--145}.

\bibitem[Zouzoulas \& Rott(1976)]{ZouzoulasR_ZAMP_1976}
{\sc \au{Zouzoulas, G.} \& \au{Rott, N.}} \yr{1976}  \at{{Thermally driven
  acoustic oscillations, part V: Gas-liquid oscillations}}.  \jt{Z. Angew.
  Math. Phys.}  \bvol{27},  \pg{325 -- 334}.

\end{thebibliography}
